\documentclass[aps,prb,twocolumn,showpacs,amsmath,amssymb,superscriptaddress,floatfix,nofootinbib]{revtex4-2}
\usepackage{amssymb}
\usepackage{amsbsy}
\usepackage{amsmath}
\usepackage{mathtools}
\usepackage{mathrsfs}
\usepackage{epsfig}
\usepackage{graphicx}
\usepackage[notcite,notref,final]{showkeys}
\usepackage{xcolor}
\usepackage{braket}
\usepackage{hhline}
\usepackage{dsfont}
\usepackage{comment}
\usepackage{bm}
\usepackage{hyperref}
\usepackage[justification=raggedright,font=small]{caption}
\usepackage[titletoc,toc,title]{appendix}
\usepackage[normalem]{ulem}
\usepackage{cleveref}
\hypersetup{
	pdffitwindow=false,            % window fit to page
	pdfstartview={Fit},            % fits width of page to window
	colorlinks=true,               % coloured links, not boxed
	linkcolor=blue,      % colour of internal links
	citecolor=blue,       % colour of links to bibliography
	urlcolor=magenta           % colour of external links
}
\newcommand{\tr}{\text{Tr}}
\newcommand{\eq}[1]{\begin{equation} #1 \end{equation}}
\newcommand{\eqsp}[1]{\begin{split} #1 \end{split}}
\newcommand{\sket}{\rangle \rangle }
\newcommand{\sbra}{\langle \langle}

\begin{document}
%\preprint{}
\title{Open-system dynamics in local Lindbladians with chaotic spectra}

\author{Sanket Chirame}
\email{chira012@umn.edu}
\author{Fiona J. Burnell}
\affiliation{School of Physics and Astronomy, University of Minnesota, Minneapolis, Minnesota 55455, USA}
\date{\today}
\begin{abstract}
We investigate the physical consequences of having a spectrum that satisfies random matrix theory (RMT) for generic Lindbladians, and compare its implications for spatially local and completely random Lindblad dynamics in one spatial dimension. We find that Lindbladians whose spectrum is described by RMT exhibit quasiuniversal early-time dynamics for quantities nonlinear in the density matrix, in the sense that for generic, highly entangled initial states, the early time evolution is independent of the choice of initial state. We numerically investigate how locality generically imposes constraints on the size-dependence of Lindblad eigenoperators. This size dependence implies that linear observables, such as expectation values of local operators, are highly sensitive to eigenmodes outside the bulk of the spectrum in the thermodynamic limit, and plays a central role in limiting operator growth in the presence of dissipation. We find that when single-site dissipation dominates, an operator's decoherence scales approximately linearly with its Pauli weight, even in the presence of two-site jump operators. When two-site only dissipation dominates, however, this generic trend in operator size can be violated for numerically accessible system sizes, leading to long-lived high Pauli-weight operators.
\end{abstract}
\maketitle

\section{Introduction}
In closed quantum systems, random matrix theory~\cite{wigner} (RMT) has long been known to be intimately connected to the emergence of equilibrium quantum statistical mechanics from the microscopic reversible unitary dynamics.   Notably, the Eigenstate Thermalization Hypothesis (ETH) \cite{bohigas1984bgsconjecture}, which predicts that the finite energy-density eigenstates of generic Hamiltonians look thermal with respect to local observables \cite{deutsch1991eth,srednicki1994eth,srednicki1999}, rests on the observation that the spectra of generic Hamiltonians strongly resemble those of random Hermitian matrices of the correct symmetry.  This resemblance to random matrices is also reflected in the structure of eigenvectors at finite energy density, and explains how local observables and their correlation functions rapidly thermalize in typical closed quantum many-body systems \cite{rigol2008thermalization,rigolETHreview}.   RMT-like fluctuations in the spectrum are also linked to the emergence of quantum chaos \cite{bohigas1984bgsconjecture,mullerAltland2004semiclassicalChaos,kosProsen2018analyticRMT}.

The ETH is a powerful tool for describing the behavior of generic interacting quantum systems at finite energy density, which reduces an extremely complex ergodic dynamics to a set of generic predictions about how local operators thermalize. Moreover, understanding the connection between generic Hamiltonians and RMT has also provided a base point from which to explore non-thermal dynamical phases of matter, including many-body localized phases \cite{abanin2019mblReview,basko2006metalInsulator,nandkishoreHuse2015mblReview}, integrable systems \cite{vidmarRigol2016GGEintegrable}, and quantum many-body scars \cite{turner2018manyBodyScars,moudgalya2022scarsReview,chandran2023scarsReview,odea2020scarsLieAlgebra}.

Rapid progress in platforms for quantum computing \cite{fossYao2024trappedIonReview,jiang2025superconductQubitAdvance,cornish2024ultraColdReview} has made developing a similar understanding of the dynamics of open quantum systems a problem of great interest. It is natural to ask whether there are open quantum systems where a similar application of random matrix theory can be used to extract simple physical conclusions from complex many-body dynamics.  The obvious target for this approach is systems governed by Lindblad dynamics \cite{lindblad1976,gorini1976,breuer2002theory}: as for Hamiltonian dynamics, in these systems the time evolution is described by exponentiating an appropriate non-Hermitian operator, whose spectrum and eigenvectors thus dictate the time evolution.  In the present work, we seek to address the question of what random matrix theory can tell us about the dynamics of systems undergoing generic, local Lindbladian dynamics. We specifically focus on the RMT the complex-valued eigenspectrum of the Lindbladian that controls the transient dynamics of the open systems. However, the late-time steady-states of the open systems can also exhibit dissipative chaos and transitions to non-trivial steady phases, independent of the RMT nature of the full spectrum \cite{prosen2013steadyStateLevelSpacing,richterSa2025integrableNESS,rufo2025semiclassical,sa2020spectral}.

A first step in connecting Lindblad dynamics to RMT was made through GHS conjecture, which states that the spectra of chaotic open quantum systems exhibit random matrix correlations from the non-Hermitian Ginibre ensembles \cite{grobe1988quantum,grobe1989cubic}. This conjecture was further substantiated in numerical investigations of a variety of open many-body quantum systems \cite{akemann2019}. These observations were followed by systematic classification of the possible universality classes based on the symmetry of dynamics, both for many-body Lindbladians \cite{altland2021fermionLindbladSymmetry,lieu2020tenFold,sa2023symmetry,kawabata2023symmetryLindblad}, and non-Hermitian Hamiltonians\cite{hamazaki2020nonhermitianUniversality} (which describe a particular subset of trajectories of Lindblad dynamics, but are not completely positive trace-preserving maps). These findings have led to a wide range of activity in the burgeoning field of dissipative quantum chaos \cite{li2021dsff,sa2020csr,sa2020randomKraus,sa2020spectral,peyruchat2025landau,ferrari2025trajectories,villasenor2024breakdownCSR,villasenor2024opendicke,mondal2025transient,rubio2022integrability,almeida2025openETH,ferrari2025openStripeETH,hamazaki2019nonHermitianMBL,tzortzakakis2020nonHermitianDisroder,huang2020andersonNonHermitian}.\\

The identification of RMT-like spectra exhibiting level repulsion in generic Lindbladians has inspired a number of developments in our understanding of Lindbladian dynamics.  For example, connections to classical dissipative chaos have been made via an analog of the Berry conjecture for open systems \cite{li2025opensystemBerryConjecture,mondal2025transient,villasenor2024breakdownCSR}. Time evolution of the purity in ensembles of (non-local) Ginibre matrices has been shown to admit a simple description characterized by the Lindblad superoperator density \cite{campo2024purity}.  Moreover, several interesting examples of Lindbladians that fail to behave generically have been investigated, including systems that exhibit Lindbladian analogs of many-body localization \cite{hamazaki2022mbl}, scars \cite{garciaVerbaarschot2025lindbladScars}, and integrability \cite{rubio2022integrability}.

However, the implications of RMT spectra on the dynamics of systems undergoing Lindbladian evolution remain relatively underexplored, particularly in the case where the Lindbladian is spatially local.  It is clear that the resulting physics is fundamentally different than ETH, which predicts (among other things) that local observables will thermalize at long times.  In Lindbladian dynamics, the long-time behavior is dictated by the steady state and the longest-lived eigenvectors; in typical local Lindbladians, these are located far from the bulk of the spectrum and thus not governed by RMT.  Moreover, there are indications that the eigenoperators of random Lindbladians are generically not random operators: For example, Ref. \cite{wang2020hierarchy} studied the structure of eigenoperators for both totally random and random $k$-local Lindbladians in the limit of very strong dissipation (for which the Lindbladian spectra fragment into discrete sectors of differing decay rates), finding that locality imposes stringent constraints on their size. These stark contrasts with the ETH as it applies to closed systems raise basic questions: which physical quantities are sensitive to RMT statistics in the bulk of the Lindblad spectrum? Does the resulting dynamics ever admits a simple description in terms of a modified, ETH-like ansatz?  What is the generic size structure of Lindbladian eigenoperators, and how does it dictate properties of physical interest, such as the dynamics of local operators?

This work presents a detailed analysis connecting generic properties of the spectra and eigenvectors of local Lindbladians to their implications for the resulting dynamics.  
Our key results are as follows: First, we identify a generic size-dependence of Lindbladian eigen-operators.  This size dependence implies that the dynamics of local operators is controlled predominantly by eigenmodes outside the bulk of the spectrum, while non-local observables such as the purity and nonlinear correlators display universal short-time dynamics controlled by the bulk eigenmodes. Second, we explore the implications of this size dependence on operator growth and spreading, complementing numerous other approaches \cite{schuster2023operator,weinstein2023radiativeRUC,zhangHunagChen2019} to this problem. Unlike much of the existing literature, we focus on models with non-Hermitian jump operators (or non-unital time-evolution), and hence do not flow to an infinite temperature steady state.  This allows us to sharply distinguish between the time evolution of operators and states through the different size distributions of left and right eigenoperators. The evidence presented here is primarily numerical, but where possible, is supported by simplified theoretical models and finite-size scaling considerations, which allow us to infer which features of the behavior at finite system sizes can be expected to remain in the thermodynamic limit. 

In more detail, we analyze the model-independent structure of the eigenoperators of a family of generic, spatially local Lindbladians. Several recent works \cite{hamazaki2022mbl,almeida2025openETH,ferrari2025openStripeETH} have focused on characterizing the distributions of overlaps between eigenvectors and local operators. Here, we explore operator overlaps with Pauli strings of all sizes.  We find strong correlations between an eigenoperator's decay rate and its overlap with Pauli strings of different weights, with faster-decaying eigenmodes in the bulk of the spectrum heavily concentrated on high-weight Pauli strings. Because of this strong size dependence, eigenoperators of local Lindbladians in the bulk of the spectrum are {\it not} random in the Pauli string basis.  
However, we show numerically that within a given size sector, the bulk eigenoperators exhibit near-maximal scrambling over Pauli strings, and in this sense are as random as possible given the size constraints.  
This allows us to identify universal early-time dynamics associated with the bulk of the spectrum: non-linear functions of the initial density matrix, such as purity and non-linear correlators, decay with a universal rate independent of the initial state. We attribute this universality to the delocalization of generic initial states over the bulk part of the Lindblad spectrum.  We also point out that because the initial rate of decoherence of a pure state increases with increasing system size \cite{campo2019extreme}, the time window for which this generic dynamics occurs vanishes in the large system size limit. 

We also present a detailed analysis of how this size dependence controls dissipative aspects of operator growth and spreading, such as the decay of an operator's norm and the increase of its trace.  Local operators overwhelmingly overlap with slow-decaying eigenmodes outside of the bulk of the spectrum, as they must in order for their decay rates to remain independent of system size.  We highlight that the distribution of slow-decaying left eigenvectors over Pauli strings controls the long-time deviations of operators from their steady state values, and find that when dissipation is dominated by single-site terms, these left eigenoperators are dominated by short Pauli strings. We also find that when dissipation is dominated by two-site terms,  in some models at the finite system sizes studied here, the slowest decaying eigenoperators are predominantly composed of Pauli strings of weight comparable to the system size. We find that the spectral gap of these high-weight, slowly decaying eigenoperators increases with system size (whereas that of short eigenoperators remains approximately constant), indicating that fine-tuning is required for this effect to persist in the thermodynamic limit.

We note that Lindblad dynamics is limited in scope, and captures only a small set of the possible dynamical regimes of open quantum systems.  Nevertheless, Lindbladians have the advantage that they always generate a physically meaningful, probability-conserving time evolution, and are uniquely amenable to the kind of RMT analysis we apply here.
The traditional derivation of the Lindblad equation from microscopic interactions between a system and a bath generally involves a set of approximations that are highly suspect in many-body systems, and lead to non-local Lindbladians.  While methods do exist to address these deficits \cite{rudner2020universalLindblad,shiraishi2025manybodyLindblad}, the Lindbladians we study are most appropriately interpreted as short-time, weak-coupling evolutions of quantum channels \cite{alickiLendi2011book}, which naturally leads to the kind of local Lindbladian dynamics we consider here.

The rest of the paper is organized as follows: In section ~\ref{sec:background}, we review the relevant details of Lindbladians and open system dynamics. In section~\ref{sec:eigenvalues}, we numerically verify that the eigenvalue spectrum of models considered here exhibits the RMT correlations. In section~\ref{sec:eigenvecs}, we show that the size distribution of the eigenoperators is strongly correlated with their eigenvalues and present a detailed analysis of a non-interacting toy model, which qualitatively captures this dependence. Having identified the generic spectral features, we discuss the universal early time decoherence of initial states in section~\ref{sec:state_dynamics}. In section~\ref{sec:operator_dynamics}, we present the results on the implications of local dissipation for the operator dynamics. Finally, in section~\ref{sec:conclusions}, we summarize the conclusions and discuss future directions based on this work. Additional numerical results on eigenvalue statistics, non-local Lindblad models, the matrix elements of superoperators, initial state matrix elements, and anomalous operator growth are relegated to appendices.

%% =================================================
\section{Background: Lindbladian Systems}
\label{sec:background}
We begin with a brief overview of essential properties of the Lindbladians and the time evolution equations for open system dynamics. This is followed by the discussion of the complex spacing ratio, which is the metric that we will use to diagnose the non-Hermitian universality in the spectrum of generic Lindbladians. Along the way, we introduce the superoperator notation that will be used throughout this paper. 

\subsection{Lindbladian superoperators}
\label{sec:lindblad}
Let us consider a quantum system represented by a density matrix $\rho$. The most general evolution that maps this state to a new valid density matrix can be written as
\eq{\label{eq:cptp}\rho\rightarrow\mathcal{E}[\rho] = \sum_{a}K_a \rho K_a^\dagger,}
where $\mathcal{E}$ is a completely positive trace preserving (CPTP) map and $K_a$ are Kraus operators that furnish its operator sum representation. The Kraus operators obey normalization condition $\sum_{a}K_a^\dagger K_a=1$, which ensures that the trace of the density matrix remains unchanged. Such evolution is a result of entangling the system $\rho$ with an auxiliary quantum system using some unitary operation followed by tracing out the auxiliary degrees of freedom. These CPTP maps naturally describe the quantum dynamics of monitored quantum circuits, where unitary gates are combined with measurements and results are averaged over measurement outcomes. If the applied map is close to the identity map in a given time step, then one can formulate a time-continuous master equation for the dynamics. This can arise, for example, when the applied unitary operations are the result of short-time Hamiltonian evolution, and if the measurements only weakly disturb the state of the system. Such a situation can be represented in terms of Kraus operators that have the following form:
\eq{K_0 = \mathbb{I} + dt \ L_0 ,\qquad K_{a>0} = \sqrt{2 dt}L_{a} }
such that $\rho\rightarrow \rho$ in the limit $dt \rightarrow 0$. The normalization condition $\sum_{a}K_a^\dagger K_
a=1$ is satisfied up to order $dt$, if we impose the condition
\eq{L_0 + L_0^\dagger = -2\sum_{\alpha>0}L_\alpha^\dagger L_\alpha.}
The general form of an operator that obeys this condition is given by $L_0 = -i H - \sum_{a>0}L_a^\dagger L_a$, where $H$ is a Hermitian operator. Hence, the state of the system after time $dt$ in terms of the redefined Kraus operators is obtained using Eq.~\eqref{eq:cptp} as
\eq{\eqsp{\rho(t+dt) = \rho(t) + &dt(-i[H,\rho] + \\
&\sum_a (2 L_a\rho L_a^\dagger - \{L_a^\dagger L_a, \rho\}) ) + \mathcal{O}(dt^2).}}
If the applied dynamical map is time-independent and the auxiliary system acts as a Markovian environment, then the state $\rho(t)$ of the system evolves in time according to the Lindblad master equation \cite{gorini1976,lindblad1976}
\eq{\label{eq:lindblad}\frac{d}{dt}\rho(t) = -i[H,\rho] + \sum_a 2 L_a\rho L_a^\dagger - \{L_a^\dagger L_a, \rho\} :=\mathcal{L}[\rho(t)]}
Here $H$ is a Hermitian Hamiltonian operator that describes the unitary evolution and $L_a$ are dissipative jump operators, which can in general be non-Hermitian. The Lindbladian $\mathcal{L}$ is a superoperator, meaning that it acts on the operators and transforms them into a different set of operators. The general solution of Eq.~\eqref{eq:lindblad} for a time-independent Lindbladian and initial state $\rho_0$ can be written as $\rho(t)=e^{\mathcal{L}t}\rho_0$.

The jump operators $L_a$ presented above are general operators without any constraints. However, in this work, we will consider Lindbladians where both the Hamiltonian and jump operators are spatially local. This is an appropriate model for errors in gate-based quantum computers \cite{nielsenChuang}. Here, the gates act on spatially local regions, and the jump operators correspond to the errors in their operation, which naturally have spatially local support. A more conventional derivation in terms of the system-bath coupling results in non-local Lindbladians when applied to many-body systems. However, alternative approximation methods that result in purely local jump operators have also been formulated recently \cite{rudner2020universalLindblad,shiraishi2025manybodyLindblad}.

Since the dynamics is generated by a superoperator, it is convenient to introduce a notation where operators form a Hilbert space endowed with the Hilbert-Schmidt inner product. Let $\mathcal{H}$ be the Hilbert space with $\text{dim}(\mathcal{H})=\mathcal{D}$. Then an $\mathcal{D}\times\mathcal{D}$ operator $A$ acting on this space is a $\mathcal{D}^2$ dimensional column-vector, represented as a superket $|A\sket$. The inner product of two operators is given by the Hilbert-Schmidt product
\eq{\sbra A|B\rangle\rangle = \tr{(A^\dagger B)},}
where a superbra $\sbra A|$ is a row vector. A general transformation such as right and left multiplication can be equivalently written as
\eq{\label{eq:opvec} O_1AO_2 \longleftrightarrow (O_1\otimes O_2^T)|A\sket.}

Using this representation of the operator space, the time evolution superoperator $e^{\mathcal{L}t}$ can be studied by expressing it in terms of the eigenoperators of the generator, similar to the study of Hamiltonian dynamics in terms of its stationary energy eigenstates. We proceed by writing the Lindblad superoperator in the matrix representation as
\eq{\label{eq:matrep}\eqsp{ \mathcal{L} = &-i(H\otimes\mathbb{I} - \mathbb{I}\otimes{H^T}) \\&+\sum_{a} [\ 2L_a\otimes L_a^* -L_a^\dagger L_a\otimes\mathbb{I} - \mathbb{I}\otimes(L_a^\dagger L_a)^T],}}
where we have used Eq.~\eqref{eq:opvec} to represent the left and right operator products.

Here, we briefly comment on algebraic properties of this superoperator relevant for our work (see \cite{ueda2020review} for additional details on non-Hermitian matrices). In general, $\mathcal{L}$ is a non-normal matrix, meaning $\mathcal{L}^\dagger\mathcal{L}\neq \mathcal{L}\mathcal{L}^\dagger$, and hence can not be diagonalized by a unitary transformation. Instead, it satisfies the eigenvalue equation given by \cite{ueda2020review}
\eq{\mathcal{L}|r_j\sket = \lambda_j|r_j\sket ,\qquad \sbra l_j| \mathcal{L} = \sbra l_j| \lambda_j}
where $\lambda_j$ are complex eigenvalues with distinct right ($r_j$) and left ($l_j$) eigenoperators. In the absence of any strong symmetries, we expect the eigenvalues $\lambda_j$ to be distinct, and hence the Lindbladian can be diagonalized by a general similarity transformation. As a result, eigenoperators obey mutual bi-orthogonality condition,
\eq{\label{eq:bi-orthogonal}\sbra l_j | r_k \sket = \delta_{j,k}\frac{1}{\alpha_k},}
where $\alpha_k$ are in general complex coefficients. For a Hermitian matrix, one typically chooses $\alpha_k = 1$.  Here, however, we will normalize these eigenoperators such that $\sbra l_k | l_k \sket = \sbra r_k | r_k \sket =1$. Since for non-Hermitian $\mathcal{L}$ the right and left eigenvectors are not equal, this choice forces $\alpha_k \neq 1$ in general \cite{ueda2020review}. Moreover, the non-normal nature of $\mathcal{L}$ generally implies that the both $\sbra r_j|r_k\sket$ and $\sbra l_j|l_k\sket$ are non-vanishing for $j\neq k$.

The identity super-operator acting on this space can be resolved as $\mathcal{I} = \sum_{j=0}^{\mathcal{D}^2-1}\alpha_j|r_j\sket\sbra l_j|.$
We can use these relations to write the formal solution of the master equation~\eqref{eq:lindblad} as 
\eq{\label{eq:dyn} |\rho(t)\sket = e^{t\mathcal{L}}|\rho(t=0)\sket=\sum_{j=0}^{\mathcal{D}^2-1}e^{\lambda_j t}\alpha_j|r_j\sket\sbra l_j|\rho(t=0)\rangle\rangle.}
The dynamics generated by this equation is trace preserving and hence there is always at least one eigenvalue $\lambda=0$. The corresponding right eigenoperator $r_0$ is proportional to the steady state of the dynamics.  

The Lindblad superoperator preserves the hermiticity of the operators, i.e. $\mathcal{L}(O)^\dagger=\mathcal{L}(O^\dagger)$. This implies that both $\lambda$ and its complex conjugate $\lambda^*$ are the eigenvalues with eigenoperators $r$ and $r^\dagger$ respectively. In the case of the unitary evolution (where $L_a=0$), the real parts of the eigenvalues are strictly zero, and the time evolution only leads to the modulation of relative phases among the energy eigenstates. In general, however, when $L_{a}$ are non-zero, Lindbladian evolution causes pure states to decay to mixed states. This change in the purity results from the negative real parts of the eigenvalues of $\mathcal{L}$. 

The equivalent master equation for the time evolution of an observable $A$ in the Heisenberg picture can be written in terms of the adjoint Lindbladian 
\eq{\label{eq:heisen} \frac{d}{dt}\hat{A}(t) = i[H,\hat{A}] + \sum_a 2 L_a^\dagger\hat{A} L_a - \{L_a^\dagger L_a, \hat{A} \}:=\mathcal{L}^\dagger[A].}
The formal solution of the operator evolution equation can be similarly expressed in terms of the eigendecomposition as 
\eq{\label{eq:op-dyn} A(t) :=e^{t\mathcal{L}^\dagger}[A] = \sum_{j=0}^{\mathcal{D}^2-1} e^{\lambda^*_j t}\alpha^*_j|l_j\sket\sbra r_j|A_0\sket,}
where $A_0$ is operator at initial time $t=0$. From Eq.~\eqref{eq:heisen}, we observe that $\mathcal{L}^\dagger[\mathbb{I}]=0$ and hence the identity operator remains unchanged in the Heisenberg picture. As a result, the left eigenoperator $l_0$ corresponding to the steady state eigenvalue $\lambda=0$ is always the identity operator. This is the result of the trace-preserving nature of Lindblad dynamics, because the trace of the density matrix at time $t$ is nothing but the expectation value of the time-evolved identity operator in the Heisenberg picture.

%%=============================================================================
\subsection{Random matrix theory of Lindbladian systems}
\label{sec:CSR}
We conclude this overview of Lindbladians with a discussion of universal eigenvalue repulsion observed in the random non-Hermitian matrices.

For open quantum systems, generic Lindbladians have been conjectured to be statistically similar to the matrices from the complex Ginibre ensemble (GinUE) \cite{grobe1988quantum}. This is an ensemble of non-Hermitian random matrices, where each matrix element is a complex random variable independently sampled from the standard normal distribution. The statistical correlations in the spectrum can be quantified using the complex spacing ratios (CSR) introduced in \cite{sa2020csr}. This quantity is an extension of level spacing ratios \cite{oganesyan2007levels,atas2013levels} used for real eigenvalues of closed quantum systems. The CSR $z_j$ corresponding to eigenvalue $\lambda_j$ is defined as
\eq{\label{eq:zz} z_j = \frac{\lambda_j^{NN}-\lambda_j}{\lambda_j^{NNN}-\lambda_j},}
where $\lambda_j^{NN}$ and $\lambda_j^{NNN}$ are nearest and next-nearest neighbors of $\lambda_j$ in the complex plane, respectively. The CSR is independent of the non-universal local density of eigenvalues, which generally depends on specific models \cite{akemann2019}. Hence, it readily captures the universal fluctuations in the eigenvalue spectrum. In a similar spirit to the level statistics of generic Hamiltonians, it has been shown to be a reliable indicator of the genericness of the Lindbladians. The distribution of CSR has been used to compare spin systems with local dissipation \cite{sa2020csr,prasad2022dissipative,hamazaki2022mbl,hamazaki2020nonhermitianUniversality}, and random Lindbladians with non-local interactions \cite{denisov2019universal,sa2022lindbladian,tarnowski2021random} with relevant Ginibre ensembles. Changes in the CSR can also be used to characterize the dissipative transition from a  Ginibre-like spectrum to a regular spectrum with Poisson statistics \cite{sa2023symmetry,rubio2022integrability}.

\section{Lindblad models with Ginibre statistics}
\label{sec:eigenvalues}
\subsection{Models}
\label{sec:model}
In the following, we will study two types of local Lindblad models that have Ginibre-like statistics, as quantified by the CSR. Both of these models have a spin $1/2$ particle at each site on a length $N$ chain with open boundary conditions. These Lindbladian superoperators obey the relation $[\mathcal{L},R\otimes R^*]=0$, where $R$ corresponds to the reflection symmetry about the midpoint of the chain. This commutation relation implies that the Lindbladians have weak symmetry \cite{buca2012note}, and they can be block-diagonalized into reflection-even and reflection-odd sectors. The models are chosen to be translationally invariant in the bulk to avoid localization effects. In this work, we avoid additional strong symmetries, which lead to degenerate steady-states \cite{buca2012note,albert2014symmetry}. In the presence of strong symmetries, each independent symmetry sector would be expected to behave generically. Moreover, we will focus on the case where the steady state is not the identity matrix (i.e. $\hat{r}_0\neq\mathbb{I}$), which allows us to better differentiate between generic time evolution of operators versus generic time evolution of states. A necessary (but not sufficient) condition for this asymmetry is to include non-Hermitian jump operators.

\subsubsection{Ising model with dissipation}
Let us consider a spin $1/2$ chain with $N$ sites and open boundary conditions. The unitary part of the evolution is governed by the nearest-neighbor Ising Hamiltonian 
\eq{\label{eq:lindblad-ham} H = -J\sum_{j=1}^{N-1} \sigma^z_{j}\sigma^z_{j+1} -h_x\sum_{j=1}^N\sigma^x_{j} -h_z\sum_{j=1}^N\sigma^z_{j},}
where $h_x$ and $h_z$ represent the transverse and longitudinal magnetic fields, respectively. This Hamiltonian is known to be robustly non-integrable \cite{banuls2011isingmodel,kim2014testing,chiba2024isingProof} when both $h_x$ and $h_z$ are non-zero. The Hamiltonian parameters are fixed at $J=1, h_x=1.3, h_z=1.2$. The local dissipation is introduced using the following set of jump operators
\eq{\label{eq:lindblad-lind} \eqsp{ 
L_{1,j}  &= \sqrt{\gamma}\sigma^+_j  \quad j=1\ldots N\\
L_{2,j}  &=  \frac{\sqrt{\gamma}}{2} \sigma^z_j \quad  j=1\ldots N  \\
L_{3,j} &= \frac{\sqrt{\gamma}}{4}(\mathbb{I} + \sigma^x_{j})(\mathbb{I} + \sigma^x_{j+1}) \quad j=1\ldots N-1\\
L_{4,j}  &= \frac{\sqrt{\gamma}}{4}(\mathbb{I} + \sigma^x_{j})(\mathbb{I} + \sigma^x_{j+2}) \quad j=1\ldots N-2.
}}
where $\gamma\geq 0$ parameterizes the dissipation strength. Here, non-Hermitian jump operators $L_1$ represent amplitude damping processes (e.g., resulting from spontaneous emission), and $L_2$ correspond to dephasing in $\sigma^z$ basis at each site. The additional inclusion of nearest and next-nearest neighbor jump operators ensures that the dissipative terms remain non-integrable in the limit of weak Hamiltonian interactions.

%%==========================================================================

\subsubsection{Random Lindblad model}

To verify which features are generic, we also study translationally invariant Lindbladians with random local interactions and local jump operators \cite{liProsenChan2024MultipleModelDsff}. Let us denote the Pauli operators acting on site $k$ by $\sigma^{1,2,3}_k=\sigma^{x,y,z}_k$. The Hamiltonian is given by
\eq{ \label{eq:RL3-ham} H = \frac{J_1}{3}\sum_{k=1}^N\sum_{\alpha=1}^3 Q_\alpha \sigma^\alpha_k + \frac{J_2}{9}\sum_{k=1}^{N-1}\sum_{\alpha,\beta=1}^3 (R_{\alpha\beta} + R_{\beta\alpha})\sigma^\alpha_k\sigma^\beta_{k+1}, }
where $Q_\alpha$ and $R_{\alpha\beta}$ are real-valued random variables independently drawn from the normal distribution with zero mean and unit variance. The real valued-ness ensures that the Hamiltonian remains Hermitian. The first term corresponds to a uniform field of strength $\sim J_1$ polarized in an arbitrary direction, and the second term represents the nearest-neighbor interaction of order $\sim J_2$. Unless stated otherwise, we set $J_1=J_2=J$. The dissipation is modeled by single-site and two-site jump operators defined as
\eq{\label{eq:RL3-lind} \begin{aligned}
L^I_{j} &= \sqrt{\frac{\gamma_1}{3}} \sum_{\alpha=1}^{3} K_{\alpha} \sigma^\alpha_j , \text{  for  } j=1,2,\ldots,N\\
L^{II}_j &= \frac{1}{2}\sqrt{\frac{\gamma_2}{9}}\sum_{\alpha,\beta=1}^{3} (D_{\alpha\beta}+D_{\beta\alpha})\sigma^\alpha_j\sigma^\beta_{j+1}, \\
\hfill &\text{  for } j=1,2,\ldots,N-1. 
\end{aligned}
}
Here, the coefficients $K_{\alpha}$ and $D_{\alpha\beta}$ are complex random variables, which makes the jump operators non-Hermitian. Their real and imaginary parts are independently drawn from the normal distribution with zero mean and unit variance. The strength of single-site (two-site) jump operators is parameterized by $\gamma_1(\gamma_2)$. The symmetrization $(R+R^T)$ and $(D+D^T)$ ensures that the Lindbladian obeys the weak reflection symmetry similar to the dissipative Ising model defined in Eq.~\eqref{eq:lindblad-ham} and \eqref{eq:lindblad-lind}. We stress that the randomness is only present in the choice of operators acting on a local region of two sites. Once these operators are chosen, they identically act on neighboring pairs of spins across the length of the spin-chain. The numerical values of the model parameters for all of the independently sampled realizations studied here are listed in \cref{tab:h1,tab:h2,tab:g1,tab:g2}.
%%==========================================================================
\subsection{Spectra and Ginibre level statistics}
\label{sec:spectrum}

\begin{figure}
\includegraphics{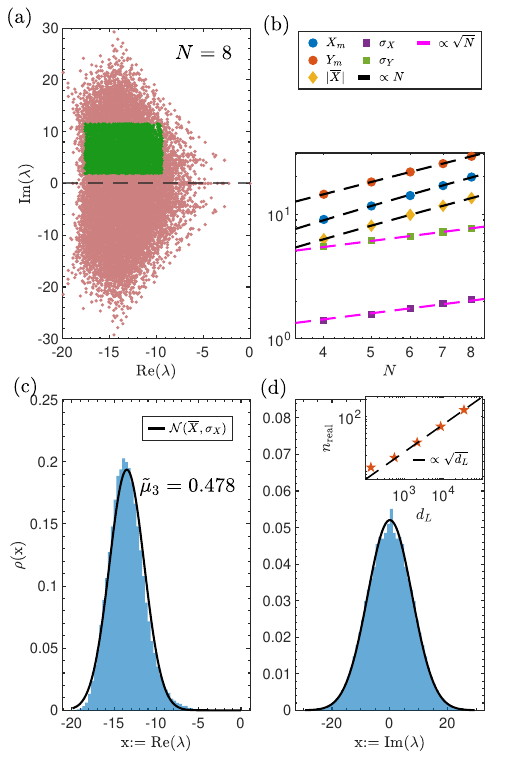}%
\caption{\label{fig:ee4} The eigenvalue spectrum of dissipative Ising model: The spectral data is shown for $J=1,h_x=1.3,h_z=1.2$, and $\gamma=0.8$ in the weak reflection symmetry sector labeled by $\mathcal{R}=1$. (a) The scatter plot of complex eigenvalues $\lambda_j$ is shown for an $N=8$ site one-dimensional spin chain. The green colored patch shows the eigenvalues used to compute CSR statistics in Fig.~\ref{fig:zzPro}. (b) The values of the largest absolute real part $(X_m)$, the largest imaginary part $(Y_m)$, and the average value of the real parts $(\overline{X})$ are plotted as a function of the system size $N$. They approximately fall on the trend-line $\sim N$ indicated by dashed black lines. The spread of the bulk of the spectrum along real(imaginary) direction computed in terms of $\sigma_X$ ($\sigma_Y$) increases as a function of system size and is proportional to $\sqrt{N}$ indicated by magenta colored lines (see~\cref{eq:xm-ym,eq:xbar,eq:sigmaXY} for details). (c) The marginal distribution of the eigenvalue, $\rho_X = \int [\text{d Im}(\lambda)]\ \rho(\lambda)$, is shown as a function of the real part of the eigenvalues for system size $N=8$. The black line shows the Gaussian distribution with mean $\overline{X}$ and standard deviation $\sigma_X$. (d) The marginal distribution $\rho_Y=\int [\text{d Re}(\lambda)]\ \rho(\lambda)$ is plotted as a function of the imaginary part of the eigenvalues. The overlaying curve shows the Gaussian distribution with zero mean and $\sigma_Y$ standard deviation. The distributions are normalized such that $\int \text{dx} \rho(\text{x})=1$. The inset shows the number of real eigenvalues ($n_{\text{real}}$) as a function of the dimension of the operator space $d_L$.} 
\end{figure}

We begin by examining the generic features of the eigenvalues of these model Lindbladians. The models considered here only consist of spatially local and bounded terms, which, as discussed below, lead to universal scaling of the global spectral features as a function of the system size. Furthermore, we show that these models exhibit universal Ginibre-like level repulsion and local correlations in their eigenvalue spectra.

In the following, we present a detailed analysis of the eigenvalue spectrum of the dissipative Ising model. The numerically computed spectrum for parameters $J=1,h_x=1.2,h_z=1.3,$ and $\gamma=0.8$ is shown in Fig.~\ref{fig:ee4}(a) for the weak symmetry sector labeled by $\mathcal{R}=+1.$ The dimension of the operator space in this sector is given by 
\eq{ d_L = \frac{4^N}{2}(1+4^{\lceil N/2\rceil - N })\approx 2^{2N-1},}
which increases exponentially with the number of sites $N$ in the system. In Fig.~\ref{fig:ee4}(b), we numerically analyze the shape of this spectrum in the complex plane. The extent of the boundary of the spectrum along the real and the imaginary direction respectively given by 
\eq{\label{eq:xm-ym} X_m = \text{max}_{\lambda_j} |\text{Re}(\lambda_j)| \quad , \text{and}  \quad Y_m = \text{max}_{\lambda_j} \text{Im}(\lambda_j),}
increases linearly with increasing system size $N$. Similarly, the location of the center of the spectrum computed in terms of the average of real part 
\eq{\label{eq:xbar}\overline{X} = \frac{1}{d_L}\sum_{j=1}^{d_L} \text{Re}(\lambda_j) }
is also proportional to $N$. As mentioned in section~\ref{sec:lindblad}, the eigenvalue spectrum is symmetric about the real axis, which leads to $\overline{Y}=0$. Finally, the spread of the bulk of the spectrum along the real and imaginary direction is characterized in terms of the standard deviations
\eq{\label{eq:sigmaXY}\sigma_X = \sqrt{ \frac{\sum_{j=1}^{d_L}[ \text{Re}(\lambda_j)-\overline{X}]^2}{d_L}}, \quad \sigma_Y = \sqrt{ \frac{\sum_{j=1}^{d_L} [\text{Im}(\lambda_j)]^2}{d_L}}.} 
respectively. Both $\sigma_X$ and $\sigma_Y$ approximately grow as $\sqrt{N}$.

We now provide physical arguments to justify these numerically observed scalings of the spectrum in Fig.~\ref{fig:ee4}(b). The Lindbladian of both of the models considered here can be decomposed as $\mathcal{L} = \sum_{j=1}^N \mathcal{L}_j$. Here, $\mathcal{L}_j$ is a local superoperator that acts on the spins in the local neighborhood of site $j$ (eg, up to next-nearest neighbor in the dissipative Ising model). Each of these local superoperators has a bounded spectrum, with complex-conjugate pairs of eigenvalues satisfying Re$(\lambda_j)<0$, and an eigenvalue of maximal modulus. This implies that the largest eigenvalue of the Lindbladian should increase at most linearly as a function of the system size $N$, as observed above. Heuristically, we can treat the contribution of each bounded term $\mathcal{L}_j$ to the total eigenvalue as a random variable with a finite mean and variance. Since there are $\mathcal{O}(N)$ such terms in a system with $N$ sites, the average value of the $\lambda$ should be proportional to $N$. Since each term has spatially local support, the contributions from spatially well-separated terms towards random eigenoperators will only have weak correlations. As a result, each eigenvalue can be approximated as a sum of independent random variables. The fluctuations of these random contributions around their average value, estimated in terms of the standard deviation, only grow $\propto \sqrt{N}$ (see Refs.~\cite{rigol2008thermalization,torres2016realistic} for similar arguments in closed systems). In practice, this implies that an eigenmode with finite density of excitations with respect to the steady state will have an eigenvalue $|\lambda| \propto N$. As a result, the eigenvalues of most of the eigenmodes of local Lindbladians are well separated from the steady-state ($\lambda=0$) in the thermodynamic limit.

We further illustrate the detailed distribution of the exponentially many eigenvalues in the complex plane by evaluating their marginal distributions as
\eq{\label{eq:marginal} \rho_X(x) = \int dy \rho(x,y)  , \qquad\rho_Y(y) = \int dx \rho(x,y), }
where we write the complex eigenvalues $\lambda=x+iy$ in terms of their real and imaginary part. $\rho(x,y)$ is the full density of states normalized such that $\int dx dy \rho(x,y)=1$. 
If the individual terms in the Lindbladian were truly independent of each other, following the central limit theorem, the eigenvalues would have a  Gaussian distribution (as is the case for local density of states for generic closed quantum systems \cite{rigol2008thermalization,torres2016realistic}).
To test this, the marginal distribution $\rho_{X}(x)$, together with a Gaussian curve $\mathcal{N}(\overline{X},\sigma_X)$, is shown in Fig.~\ref{fig:ee4}(c). While the Gaussian curve qualitatively captures the shape of the spectral density, the Lindblad spectrum is not symmetric about its mean due to important physical differences between eigenoperators with larger and smaller values of Re$(\lambda)$. We estimate this asymmetry using the skewness parameter as
\eq{ \label{eq:skew} \tilde{\mu}_3 = \frac{1}{\sigma_X^3} \int \text{d} x \rho_X(x) (x-\overline{X})^3.}
For the dissipative Ising model, we obtain $\tilde{\mu}_3 =0.478$. Notwithstanding these differences, the spectrum has a single sharp peak with the majority of the eigenvalues contained within the first 2 standard deviations of the Gaussian envelope $\mathcal{N}(\overline{X},\sigma_X)$, which we use to define the bulk of the spectrum. In Fig.~\ref{fig:ee4}(a), it is apparent that there are eigenvalues that are not part of this bulk of the spectrum, as they lie at values of $\text{Re}(\lambda)>\overline{X}+2\sigma_X$, where the Gaussian curve in Fig.~\ref{fig:ee4}(c) is extremely small. These eigenvalues and the corresponding eigenoperators determine the longest time dynamics of our system, which may be quite different from the short-time dynamics dictated by the bulk of the spectrum. While the fraction of the eigenvalues found in these regions vanishes in the thermodynamic limit, it is difficult to clearly specify the location of this crossover for the numerically accessible system sizes that are analyzed here.

In Fig.~\ref{fig:ee4}(d), we show the marginal distribution $\rho_Y(y)$ as a function of the imaginary parts of the eigenvalues. It qualitatively matches the Gaussian distribution $\mathcal{N}(0,\sigma_Y)$ for values away from the real axis. An enhanced peak that deviates from the Gaussian behavior at $\text{Im}(\lambda)=0$ can be attributed to $\mathcal{O}(\sqrt{d_L})$ number of eigenvalues with exactly zero imaginary part (see inset in Fig.~\ref{fig:ee4}(d)), expected for real Ginibre matrices \cite{edelman1994many}. 

In appendix~\ref{app:RL-spectrum}, we show the analogous data for the shape of the spectrum observed in different realizations of the random Lindblad model. As for the dissipative Ising model, the width of the spectrum in our random Lindbladian models consistently scales as $\sqrt{N}$, while its mean is on the real axis and scales with $N$, allowing us to define the bulk of the spectrum as those eigenvalues that are not more than 2 standard deviations from the mean. We note that other details of the spectrum's shape, such as the exact shape of its border or the aspect ratio, and the precise shape of eigenvalue densities $\rho_X,\rho_Y$ vary across different models as seen in Figs.~\ref{fig:RL-sd-4-spectrum} and ~\ref{fig:RL-fragment-spectrum}. Indeed, the analytical expressions for the shape of the spectrum in certain classes of random Lindblad ensembles have been found to depend on the specific details of the model, such as the number of jump operators, the relative strength of dissipation compared to Hamiltonian interactions, and the degree of spatial locality \cite{denisov2019universal,sa2020spectral,can2019midgap,can2019random,wang2020hierarchy}. 

\begin{figure}
\includegraphics{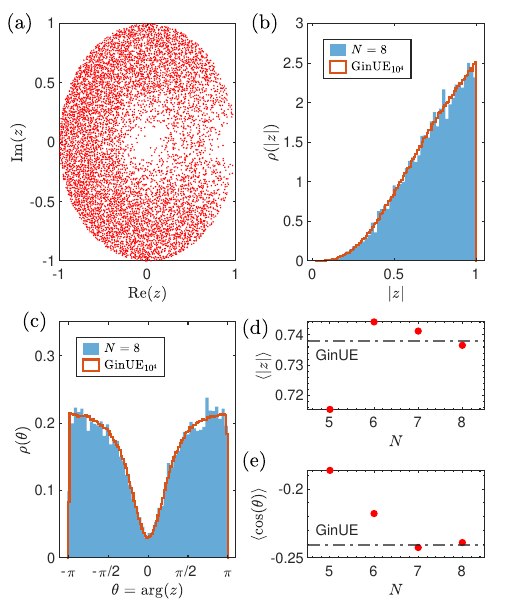}%
\caption{\label{fig:zzPro} Statistics of complex spacing ratios (CSR) in dissipative Ising model ($J=1,h_x=1.2,h_z=1.3,$ and $\gamma=0.8$): (a) The CSR $z_j=|z_j|e^{i\theta_j}$ is defined for each eigenvalue $\lambda_j$ according to Eq.~\eqref{eq:zz}. The 2d distribution of the $z_j$ in the complex plane is shown for an $N=8$ size system. The data is shown for the eigenvalues in the upper-half of the complex plane away from the real axis ($|\text{Re}(\lambda)-\overline{X}|\le2\sigma_X$, and $\sigma_Y/4\le\text{Im}(\lambda)\le3\sigma_Y/2$ --- see Fig.~\ref{fig:ee4}(a)). The marginal probability distribution of (b) the absolute value $|z|$ and (c) the angular variable $\theta$ are shown in blue colored histograms. The orange curves in each panel show the corresponding distribution for the non-Hermitian Ginibre ensemble. The Ginibre results are numerically computed by diagonalizing 10 independent samples of $10^4\times 10^4$ sized non-Hermitian random matrix with complex-valued Gaussian matrix elements (GinUE ensemble). The distributions are normalized as $\int dx \rho(x)=1$. (d) The mean of the absolute values $|z|$ is shown as a function of system size $N$. It approaches $|z|_{\text{GinUE}}$ (black dashed line) with increasing system size. (e) The angular average $\langle\cos{\theta}\rangle$ is plotted for different systems sizes. The angular repulsion is captured by the negative value of $\cos(\theta)$ (i.e., large angular separation between neighbors), where the average value approaches the Ginibre ensemble value with increasing system size.}
\end{figure}

Having identified the bulk of the spectrum, we now show that the eigenvalue spectra of our models have correlations similar to Ginibre matrices by analyzing the statistics of the CSR $z_j:=|z_j|e^{i\theta_j}$ defined in Eq.~\eqref{eq:zz}. The two-dimensional distribution in the complex plane is shown in Fig.~\ref{fig:zzPro}(a) for the dissipative Ising model. Level repulsion is evident from the lower density of data points close to $|z|\sim 0$. We further illustrate this in Fig.~\ref{fig:zzPro}(b) by plotting the marginal radial distribution of $z$ defined as $\rho(|z|) = \int d\theta \ |z| z$. The distribution matches well with the numerically computed distribution for the random matrices from the complex Ginibre ensemble. This agreement becomes better with increasing system size as shown in Fig.~\ref{fig:zzPro}(d), where we see that the mean of the distribution approaches $\langle|z|\rangle_{\mathrm{GinUE}}\approx0.738$ rapidly with increasing system size. Level repulsion in the complex plane also implies the presence of the angular repulsion among the neighboring eigenvalues. Specifically, using the definition in Eq.~\eqref{eq:zz}, we can define $\theta:=\text{arg}(z)$ as the angle between two arrows that point towards the nearest and next-nearest neighbors of the given eigenvalue. The angular repulsion implies that this angle $\theta$ should typically have large values. In Fig.~\ref{fig:zzPro}(c), we show that the marginal angular distribution defined by $\rho(\theta)=\int d|z|\ |z|e^{i\theta}$ has a dip near $\theta=0$ and agrees with the result for the Ginibre ensemble. The absence of small angles can be quantified in terms of $\langle \cos{\theta}\rangle$ \cite{sa2020csr}, which takes negative values (see Fig.~\ref{fig:zzPro} (e)) and approaches the Ginibre result $\langle \cos{\theta}\rangle_\mathrm{GinUE}\approx-0.24$ for increasing $N$.

In summary, analyzing the statistics of CSR confirms that the eigenvalue spectrum of the dissipative Ising model shows generic level repulsion characterized by the random Ginibre matrices. Similar plots demonstrating the genericness of the random Lindblad models studied in this work are shown in Appendix \ref{app:RL-spectrum}. 

%%=====================================================================
\begin{figure}
\includegraphics{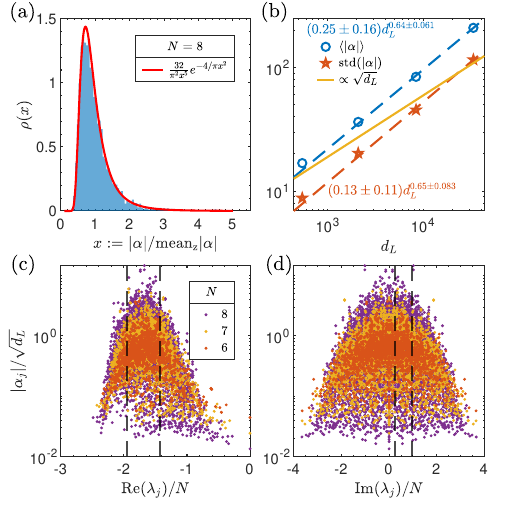}%
\caption{\label{fig:alpha}Eigenoperator overlaps in dissipative Ising model ($J=1,h_x=1.2,h_z=1.3,$ and $\gamma=0.8$): (a) The probability distribution function of overlaps $|\alpha_j|/\text{mean}_z(|\alpha|)$, where $\alpha_j=\frac{1}{\sbra l_j|r_j\sket}$ and $\text{mean}_z(|\alpha|)$ is average of their absolute value (only includes the eigenmodes inside $\lambda\in \Omega$). The blue histogram shows data for the dissipative Ising model, and the red line shows the complex Ginibre ensemble. The data is gathered for the eigenmodes in the window $\Omega=\{\lambda_j|\ |\text{Re}(\lambda_j)-\overline{X}|\le \sigma_X, \frac{\sigma_Y}{4}\le\text{Im}(\lambda_j)\le\sigma_Y\}$, where $\overline{X}$ is the center and $\sigma_{X(Y)}$ are standard deviations of the spectrum in $X(Y)$ direction. (b) The mean and standard deviation evaluated for the eigenmodes inside the window $\Omega$ are plotted as a function of increasing dimension of the operator space $d_L\approx 2^{2N-1}$. The dashed lines show the best fits to a power law function $a d_L^b$. The resulting fits to mean and standard deviation are shown in the respective colors. The yellow line depicts the trivial scaling $\propto\sqrt{d_L}$ for reference. (c-d) The scatter plot of $|\alpha_j|/\sqrt{d_L}$ is shown as a function of real (panel (c)) and imaginary part (panel(d)) of the eigenvalues $\lambda_j/N$ where $N$ is the total number of sites in the system. The dashed black lines show the boundary of the window $\Omega$ used in panel (a--b) for $N=8$.}
\end{figure}
%%=====================================================================

\subsection{Overlap of eigenoperators}

The time evolution described in Eq.~\eqref{eq:dyn} further depends on the overlap between the left and right eigenoperators of the eigenvalue $\lambda_j$ defined by $\alpha_j:=\frac{1}{\sbra l_j | r_j\sket}$. For eigenvectors of random Ginibre matrices, the average value of analogous overlaps depends on the corresponding eigenvalue \cite{chalkerMehlig1998prlEigenvectors}. If $\text{mean}_{z}(|\alpha|)$ is the average of the $|\alpha|$ conditioned on eigenvalue $\lambda=z$, then the rescaled variables $x:=|\alpha|/\text{mean}_{z}(|\alpha|)$ for eigenvectors of Ginibre matrices are distributed according to $\rho_G(x) = \frac{32}{\pi^2x^5}e^{-4/\pi x^2}$ \cite{hamazaki2022mbl,fyodorov2018statistics,bourgade2020distribution}. Since $\text{mean}_z(|\alpha|)$ varies across the eigenvalue spectrum, it is necessary to focus on a small range of $z$ to observe this distribution.  In Fig.~\ref{fig:alpha}(a), we show this distribution for the eigenoperators of the dissipative Ising model, %. Since $\text{mean}_z(|\alpha|)$ varies across the eigenvalue spectrum, we gather data for
 with $|\alpha|$ drawn from a small window of eigenvalues near the bulk of the spectrum, and compute the $\text{mean}_z(|\alpha|)$. We observe that the probability distribution for the Ising model approximately matches that of the Ginibre matrix, depicted by the red curve. Next, we are interested in characterizing the system size dependence of these overlaps. To this end, let us consider an orthonormal basis of Pauli strings $|F_\mu\sket$ obeying $\sbra F_\mu|F_\nu\sket=\delta_{\mu,\nu}$ (see ~\cref{eq:pauli-string,eq:ref-pauli,eq:pauli-norm} for details). For a generic model, the eigenoperators written in this basis are expected to be essentially uncorrelated entries that can be treated as independent and identically distributed random variables. For $d_L$-dimensional vectors normalized such that $\sbra r_j|r_j\sket=\sbra l_j|l_j\sket=1$, these entries will be of order $\frac{1}{\sqrt{d_L}}$. Then the overlap between two non-orthogonal operators corresponds to 
\eq{ \label{eq:rr-scaling} \sbra r_j |l_j \sket =\sum_{\mu=1}^{d_L} \sbra r_j|F_\mu\sket\sbra F_\mu | l_j\sket  \approx  \frac{\sqrt{d_L}}{\sqrt{d_L} \sqrt{d_L}} = \frac{1}{\sqrt{d_L}} ,}
where the $\sqrt{d_L}$ factor in the numerator accounts for addition of $d_L$ numbers with fluctuating complex phases. This suggests that uncorrelated random eigenoperators should exhibit $|\alpha|\sim \mathcal{O}(\sqrt{d_L})$ scaling, which is also the case for the eigenvectors of the Ginibre matrices \cite{chalkerMehlig1998prlEigenvectors}. In Fig.~\ref {fig:alpha}(b), we fit the data for the average and standard deviation of $|\alpha|$ with a functional form $a d_L^b$. The fit shows a slightly faster scaling $\approx d_L^{0.64}$ with increasing system size, suggesting that left- and right- eigenvectors are not uncorrelated random vectors in our local model. However, one should not read too much into this result, given the uncertainty in the pre-factors and the availability of only four data points. In appendix~\ref {app:RL-spectrum}, we show similar scaling for the eigenoperators of the random Lindblad model, which similarly hints at a slightly higher exponent than for the Ginibre case, but is not inconsistent with an exponent of $0.5$. Finally, in Fig.~\ref {fig:alpha}(c) and (d), we show the absolute value of $|\alpha|/ \sqrt{d_L}$ for the entire spectrum as a function of the real and imaginary parts of the eigenvalue, respectively, for a range of system sizes. 

\section{Locality structure of eigenoperators}
\label{sec:eigenvecs}

Having established the genericness of the Lindblad eigenvalue spectrum of our models, we now turn to their eigenoperators. We identify the generic features in the eigenoperators of local Lindbladians that influence the resulting open-system dynamics. We discuss the impact of these on early-time decoherence of initial states in section~\ref{sec:state_dynamics}, and on the time evolution of local operators in section~\ref{sec:operator_dynamics}.

The most striking difference between eigenoperators of local Lindbladians and those of their Hamiltonian counterparts is that for Lindbladian eigenoperators, we find a strong correlation between the effective size (see Eq.~\eqref{eq:size-dist}) of the Lindblad eigenoperators and the real part of the corresponding eigenvalue, which controls the decay rate of the corresponding eigenmode. Thus, Lindbladian eigenoperators have more structure in the Pauli basis than their Hamiltonian counterparts. We also find that bulk Lindbladian eigenoperators are well-scrambled within the relevant size-sector of the Pauli string basis.

This correlation between decay rate and operator size has important physical consequences: under local Lindbladian dynamics, operators generically decay at a rate that is proportional to their size. Similar ideas were explored in \cite{wang2020hierarchy,hartmann2024} by analyzing random Lindblad models with few-body but spatially non-local Lindblad operators in the limit of strong dissipation. The jump operators in this family of models are spatially long-ranged but few-body operators. It was observed that in this strong dissipation limit, the Lindblad spectrum fragments into well-separated blobs, each of which only contains eigenoperators composed of Pauli strings of identical size. This correspondence between the size and decay rate of the eigenoperators leads to the operator size-dependent decay of local observables. The models we consider here consist of generic local interactions, including both dissipative and unitary dynamics, and thus exhibit the generic (non-fragmented) spectra described in the previous section. Nevertheless, the strong correspondence between the size and decay time of operators persists.\\

\subsection{Size distribution of eigenoperators}
We begin by defining the ``size'' of an operator. Consider a Pauli string operator defined as
\eq{\label{eq:pauli-string}S_{\mathbf{m}} = \prod_{j=1}^N \sigma^{\mathbf{m}_j}_j  ,} 
where, $\mathbf{m}=\{0,1,2,3\}^{\otimes N}$ is a length-N bit string where the $j^{\text{th}}$ entry denotes the Pauli operators $\sigma^0_j = \mathbb{I}_{2\times2},\sigma_j^1=\sigma_j^x, \sigma_j^2=\sigma_j^y,$and $ \sigma_j^3=\sigma_j^z$ acting on site $j$. In what follows, owing to the weak reflection symmetry of the dynamics, we focus on the reflection symmetric subspace spanned by the operators that obey $R \hat{O} R=\hat{O}$, where $R$ is the reflection operator. An orthonormal basis for this subspace is given by:
\eq{\label{eq:ref-pauli}F_{\mathbf{m}} = \begin{cases}
	\frac{1}{2^{N/2}}S_{\mathbf{m}} & \text{if  } S_{\mathbf{m}}=R\ S_{\mathbf{m}}\ R\\
	\frac{1}{2^{(N+1)/2}}(S_{\mathbf{m}} + R \ S_{\mathbf{m}} \ R) & \text{otherwise	}
\end{cases}.}
The total dimension of the symmetric subspace is given by $d_L = \frac{4^N}{2}(1+4^{\lceil N/2\rceil - N })\approx \frac{4^N}{2}$. These operators satisfy the orthonormality condition given by
\eq{\label{eq:pauli-norm}\sbra F_{\mathbf{m}}|F_{\mathbf{m'}}\sket=\delta_{\mathbf{m},\mathbf{m'}}.} A general reflection symmetric operator $|A\sket$ can be expanded as 
\eq{\label{eq:pauli-expansion} |A\sket = \sum_{\mathbf{m}}\sbra F_{\mathbf{m}}|A\sket |F_{\mathbf{m}}\sket.}
Each of the basis operators has a fixed number of sites that are acted on by a non-identity Pauli operator. Hence, we can define the size $\mathtt{S}[\mathbf{m}]$ of the basis operator associated with the string $\mathbf{m}$ as 
\eq{\mathtt{S}[\mathbf{m}] = N-\sum_{j=1}^N\delta_{\mathbf{m}_j,0},}
which counts the number of non-identity Pauli operators present in $F_{\mathbf{m}}$. The classification of the basis strings in terms of their size then allows us to define the operator size distribution \cite{qi2019quantum} of an operator $A$ as
\eq{ \label{eq:size-dist} p_{s_0}(A) = \frac{1}{\sbra A|A\sket}\sum_{\mathbf{m}} |\sbra F_{\mathbf{m}}|A\sket|^2 \delta_{\mathtt{S}[\mathbf{m}],s_0} , \quad 0\le s_0 \le N}
where $p_{s_0}(A)$ is the total weight of the operator $A$ that is concentrated on basis operators of size $s_0$. The term in the denominator ensures the normalization condition $\sum_{s=0}^N p_{\mathtt{s}}(A)=1$. 

\begin{figure}
    \includegraphics{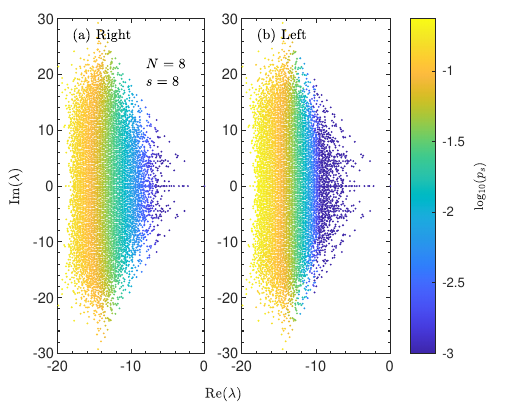}
    \caption{The complex eigenvalue ($\lambda$) spectrum for the dissipative Ising model is shown for $N=8$ system size ($J=1,h_x=1.2,h_z=1.3,$ and $\gamma=0.8$). Each eigenvalue $\lambda$ is colored according to the $\log_{10}$ of size distribution (a) $p_{s=8}(\hat{r})$ of its right, and (b)  $p_{s=8}(\hat{l})$ of its left eigenoperator as defined in Eq.~\eqref{eq:evec-size-dist}.} 
    \label{fig:ising-evecsize-scatter} 
\end{figure}

We can now apply this framework to the eigenoperators of the Lindbladian. The size distributions of the right and left eigenoperators that correspond to eigenmode $\lambda_j$ are given by
\eq{\label{eq:evec-size-dist} \eqsp{p_{s_0}(r_j) &= \sum_{\mathbf{m}} |\sbra F_{\mathbf{m}}|r_j\sket|^2 \delta_{\mathtt{S}[\mathbf{m}],s_0} ,\\
p_{s_0}(l_j) &= \sum_{\mathbf{m}} |\sbra F_{\mathbf{m}}|l_j\sket|^2 \delta_{\mathtt{S}[\mathbf{m}],s_0}.
}}
Here we have used the normalization condition $\sbra r_j|r_j\sket=\sbra l_j|l_j\sket=1$. We numerically observe that the weights $p_s$ do not show large variations as a function of the imaginary part of the eigenvalue for both the right and left eigenoperators. In Fig.~\ref{fig:ising-evecsize-scatter}, we demonstrate this for the weights on $s=8$ size basis operators in a system of $N=8$ sites. The eigenvalues are colored according to the weights of the eigenoperator on size-8 basis operators. Although the color proportional to the operator weight $\log_{10}p_s$ varies as a function of the real part of the eigenvalue $\lambda$, it remains constant as the imaginary part is varied for a fixed value of $\text{Re}(\lambda)$. We will thus find it useful to analyze the coarse-grained weights as a function of the real part of the eigenvalues. These coarse-grained weights of right eigenoperators are defined as 
\eq{ \label{eq:ps_tilde} \widetilde{p}_{s}^{R}(x_0) = \frac{1}{n_{\Delta}(x_0)}\sum_{\{j:|\text{Re}(\lambda_j)-x_0|\leq\Delta/2\}} p_{s}(r_j).} 
Here, the summation runs over all eigenmodes in a vertical strip of width $\Delta$ around $\text{Re}(\lambda_j)=x_0$ and $n_{\Delta}(x_0)$ is the number of eigenmodes in this window. The coarse-grained weights $\widetilde{p}_{s}^{L}(x_0)$ associated with the left eigenoperators are defined analogously.

\begin{figure}
\includegraphics{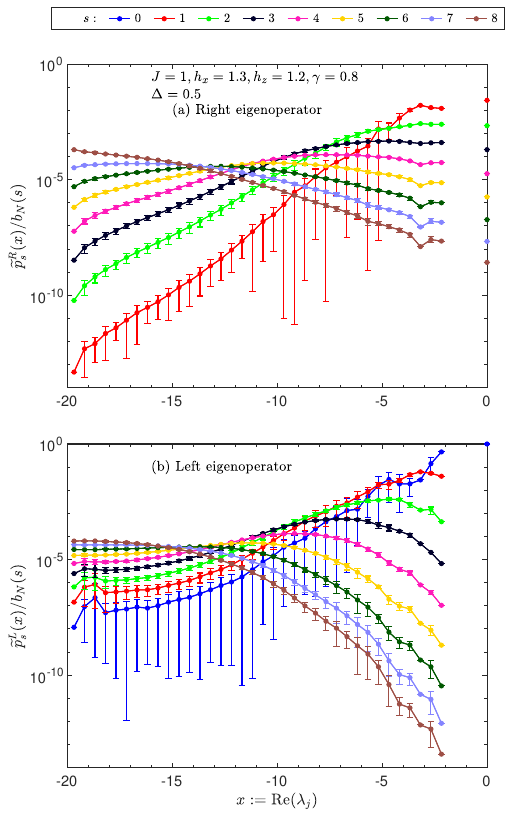}
\caption{\label{fig:secAvg} The coarse-grained weights of eigenoperators on basis-operators of size $s$ (see Eq.~\ref{eq:ps_tilde}) are plotted as a function of the real part of eigenvalues $\lambda$. The data is shown for the (a) right and (b) left eigenoperators of $N=8$ size dissipative Ising model with parameters $J=1,h_x=1.3,h_z=1.2$, and $\gamma=0.8$. The coarse-graining scale is set to $\Delta=0.5$. The weight on $s=0$ basis operator is identically zero for the right eigenoperators with $\lambda\neq 0$ (panel (a)), since they are traceless due to the biorthogonality relation. The error bars show the variation of the weights of eigenmodes in the window $\Omega(x_0):=\{\lambda_j:|\text{Re}(\lambda_j)-x_0|\leq\Delta/2\}$ about its mean value $\tilde{p}_{s}(x_0)$. The value of top error bar is given by $e_t(x_0)=\text{min}(\text{max}_{j\in \Omega(x_0)}p_s(x_j),\sigma_{j\in \Omega(x_0)}p_s(x_j))$ and the bottom error bar is given by $e_b(x_0)=\text{min}(\text{min}_{j\in \Omega(x_0)}p_s(x_j),\sigma_{j\in \Omega(x_0)}p_s(x_j))$ where $\sigma$ is standard deviation of values within given window.} 
\end{figure}

Fig.~\ref{fig:secAvg} shows $\widetilde{p}_{s}^{R}$ and $\widetilde{p}_{s}^{L}$  as a function of the real part of the eigenvalue of the dissipative Ising model. We observe that the eigenmodes for which the corresponding eigenvalues have a small $|\text{Re}(\lambda)|$ exhibit an operator size distribution similar to the corresponding steady state. In the case of right eigenoperators (panel (a)), this implies that the size distribution depends on the purity of the steady state. Meanwhile, for left eigenoperators (see panel(b)), the steady state operator is always the identity operator, and hence, the weight of larger basis operators is highly suppressed. The decay rate of an operator is generally proportional to its size \cite{wang2020hierarchy,schuster2023operator,shirai2024accelerated} (also see the discussion in section~\ref{sec:operator_dynamics}) when the Lindblad dynamics is local. According to Eq.~\eqref{eq:heisen}, this will hold if slowly decaying left eigenoperators are mostly concentrated on shorter basis operators. Moreover, the overlaps with the right eigenoperators $\sbra r_j|O\sket$ determine which part of the initial operator follows this slow evolution. Intuitively, this suggests that the small weight operators $\hat{O}$ should only overlap with the right eigenoperators close to the steady state. This expectation is confirmed in Fig.~\ref{fig:secAvg}(a), where the overlap of the right-eigenoperators in the bulk of the spectrum with the $s=1,2$ sized basis operators is exponentially suppressed relative to eigenmodes near the steady state. A similar exponential suppression of the overlap of right eigenoperators in the bulk of the spectrum was also observed in \cite{hamazaki2022mbl}, where the authors used the fact that the Lindblad superoperator $\mathcal{L}^\dagger$ with local terms only modifies initially local operators to a few nearby sites. Then using this along with techniques from \cite{abanin2015floquetPrethermal,mori2016floquetBound} for bounds on growth of operator sizes as a result of action of local Hamiltonians, it was shown that  $|\sbra \hat{O}|r_j\sket|\le ||O||e^{-|\lambda_j|/(c\gamma \kappa)}$ , where $\kappa$ is the range of interactions and $c$ is an $\mathcal{O}(1)$ constant. The size distributions of the eigenoperator in smaller size sectors $s\le 1$ show a higher degree of statistical fluctuations due to relatively fewer number of basis operators at these sizes, which is also observed for the non-local Lindblad model discussed in Fig.~\ref{fig:nonLoc-size}.

\begin{figure*}
\includegraphics{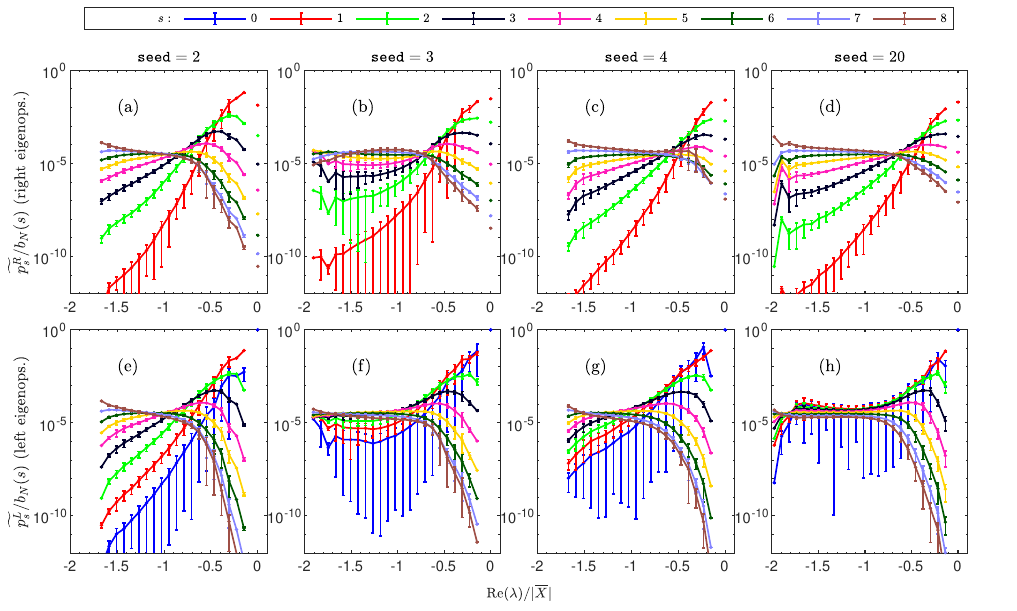}
\caption{\label{fig:RL-secAvg} Eigenoperator size distribution for random Lindblad model: The model parameters are set to $J=1$ and $\gamma_1=\gamma_2=\gamma=0.25$ with total $N=8$ sites in the system. Each column shows the data for one independently sampled realization (see~\cref{tab:h1,tab:h2,tab:g1,tab:g2} in Appendix \ref{app:RL-spectrum} for specific model parameters). The averaged distribution $\tilde{p}_s$ (shown using dots), defined in Eq.~\eqref{eq:ps_tilde}, is plotted as a function of the real part of the eigenvalue. The y-axis is rescaled by $b_N(s)$, which equals the total number of basis operators with size $s$. The data is shown for (top row) right and (bottom row) left eigenoperators. The coarse-graining window is set to $\Delta=0.08\times|\overline{X}|$. The error bars are as for Fig.~\ref{fig:secAvg}. The lines joining the data points are guides to the eyes.}
\end{figure*}

In Fig.~\ref{fig:RL-secAvg}, we similarly analyze the size distribution of the eigenoperators of the random Lindblad model. Here we show four independently sampled realizations of this model with parameters $J=1$ and $\gamma_1=\gamma_2=\gamma=0.25$. In Fig.~\ref{fig:RL-secAvg} (a-d), we observe that the right eigenoperators have a qualitatively similar operator size dependence with the real part of their eigenvalues. For large values of the $|\text{Re}\lambda|$, the weight on the shorter basis operators is exponentially suppressed compared to the longer basis operators. However, the degree of such suppression varies across realizations. In the following subsection, we provide an intuitive picture of why such suppression is generic for local systems by analyzing a non-interacting model. The size distribution of the left eigenoperators, on the other hand, qualitatively differs across different realizations. In the realization labeled by $\mathtt{seed}=2$ (see Fig.~\ref{fig:RL-secAvg} (e)), we observe that the size distribution of the left eigenoperators is similar to that of the corresponding right eigenoperators. Whereas, the left eigenoperators of the realization $\mathtt{seed}=20$ have a size distribution similar to a random mixture of operators for large values of $|\text{Re}(\lambda)|$. Several other realizations interpolate between these two extreme features as seen in Fig.~\ref{fig:RL-secAvg}(f) and (g). In the following, we identify the steady state purity as one of the factors that can account for some of these variations.

\subsection{Non-interacting model} 
To better understand these qualitative trends in operator size, we consider a non-interacting model in this section by setting $J_2=0$ and $\gamma_2=0$ in Eqs.~\eqref{eq:RL3-ham},\eqref{eq:RL3-lind}. In this limit, each spin evolves independently under the Lindbladian $\mathcal{L}_i$ that only includes a single site Hamiltonian and jump operator term. The total Lindbladian for the system is then simply given by $\mathcal{L} = \sum_{i=1}^N\mathcal{L}_i$. The spectrum of the $N$-site Lindbladian $\mathcal{L}$ can be constructed in terms of the local spectrum of $\mathcal{L}_i$ at each site. 

\subsubsection{Right eigenoperators} Let us first write down the general form of eigenoperators of single-site Lindbladian $\mathcal{L}_i$ acting on site labeled by $i$. In general, there will be a single eigenvalue $\lambda=0$ that corresponds to the local steady state with its right eigenoperator given by 
\eq{\label{eq:NI-r0-before}\hat{r}_i^{(0)}= \frac{1}{\sqrt{2}}(\phi \mathbb{I}_i + \vec{\beta}.\vec{S}_i), }
where $\vec{S_i} = \{X_i,Y_i,Z_i\}$ are the Pauli operators acting at site $i$ and $\phi,
\vec{\beta}=\{\beta_x,\beta_y,\beta_z\}$ are corresponding coefficients. The normalization $\sbra r|r\sket=1$ implies that the coefficients obey the condition $|\phi|^2+|\vec{\beta}|^2=1$. At late times, the state of the system reaches the steady state $\hat{r}^{(0)}_i/\tr{(\hat{r}^{(0)}_i)}$. The purity $p_{ss}$ of this steady state is given by
\eq{\label{eq:non-int-purity} p_{ss} = \frac{\tr{(\hat{r}^{(0)}_i \hat{r}^{(0)}_i)}}{ [\tr{(\hat{r}^{(0)}_i)}]^2 } = \frac{1}{2\phi^2}. }
Using this expression and the normalization condition, the right eigenoperator is expressed in terms of the steady state purity as 
\eq{\label{eq:NI-r0} \hat{r}_i^{(0)} = \frac{1}{\sqrt{2p_{ss}}} \frac{\mathbb{I}_i}{\sqrt{2}} + \sqrt{1-\frac{1}{2p_{ss}}}\  \frac{\hat{\beta}.\vec{S_i}}{\sqrt{2}},}
where $\hat{\beta}=\vec{\beta}/|\vec{\beta}|$ is a unit vector. As noted in Eq.~\eqref{eq:bi-orthogonal}, the bi-orthogonality condition implies that the right eigenoperators of non-zero eigenvalues must be traceless. Hence, the general expressions for eigenoperators $\hat{r}_i^{(k)}$ for remaining non-zero eigenvalues $\lambda_k$ are given by 
\eq{\label{eq:NI-rk} \hat{r}_i^{(k)} = \vec{\eta}^{(k)}.\frac{\vec{S_i}}{\sqrt{2}}, \quad \text{for } k=1,2,3.  }
The normalization $\sbra r_i^{(k)}|r_i^{(k)}\sket=1$ implies that the coefficients obey the condition $|\eta_x|^2+|\eta_y|^2+|\eta_z|^2=1$ for all values of $k$. If two of the eigenmodes are complex conjugates of each other, i.e. $\lambda_k=\lambda_{k'}^*$, then $\hat{r}^{(k)}=(\hat{r}^{(k')})^{\dagger}$. 

The eigenspectrum of the model with $N$ sites can now be composed by adding independent contributions from each site. Specifically, the \emph{many-body} eigenvalue $\lambda$ and the corresponding eigenoperator $\hat{r}$ can be written as 
\eq{\lambda = \sum_{i=1}^N \lambda^{(k_i)} ,\qquad \hat{r}=\otimes_{i=1}^N \hat{r}_i^{(k_i)} }
where $k_i\in\{0,1,2,3\}$ represents the local eigenmode at site $i$. 

In order to get a qualitative understanding, let us consider a simplified model where three non-zero eigenvalues at each site are identical, i.e, $\lambda^{(1)}=\lambda^{(2)}=\lambda^{(3)}=\lambda_0$. Then the many-body spectrum can be mapped out in terms of the number of \emph{excitations} alone. Specifically, a possible right eigenoperator of eigenvalue $M\lambda_0$ will be given by 
\eq{\hat{r} = (\otimes_{i=1}^{M}\hat{r}^{(1)}_i) \ ( \otimes_{j=M+1}^N \hat{r}_j^{(0)}) ,}
where the first $M\le N$ sites are excited, i.e., in a non-steady eigenmode, while the remaining sites are in their steady-state. We want to find the total weight of this operator on the basis operators of size $s$ as defined in Eq.~\eqref{eq:size-dist}. Since $\hat{r}^{(k)\neq 0}$ are traceless operators (see Eq.~\eqref{eq:NI-rk}), the first $M$ sites will always contribute a non-identity Pauli string. The leftover $s-M$ non-identity operators have to be chosen from the remaining $N-M$ sites, which can be done in $\binom{N-M}{s-M}$ different ways. Combining this with Eq.~\eqref{eq:NI-r0-before} and \eqref{eq:NI-rk}, we obtain the weight of $\hat{r}$ on size-$s$ basis operators to be 
\eq{p^R_s(\lambda=M\lambda_0) = \binom{N-M}{s-M} (|\eta|^2)^s(|\vec{\beta}|^2)^{s-M}(\phi^2)^{N-s}.}
Using Eq.~\eqref{eq:non-int-purity} and $|\eta|^2=1$, this expression can be written in terms of the steady state purity of the single-site Lindbladian as 
\eq{\label{eq:NI-right-ps} p^R_s(\lambda=M\lambda_0) = \begin{cases}
\binom{N-M}{s-M} \frac{(2p_{ss}-1)^{s-M}}{(2p_{ss})^{N-M}}  &,\text{ if } s\geq M\\
0 &,\text{ if } s < M.
\end{cases}}
This expression illustrates that the right eigenoperators corresponding to eigenvalue $\lambda\sim\mathcal{O}(M\lambda_0)$ consist of basis operators with at least size-$M$. The weight $p^R_s$ on shorter basis operators is exactly zero for this non-interacting toy model when $|\lambda|$ is large (i.e., faster decaying eigenmodes). In the presence of interactions, we expect these vanishing weights to grow into non-zero but exponentially suppressed contributions as observed numerically in Fig.~\ref{fig:RL-secAvg} (a)-(d). In section \ref{sec:perturb}, we provide a heuristic explanation behind origins of these modifications using perturbative arguments.

\subsubsection{Left eigenoperators} 
We now discuss the structure of the left eigenoperators of this model, starting with the single-site eigenmodes. The single-site eigenoperator $\hat{l}^{(0)}$ corresponding $\lambda^{(0)}=0$ is always $\mathbb{I}/\sqrt{2}.$ In general, the left eigenoperators of remaining non-zero eigenvalues are traceful and can be written in terms of the Pauli operators $\vec{S}_i$ as
\eq{\hat{l}_i^{(k)} = f^{(k)} \frac{\mathbb{I}}{\sqrt{2}} + \frac{\vec{g}^{(k)}.\vec{S}_i}{\sqrt{2}},  \text{ for } k=1,2,3.} 
The normalization $\sbra l^{(k)}_i|l^{(k)}_i\sket=1$ imposes condition $|f^{(k)}|^2+|\vec{g}^{(k)}|^2=1$ on the coefficients. As discussed before, a pair of complex conjugate eigenmodes $\lambda_k=\lambda_{k'}^*$ will result in $l^{(k)}={l^{(k')}}^\dagger$.

To get the expressions for coefficients $f,g$ in terms of the purity of the steady state, we make use of the bi-orthogonality relation in Eq.~\eqref{eq:bi-orthogonal}. Then the vanishing overlap of non steady-state left eigenoperator with $r^{(0)}$ in Eq.~\eqref{eq:NI-r0} results in
\eq{\frac{f^{(k)}}{\sqrt{2 p_{ss}}} + \sqrt{1-\frac{1}{2p_{ss}}} \hat{\beta}.\vec{g}^{(k)}=0, \ \text{ for } k=1,2,3. }
Using the normalization condition $|\vec{g}^{(k)}|^2=1-|f^{(k)}|^2$ and introducing parameter $\mu_k:=\hat{\beta}.\vec{g}^{(k)}/|\vec{g}^{(k)}|$, we obtain the relation
\eq{\label{eq:NI-l-coeff}|f^{(k)}|^2 = \frac{|\mu_k|^2\ (2 p_{ss}-1)}{1+|\mu_k|^2(2p_{ss}-1)} .}
Since, by definition, $|\mu_k|^2\leq 1$ we get an upper bound $|f^{(k)}|^2\leq 1-\frac{1}{2 p_{ss}}$. This expression is in contrast with the right eigenoperators, where the coefficient corresponding to $\mathbb{I}$ is strictly zero in Eq.~\eqref{eq:NI-rk}. The size distribution of the right eigenoperators relied heavily on this traceless nature of the $r^{(k)\neq0}$. In the following, we will see that the left eigenoperators being traceful will lead to quite distinct features in their size distribution. The left eigenoperators also become traceless in cases when the steady state is maximally mixed, leading to $p_{ss}=1/2$ (e.g., when all jump operators are Hermitian). This is a consequence of bi-orthogonality condition in Eq.~\eqref{eq:bi-orthogonal}, which requires all left eigenoperators with $\lambda\neq0$ to be orthogonal to the maximally mixed steady state $\hat{r}_0\propto \mathbb{I}$. However, as long as $p_{ss}$ deviates from this limiting value, at least some of the non-steady left eigenoperators become traceful.

Proceeding similarly as we did for the right eigenoperators, we can estimate the total weight of the left eigenoperator on basis operators of size $s$ for eigenvalue $\lambda=M\lambda_0$. A possible eigenoperator will be of the form $\hat{l}= \otimes_{i=1}^M\hat{l}^{(1)}_i$, where first $M$ sites host an \emph{excited} eigenmode. The weight on the basis operators of size $s$ will correspond to the operator strings in $\hat{l}$ that are precisely of size $s$. Out of $M$ non-zero eigenmodes, $s$ of them should contribute as a non-identity operator with coefficient $1-|f|^2$, resulting in
\eq{ p^L_{s}(\lambda=M\lambda_0) = \binom{M}{s} (1-|f|^2)^{s}\ (|f|^2)^{M-s}.} 
Then, using Eq.~\eqref{eq:NI-l-coeff}, the operator size distribution can be obtained as  
\eq{\label{eq:NI-left-ps} p^L_{s}(\lambda=M\lambda_0) =  \begin{cases}
0  &,\text{ if } s > M\\
\binom{M}{s} \frac{(\mu^2(2p_{ss}-1))^{M-s}}{(1+\mu^2(2p_{ss}-1))^M} &,\text{ if } s \leq M
\end{cases},
}
where we have assumed $\mu_k:=\mu$ to be the same for all three left eigenoperators.

While the expressions in Eq.~\eqref{eq:NI-right-ps} and Eq.~\eqref{eq:NI-left-ps} give a qualitative picture of the operator size distribution of eigenoperators, they are not strictly accurate even for the non-interacting model. First, the assumption that the many-body eigenvalue $\lambda=M\lambda_0$ requires exactly $M$ non-zero single-site eigenmodes holds only if the single-site eigenmodes are exactly degenerate. However, in general, they will be distinct from each other with finite gaps in between them. Then for large $M$, according to the central limit theorem, we expect that a many-body eigenmode with $M$ excitations will on average have eigenvalue $\approx M\lambda_0$ with $\mathcal{O}(\sqrt{M})$ fluctuations about this value.  Second, the coefficients of the left eigenoperators will also have variations resulting from the varying values of the parameter $\mu_k$. This will result in some variations of $p^L_s(\lambda)$, where ideally $\mu$ should be replaced by the distribution of $\mu_k$. Nevertheless, we can still get an upper bound on the size distribution by setting all $\mu_k$ to 1. In the limiting case of a maximally mixed steady state ($p_{ss}=1/2$) the size distribution simply becomes
\eq{ p^R_s(M\lambda_0)=p^L_s(M\lambda_0) = \delta_{M,s} \  \ \ \text{if } p_{ss} = 1/2 \ . }
Meaning, both the left and right eigenmodes with $M$ excitations are composed precisely of basis operators of size $M$. Away from this limit, the 
eigenoperators exhibit a more general size distribution discussed above.

\subsubsection{Perturbative corrections}\label{sec:perturb} We now consider the modifications to the size distribution of the non-interacting model upon addition of small interactions and dissipative couplings. We will briefly sketch the general argument here, leaving more detailed calculations for future work. Specifically, we clarify how the strictly vanishing contribution of the short basis operator towards the right eigenoperators in the bulk of the spectrum modifies in the presence of the two-site terms.

We consider the open system dynamics generated by $\mathcal{L} = \mathcal{L}_0 +  \varepsilon \delta \mathcal{L}$, where $\mathcal{L}_0$ is the non-interacting single-site toy model considered above, $\delta \mathcal{L}$ is an interacting piece involving terms that couple nearest-neighbor sites. The real scalar parameter $0\le\varepsilon\le 1$ is used to keep track of the order of the perturbation theory, and set to $1$ at the end of the calculations.   The interaction term can be treated as a perturbation when $||\delta \mathcal{L}||/||\mathcal{L}_0||\sim \gamma_2/\gamma_1\ll1$. Now we assume that the spectrum of the full Lindbladian can be written in terms of a power series as
\eq{\eqsp{\label{eq:pert-series}|r_n\sket &= \sum_{k=0} \varepsilon^k |r_n^{(k)}\sket , \quad \sbra l_n| = \sum_{k=0} \varepsilon^k \sbra l_n^{(k)}| ,  \\
\quad \lambda_n &= \sum_{k=0} \varepsilon^k\lambda_n^{(k)} ,}}
where the $r_n^{(0)}$ and $l_n^{(0)}$ are the eigenoperators of the unperturbed Lindbladian $\mathcal{L}_0$ with the eigenvalue $|\lambda_n^{(0)}| = n\gamma_1$.

Here we will focus on determining the perturbative correction to the size distribution of the right eigenoperators $p^R_{s=1}$ for eigenvalues in the bulk of the spectrum, where $n \gg 1$. From Eq.~\eqref{eq:NI-right-ps}, we see that $p^R_{s=1}(\lambda_n^{(0)})=0$ in the non-interacting limit, if $n>1$.  To find the perturbative correction to this, we need to compute overlaps of the form $\sbra F_1|r_n\sket$, where $F_1$ is a single-site basis operator. Since the non-interacting eigenoperators form a complete bi-othornormal basis, we can write this overlap as
\eq{ \eqsp{\sbra F_1 | r_n\sket & = \sum_{m,a}\frac{ \sbra F_1|r_{m,a}^{(0)}\sket \sbra l_{m,a}^{(0)}|r_n\sket} {\sbra l_{m,a}^{(0)}|r_{m,a}^{(0)}\sket } \\
&=cp^{R(0)}_{s=1} \frac{\sbra l^{(0)}_{m=1}|r_n\sket}{\sbra l_{1}^{(0)}|r_{1}^{(0)}\sket }, }}
where $a$ labels the eigenstates within each degenerate eigenspace, and $p^{R(0)}_{s=1}$ is the non-interacting size distribution. The last equality follows from the non-interacting result $\sbra F_1|r_{n>1}^{(0)}\sket=0$ (see Eq.~\eqref{eq:NI-right-ps}), and the proportionality constant $c$ accounts for contribution from three relevant degenerate eigenmodes at $\lambda^{(0)}_{n=1}$. In conclusion, we only need to find the component of $|r_n\sket$ in the direction of the unperturbed eigenvector of eigenvalue $|\lambda^{(0)}_1| = \gamma_1 \ll |\lambda_n^{(0)}|$. 

To proceed, we first calculate the order $k=1$ term in the perturbative expansion Eq.~\eqref{eq:pert-series} using a modification of standard non-degenerate perturbation theory to the \emph{bi}-orthogonal basis as
\eq{ \sbra l_m^{(0)}| r_n^{(1)}\sket = \frac{\sbra l_m^{(0)}|\delta \mathcal{L}| r_n^{(0)}\sket}{\lambda_n^{(0)}-\lambda_m^{(0)}} .}
Since each 2-site term in $\delta \mathcal{L}$ can reduce the size of the eigenoperator $| r_n^{(0)}\sket$ by at most by $1$,\footnote{Note that the $2$-site Lindbladian terms cannot decrease the size of an operator by $2$ due to their trace-preserving property; since for any two Pauli operators $\sigma_j^{\alpha} \sigma_{j+1}^\beta$,  $\sbra II |\delta \mathcal{L}|\sigma_j^{\alpha} \sigma_k^\beta \sket=\tr{(\delta\mathcal{L}^\dagger[II] \sigma_j^{\alpha} \sigma_{j+1}^\beta)}=0$.}  $\delta \mathcal{L}| r_n^{(0)}\sket$ is an operator of size at least $n-1$.   As a result we conclude that $\sbra l_m^{(0)}| r_n^{(1)}\sket $ is non zero only if $m\ge n-1$. Repeating this argument for higher order terms in the perturbation theory, we conclude that the local nature of the perturbation and the strict size-dependence of the non-interacting Lindblad model implies that 
\eq{\sbra l^{(0)}_m|r^{(k)}_n\sket \neq 0 \text { only if } m\ge n-k. }
From this, we see that the desired overlap $\sbra l_{m=1}|r_n\sket$ will only appear at $(n-1)^{\text{st}}$ order in the perturbation theory. 

Continuing this procedure up to $(n-1)^{\text{st}}$ order, it is straightforward to see that the leading order correction is given by 
\eq{ \sbra l_1^{(0)}|r_n\sket \sim  \prod_{m=1}^{n-1} \frac{1}{\sbra l^{(0)}_m|r^{(0)}_m\sket} \frac{ \sbra l_m^{(0)}|\delta\mathcal{L}|r_{m+1}^{(0)}\sket }{ (\lambda_{n}^{(0)}-\lambda_{m}^{(0)})} }
Recalling that $|\lambda_n^{(0)}|=n\gamma_1$, and $||\delta\mathcal{L}||\sim\mathcal{O}(\gamma_2)$, we obtain the approximate leading order correction to the size distribution as  
\eq{p^R_{s=1}(|\lambda|=n\gamma_1) \approx \frac{1}{(n-1)!}\bigg(\frac{\gamma_2}{\gamma_1}\bigg)^{n-1} .}
Similar to the standard perturbation theory procedure for Hamiltonians, we expect the remaining subleading higher-order terms to converge, leading to an exponentially suppressed weight
\eq{  p^R_{s=1}(|\lambda|=n\gamma_1) \approx \bigg(\frac{\gamma_2}{\gamma_1}\bigg)^{|\lambda/\gamma_1|} = e^{-\frac{|\lambda|}{\gamma_1}\log{(\frac{\gamma_1}{\gamma_2})}} .}

Thus perturbatively including 2-site operators to our toy model, we recover the generic numerical observation in Fig.~\ref{fig:secAvg}(a) and \ref{fig:RL-secAvg}, where the overlap of right eigenoperators in the bulk of the spectrum (i.e. $|\lambda|\propto N$) with short basis operators is exponentially suppressed with increasing absolute value of $\lambda$.

\begin{figure}
    \includegraphics{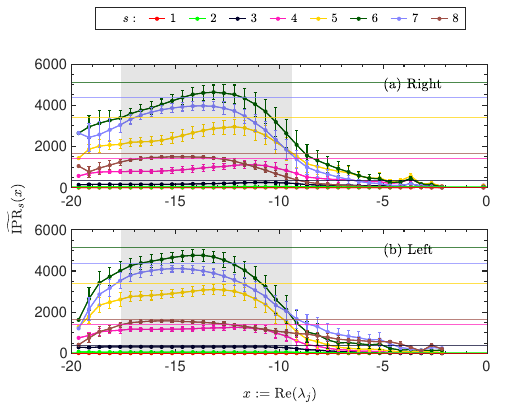}
	\caption{The coarse-grained inverse participation ratio (IPR) for the eigenoperator of the dissipative Ising model (see Eq.\eqref{eq:IPR_tilde}) plotted as a function of the real part of the eigenvalue $\lambda$. The data is shown for the (a) right and (b) left eigenoperators of $N=8$ size dissipative Ising model with parameters $J=1,h_x=1.3,h_z=1.2$, and $\gamma=0.8$. The coarse-graining scale is set to $\Delta=0.5$. The error bars show the standard deviation of the IPR computed for eigenmodes in the window of size $\Delta$. The horizontal lines correspond to the $\text{IPR}(s)$ expected for the eigenoperator of a non-Hermitian Ginibre matrix and equals $b_N(s)/2$, where $b_N(s)$ is the number of basis operators of size $s$. The shaded area depicts the extent of the ``bulk'' of the spectrum defined as $|\text{Re}(\lambda)-\overline{X}|\le2\sigma_X$ (see \cref{eq:xbar,eq:sigmaXY}). }
    \label{fig:ising-sec-IPR} 
\end{figure}

\begin{figure*}
    \includegraphics{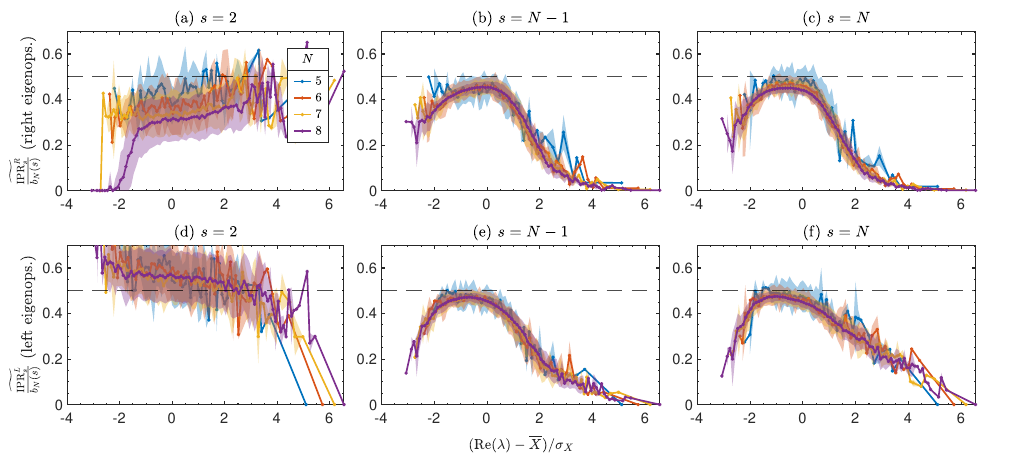}
    \caption{Coarse-grained inverse participation ratio (IPR) is shown as a function of the real part of eigenvalues for the dissipative Ising model (see Eq.\eqref{eq:IPR_tilde}). The data is shown for the (top row) right eigenoperators and the (bottom row) left eigenoperators. Each column shows the IPR in the sector of size $s$. The values of IPR are color-coded based on the total number of sites $N=5,6,7,8$. In order to compare different system sizes, the y-axis is scaled by the total number of size-s basis operators $b_N(s)$. Similarly, the x-axis is rescaled using the mean of eigenvalues $\overline{X}$, and standard deviation $\sigma_X$ of the real part of $\lambda$ such that the center and width of the spectrum for different system-sizes coincide. The shaded patches indicate the standard deviation of the IPR over the eigenmodes within the coarse-graining window $\Delta = 0.1\sigma_X$. The dashed horizontal line shows $b_N(s)/2$ --- the corresponding result for the eigenoperators of the random Ginibre matrix. } 
    \label{fig:ising-sec-IPR-sys} 
\end{figure*}

%%===============================================================
\begin{figure}
	\includegraphics{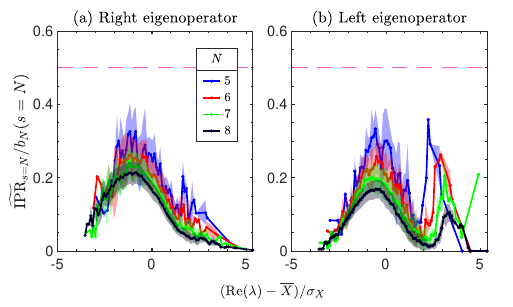}
	\caption{\label{fig:RL-10-ipr} The inverse participation ratio (IPR) of the eigenoperators of the random Lindblad model $\mathtt{seed}= 1$ ($J=1,\gamma_1=\gamma_2=0.25$): The data is shown for (a) right and (b) left eigenoperators as a function of the rescaled eigenvalue in the size sector $s=N$. Here $\overline{X},\sigma_X$ are the center and the standard deviation of the spectrum along the real axis, respectively. Here we illustrate the realization of the random model, which has considerably smaller eigenoperator-IPR compared to the Ising model in Fig.~\ref{fig:ising-sec-IPR-sys}. Several other realizations do saturate the IPR value to $\approx b_N(s)/2$ in their bulk. }
\end{figure}
%%==================================================================

\subsection{Scrambling of eigenoperators}
\label{sec:evec-ipr}
After identifying the dependence of the size distribution $p_s$ of the eigenoperators on their eigenvalue, we proceed to analyze how individual basis operators of size $s$ contribute to the weight $p_s$. In particular, we are interested in the degree of scrambling of eigenoperators within the space of basis operators of given size $s$. 

According to the random matrix theory of closed quantum systems, if the interactions are not restricted to be local, the Hamiltonian eigenstates are essentially random vectors. Along similar lines, if we sample a generic Lindbladian without any constraints from locality, its eigenoperators should also be randomly scrambled over the basis operators. In Appendix~\ref{app:non-local}, we confirm this intuition numerically by analyzing eigenoperators of a random non-local Lindbladian. We observe that the eigenoperators are equally scrambled over all basis operators, akin to eigenvectors of a random non-Hermitian matrix from the Ginibre ensemble (see Fig.~\ref{fig:nonLoc-IPR}). On the other hand, when we restrict the interactions and dissipation to be local, we observed in the previous section that the type of basis operators that compose a given eigenoperator depends on its eigenvalue. However, since the models considered here are generic beyond these locality constraints, we expect the bulk eigenoperators to be highly scrambled across eigenoperators within each size sector.

We address this question by analyzing the behavior of the inverse participation ratio (IPR) of the eigenoperators in each size sector. The IPR for a general probability distribution is defined by 
\eq{\label{eq:simple-IPR} \text{IPR} = \frac{(\sum_{j\in\Omega}P_j)^2}{\sum_{j\in\Omega}P_j^2 },}
where $P_j$ is probability of the state labeled by $j$ in some set $\Omega$ \cite{santos2010onset,rigol2010gapped}. The IPR estimates the number of states that have a non-negligible probability. If the distribution is localized on a single state $j_0$, then the IPR evaluates to 1. Whereas, if the distribution is uniform, i.e., maximally delocalized over all states, then the IPR is equal to the total number of states in $\Omega$.

We analogously define the IPR of an operator $O$ on the space of basis operators \cite{gill2025speedlimitsscramblingkrylov} by recalling that the overlap $|\sbra F_\mathbf{m} | O\sket|^2$ can be interpreted as un-normalized probability (see Eq.~\eqref{eq:size-dist}). Specifically, we are interested in the IPR within a specific size sector $s_0$. Hence, identifying $P_j$ in Eq.~\eqref{eq:simple-IPR} with the overlap $|\sbra F_\mathbf{m} | O\sket|^2$, we define IPR of an operator $O$ in size-sector $s_0$ as 
\eq{\label{eq:ipr-s}\text{IPR}_{s_0}(O) = \frac{\big(\sum_{\mathbf{m}}|\sbra F_{\mathbf{m}}|O\sket |^2 \ \delta_{\mathtt{S}[\mathbf{m}],s_0}\big)^2}{\sum_{\mathbf{n}}|\sbra F_{\mathbf{n}}|O\sket |^4 \ \delta_{\mathtt{S}[\mathbf{n}],s_0}} .}
We numerically analyze the IPR of the left and right eigenoperators of the eigenmodes in each size sector. Since there is very little variation in the IPR as a function of the imaginary part of the eigenvalues, it is convenient to compute the coarse-grained IPR defined as 
\eq{\label{eq:IPR_tilde} \widetilde{\text{IPR}}_{s}^{R}(x_0) = \frac{1}{n_{\Delta}(x_0)}\sum_{\{j:|\text{Re}(\lambda_j)-x_0|\leq\Delta/2\}} \text{IPR}_{s}(r_j),} 
where $n_{\Delta}(x_0)$ is the number of eigenvalues in the vertical strip of size $\Delta$ around $\text{Re}(\lambda)=x_0$. An analogous definition follows for the IPR of the left eigenoperators. 

The IPR of the eigenoperators of the dissipative Ising model computed across the basis operators of different sizes is shown in Fig.~\ref{fig:ising-sec-IPR}. 
The horizontal lines show the value for eigenoperators of a random non-Hermitian matrix sampled from the Ginibre ensemble, $\text{IPR}(s)\approx b_N(s)/2$, where $b_N(s)$ is the total number of size-s basis operators.  This is exactly the value expected for a random operator $\sum_\mathbf{m} c_\mathbf{m} F_\mathbf{m} \delta_{\mathtt{S}[\mathbf{m}],s}$ where $c_\mathbf{m}$ are complex random numbers with their real and imaginary sampled independently from the standard normal distribution.  
We observe that the eigenoperators of the bulk eigenmodes, near the center of the spectrum, have a high IPR that is quantitatively close to this value.  This suggests that within each size sector, bulk eigenoperators are well-approximated by a random mixture of basis operators of specific size, while their total amplitude in each sector is modulated by the eigenvalue-dependent size distribution described in section~\ref{sec:eigenvecs}.

In the case of non-local Lindbladians, the IPR for all eigenoperators saturates to the maximal Ginibre value irrespective of their eigenvalue (see Fig.~\ref{fig:nonLoc-IPR} in the appendix). In contrast, we observe that the eigenoperators of our \emph{local} Lindblad model reach a high degree of scrambling only near the center of the spectrum, where the density of states is large; the IPR of Lindblad eigenoperators becomes small in regions of low eigenmode density. Since the Lindblad models studied here only have local terms, eigenmodes with very large and small eigenvalues would depend quite a lot on the microscopic details. Hence, we expect the eigenmodes near the boundaries, i.e., the slowest and fastest decaying modes, to be more structured. For instance, if the dynamics possesses approximate conservation laws, then the slowly decaying eigenmodes are primarily concentrated over these quasi-conserved quantities, leading to lower scrambling in this part of the spectrum \cite{abanin2025integralsmotionslowmodes}. This is similar to the behavior of generic Hamiltonian eigenstates away from the bulk of the spectrum, which are typically less scrambled when the density of states is small.  In these systems, the IPR of energy eigenstates near the edges of the spectrum depends on the specific choice of basis and the nature of microscopic interactions \cite{rigol2010gapped,santos2010onset}. 

In Fig.~\ref{fig:ising-sec-IPR-sys}, we analyze the change in the $\text{IPR}(s)$ as a function of the total number of sites $N$, for $s=2, N-1$, and $N$. For each of these cases, the value of $\text{IPR}(s)$ increases with the increasing system size. Upon rescaling the y-axis by the number of basis operators $b_N(s)$, the curves for different values of $N$ collapse on each other. This implies that the IPR of the eigenoperators from the bulk part of the spectrum over the size-$s$ sector is approximately proportional to the total number of basis operators of that size. If the eigenoperators were localized, the typical number of contributing basis operators would be fixed, leading to a decrease in IPR with increasing values of $N$. This provides further evidence for the scrambled nature of the bulk eigenoperators. We note that while the scaling $\text{IPR}(\lambda) = C(\lambda)b_N(s)$ is universal in the bulk of the spectrum, the detailed functional form of $C(\lambda)$  is not universal and varies across different models. The IPR of the right eigenoperators on short basis operators, shown in Fig.~\ref{fig:ising-sec-IPR-sys}(a) for $s=2$, appears to increase more slowly than this scaling. However, the weight $p_{s=2}^R$ of these eigenoperators in the size $s=2$ sector decreases exponentially with the increasing $|\text{Re}(\lambda)|$, which suggests that characterizing further scrambling of such small contributions is not particularly meaningful.

In Fig.~\ref{fig:RL-10-ipr}, we show the IPR of the eigenoperators of the $\mathtt{seed}=1$ realization of the random Lindblad model. Here, we deliberately select this realization as it highlights a qualitatively different behavior compared to the Ising model. The bare value of the IPR does increase with the increasing system size. However, the value of $C(\lambda):=\frac{\text{IPR}(s)}{b_N(s)}$ is more strongly $\lambda$ dependent, with a maximum value closer to $0.2$ than $0.5$. Moreover, unlike the Ising model case, the value of $C(\lambda)$ appears to decrease when the system size increases $N=5,6$ to $7$. For $N=8$, it looks to saturate within the error bars, but the precise conclusion would require data for larger system sizes. While this particular class of realizations of the random model exhibits a lower degree of scrambling in the eigenoperators, the IPR of the bulk eigenoperators in the majority of 20 other realizations of the random Lindblad model considered for this work (data not shown here) do scale approximately as $b_N(s)/2$.

\begin{figure}
	\includegraphics{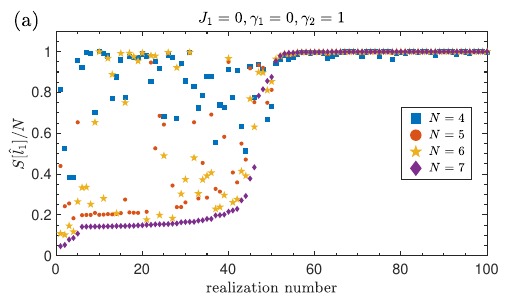}	
	\includegraphics{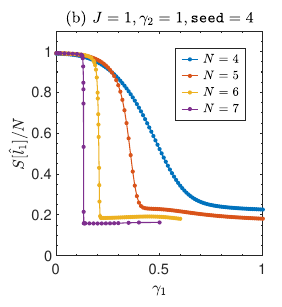}%
    \includegraphics{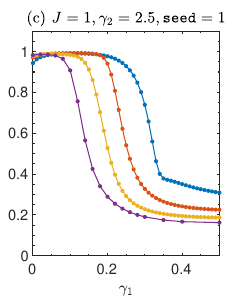}
    \caption{Unusually large size of slow decay modes: (a) Total size $S[\hat{l}_1]$ of the left eigenoperator $\hat{l}_1$ corresponding to the eigenmode with the smallest negative real part is shown for $100$ independent realizations of the random Lindblad model with only two-site jump operators. The realizations are ordered in ascending order of $S[\hat{l}_1]$ in the $N=7$ system for ease of visualization (hence should not be compared with the label $\mathtt{seed}$ used elsewhere in the text). (b) Total size $S[\hat{l}_1]$ is plotted as a function of single-site dissipation strength $\gamma_1$. The data points are shown for $J=1,\gamma_2=1$ and varying system sizes $N$ for the realization labeled by $\mathtt{seed=4}$. The lines joining the data points are guides to the eyes. (c) Analogous plot for realization $\mathtt{seed}=1$ with $J=1,\gamma_2=2.5$. The realizations in panel (b--c) are chosen such that they exhibit $S[\hat{l}_1]=N$ for all values of $N$. The y-axes of all panels are normalized by the total system size $N$, which corresponds to the largest possible size an operator can attain (see Appendix~\ref{app:unusual-op} for additional details).} 
    \label{fig:unusual-evec} 
\end{figure}

\subsection{Anomalous size distribution}
\label{sec:unusual-evec}

Based on the general expectation that the size of the operator should control its decay rate, we expect that the eigenoperator $\hat{l}_1$ corresponding to the eigenmode with the smallest non-zero absolute real part $\lambda_1$ should be a small-sized operator. This intuitive expectation is exactly true for the non-interacting model studied above, where we find that $\hat{l}_1$ is an operator of size not greater than $1$. However, we observe that in certain cases where the two-site dissipative terms dominate, the left-eigenoperator $\hat{l}_1$ is a large operator, with Pauli weight on the order of the system size. Here we present the results related to the structure of these eigenoperators; we comment on the implications for operator growth in section~\ref{sec:unusual-op-dyn}.

In Fig.~\ref{fig:unusual-evec}(a), we show the size of this slowest decaying eigenoperator $\hat{l}_1$ for 100 independently sampled realizations of random Lindblad models with only two-site dissipative terms. We explicitly set the Hamiltonian interactions $J$ and single-site dissipation $\gamma_1$ to zero. We observe that for a finite fraction of realizations, the size of the operator $\hat{l}_1$ increases linearly with increasing $N$, displaying this anomalous behavior. In some of the realizations, we observe an even-odd effect --- where $N=4,6$ exhibit anomalously large $\hat{l}_1$ operators, while the eigenoperator remains short in $N=5,7$.

To understand the robustness of this anomalous behavior beyond the $J=\gamma_1=0$ limit, let us consider a specific realization of the random Lindblad model that consistently shows $\hat{l}_1\approx N$ for all values of $N$ considered here when $J=\gamma_1=0$. In Fig.~\ref{fig:unusual-evec}(b), we observe that the size of $\hat{l}_1$ in the realization labeled by $\mathtt{seed}=4$ remains of the $\mathcal{O}(N)$ up to some finite value of single-site dissipation strength $\gamma_1^c$. Beyond this threshold value, the single-site dissipation dominates and the left-eigenoperator $\hat{l}_1$ becomes a short operator, consistent with the non-interacting model. In contrast, increasing $J$ while keeping $\gamma_1=0$ seems to have a very small effect on the size of the eigenoperator, with $S[\hat{l}_1]\sim N$ even when $J\sim 10\gamma_2$.

The sharp drop-off of the operator size as a function of $\gamma_1$ seen in Fig.~\ref{fig:unusual-evec}(b) is not observed for all models --- for some of the realizations, the size of the operator smoothly crosses over from $s=N$ to $s\sim 1$ in a manner that does not sharpen with system size. One such example, $\mathtt{seed=1}$, is shown in Fig.~\ref{fig:unusual-evec}(c). However, in all models studied here, the value $\gamma^c_1$ at which this crossover occurs decreases with increasing system size. Heuristically,  this is because the decoherence rate of an operator due to the single-site jump operators necessarily increases with the increasing size of the operator. As a result, it is reasonable to expect that, for a fixed value of $\gamma_1$, the shift in the eigenvalue of the anomalously large operator $\hat{l}_1$ would be larger for larger system sizes. Meaning, for larger system sizes, the value of $\gamma_1$ at which this single-site decoherence dominates decreases. Additionally, in appendix~\ref{app:unusual-op}, we show the dependence of the eigenvalue $\lambda_1$ as a function of system size $N$ for the $\gamma_1=0$ case. There, we observe that when the corresponding eigenoperator has size of $\mathcal{O}(N)$, the absolute real part of the eigenvalue also increases with increasing values of $N$. On the other hand, the eigenvalues with $\mathcal{O}(1)$ size eigenoperators remain independent of the system size. This suggests that for large enough system sizes, which are not numerically accessible here, the slowest decaying eigenoperators will have $\mathcal{O}(1)$ size in our generic models.

\section{Decoherence of initial states}
\label{sec:state_dynamics}

Having observed the universal features of the Lindblad spectrum, we now proceed to analyze their impact on the dynamical evolution of the system. In this section, we characterize the decoherence of the initial states by analyzing the time evolution of their purity and R\'enyi-2 correlation functions. We show that the early-time decoherence of generic entangled states is universal and independent of the specific choice of initial states. For the purity, this universality has also been established in random (non-local) ensembles of Lindbladians \cite{campo2019extreme,campo2024purity}. We explore how such universality arises as a result of the eigenoperator size distributions obtained in the previous section.

We begin by clarifying what kinds of correlation functions are expected to be sensitive to the bulk of the Lindblad spectrum. The expectation value of a local observable $\hat{A}$ evolves according to 
\eq{\label{eq:linear-corr} \tr{(\hat{A}\rho(t) )} = \sum_j e^{\lambda_j t}\frac{\sbra A|r_j\sket\sbra l_j|\rho_0\sket}{\sbra l_j |r_j\sket}, }
where $\rho(t)=e^{\mathcal{L}t}\rho_0$ is the time evolved state of the system, initialized in $\rho_0$ at $t=0$. In section~\ref{sec:eigenvecs}, we observed that the eigenoperators in the bulk of the spectrum are generally made up of large-weight Pauli operators. Specifically, the support of right eigenoperators on one or two-site operators is exponentially suppressed with increasing value of $|\text{Re}(\lambda)|$ (see Fig.~\ref{fig:secAvg}). Consequently, the contribution of the generic eigenmodes from the bulk of the Lindblad spectrum towards the dynamics of the local operator $\hat{A}$ is highly suppressed. However, the initial states $\rho_0$ with sufficiently high purity are in general highly non-local operators. As a result, correlation functions solely involving their overlap with the Lindblad eigenoperators are expected to show universal dynamical features. 

In the following, we specifically consider  a set of non-linear correlators of form 
\eq{ \label{eq:general-nonlinear} \eqsp{\sbra \rho(t)| \mathcal{A}|\rho (t)\sket = \\ &\sum_{j,k} e^{(\lambda_j^*+\lambda_k)t}\frac{\sbra \rho_0|l_j\sket \sbra r_j|\mathcal{A}|r_k\sket\sbra l_k|\rho_0\sket}{ \sbra r_j|l_j\sket \sbra l_k | r_k \sket } },}
where $\mathcal{A}:=A\otimes A^* \longleftrightarrow A (\ .\ )A^\dagger $ acts as a superoperator. Let us analyze the eigenoperator overlaps in this expression. First, the density operator $\rho_0$, being a non-local operator, has significant overlap with the left eigenoperators from the bulk of the spectrum. Unlike the linear correlator expression in Eq.~\eqref{eq:linear-corr}, overlaps of the form $\sbra r_j|\mathcal{A}|r_k\sket$ correspond to whether the two eigenoperators are related by local deformations by $A$. We expect that such overlap will be highest when the corresponding eigenvalues are close to each other, since their operator size-distribution will also be similar to each other, allowing local deformations between the pair. The overlaps of such local superoperators with the eigenvectors of random non-Hermitian matrices \cite{cipolloni2024nonHermitianViolatesETH,roy2025nonHermitianETH}, and generic Lindbladians \cite{ferrari2025openStripeETH} have been recently studied to formulate an equivalent of the ETH ansatz for non-Hermitian systems. In appendix~\ref{app:sup-ops}, we present the dependence of these overlaps on the eigenvalue differences. We show that, according to this metric, the eigenoperators of the non-local Lindblad model have correlations identical to the random Ginibre matrices. However, when we consider the local models studied here, the overlaps exhibit the anticipated suppression at large values of the difference in the eigenvalues of the corresponding eigenoperators. A systematic study of this functional dependence and its implications on the open system dynamics is left for future investigations. However, for eigenvalues $\lambda_j,\lambda_k$ that are near each other in the complex plane, the value of $\sbra r_j |\mathcal{A}|r_k\sket $ is not suppressed in local Lindbladian models, which ensures that these generic eigenmodes can contribute to the early time universal dynamics as we discuss below.

%%===============================================================
\begin{figure}
	\includegraphics{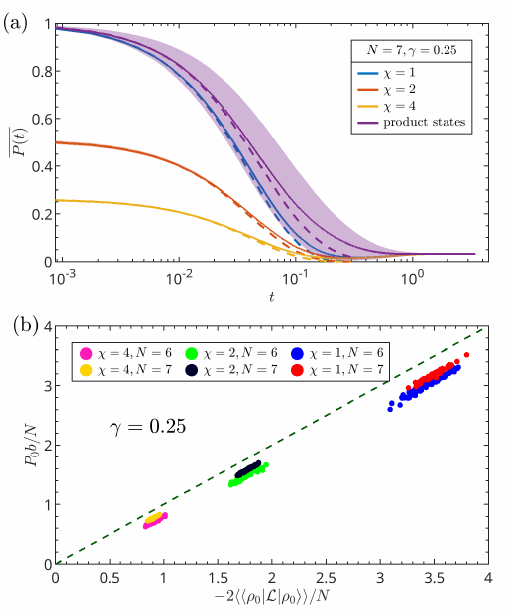}
	\caption{\label{fig:RL-10-purity} Dynamics of purity in realization of $\mathtt{seed=4}$ of the random Lindblad model $(N=7,J=1,\gamma_1=\gamma_2=\gamma = 0.25)$: (a) The purity of the time-evolved state is plotted as a function of time. The data is shown for varying values of the initial state purities parameterized using the rank of the density matrix $\chi$ (see Eq.~\eqref{eq:init_purity}). The purple curve shows the data for product states from the Haar random ensemble defined in Eq.~\eqref{eq:product-state}. The solid lines show the average value of purity, $\overline{P(t)}=\frac{1}{N_r}\sum_a \tr{(\rho^2_a(t))}$, for $100$ independent initial states from each ensemble, and the shaded patch shows the standard deviation over these initial states. The variance of initial decoherence rates for the entangled states ($\chi=1,2,4$) is much smaller, and hence the corresponding shaded patches are of the same thickness as the thickness of the line. The dashed lines show the average of analytical estimate according to $\frac{1}{N_r}\sum_{a}P_0e^{-D_a t}$, where $D_a$ is the decoherence rate of state $\rho_a$ defined in Eq.~\eqref{eq:decoherence}. (b) $P(t)/P_0$ is fitted with the function $e^{-b t}$ for every initial state shown in panel (a), where $P_0$ is the purity of the state at $t=0$. The early time data ($t$ up to $dP/dt<0$) is used to obtain the fit. The fit parameter $b$ is compared with the theoretical expectation for the decoherence rate in Eq.~\eqref{eq:decoherence}. The data approaches the $y=x$ line with the increasing system size.}
\end{figure}
%%==================================================================

\subsection{Purity}
We begin by considering the case where the local operator in Eq.~\eqref{eq:general-nonlinear} is $A = \mathbb{I}$. The resulting non-linear correlator corresponds to the purity of the system given by $P(t) = \tr{(\rho(t)^2)}$, where the density matrix $\rho(t)$ describes the state of the system at time $t$. If the system is in a pure quantum state, then its purity is equal to 1. The purity of a system in a mixed state is strictly less than 1. The smallest value it can take is $1/d$, for a system in the maximally mixed state $\frac{1}{d}\mathbb{I}$, where $d$ is the dimension of the Hilbert space \cite{nielsenChuang}. The purity of the time-evolved state quantifies the degree of decoherence resulting from the open system dynamics \cite{lidar1998dfs}, and it also serves as an experimentally accessible metric for the entropy of the system \cite{islam2015purity,kaufman2016thermalizationExperiment}. Several recent works have explored its dynamics in the context of a variety of Lindblad models \cite{campo2024purity,campo2019extreme,zhou2021renyi,hantengWang2024sykLindbladEntanglement}.  The purity of the state of the system evolving according to the Lindblad dynamics in Eq.\eqref{eq:lindblad} is given by
\eq{\label{eq:purity} P(t):=\tr{(\rho(t)^2)} = \sbra\rho_0|e^{\mathcal{L}^\dagger t}e^{\mathcal{L}t}|\rho_0\sket.}
where $\rho_0$ is the initial state of the system at time $t=0$. The open system dynamics evolves an initially pure quantum state into a mixed state, which we will characterize in terms of the change in the purity of the system.

To quantify this decoherence, we start by defining a set of initial states. As discussed in section~\ref{sec:model}, the models considered here obey a weak reflection symmetry condition leading to block diagonalization of the Lindbladian in the operator space. Hence, we consider random initial states that live in the $+1$ eigenspace of the reflection symmetry operator. Let us consider an un-normalized random matrix defined by
\eq{\label{eq:init_purity} M_\chi = \sum_{i,j=1}^{d_+} \left(\sum_{k=1}^\chi a_{i,k}a_{j,k}^*\right) |i\rangle\langle j|, }
where $\{|i\rangle,|j\rangle\}$ are the basis vectors of the $d_+$- dimensional +1 eigenspace of the reflection operator and $a_{i,k}$ are complex random variables with their real and imaginary parts sampled from the standard normal distribution. The integer $\chi$ parametrizes the purity of the resulting density matrix defined as 
\eq{\rho_\chi  = \frac{M_\chi}{\tr{(M_\chi)}}.} 
The denominator ensures the correct normalization, and it is easy to check that $\rho_{\chi}$ is a positive (semi-)definite Hermitian matrix. If $\chi=1$, then this is a density matrix of a pure entangled state. However, $\chi>1$ gives a mixed density matrix. We can interpret this as a classical mixture of $\chi$ different entangled states, or equivalently, a state obtained by tracing out a $\chi$-dimensional subsystem of a $d\otimes \chi$-dimensional pure Haar random state \cite{zyczkowski2001induced,zyczkowski2011generating}. As a result, the purity of $\rho_\chi$ is approximately given by $1/\chi$. In what follows, we will focus on both pure and moderately mixed initial states.

In Fig.~\ref{fig:RL-10-purity}(a), we show the time evolution of the purity in the realization labeled by $\mathtt{seed}=4$ of the random Lindblad model. The plots for purity in this model are qualitatively similar to those in other independently sampled realizations and the dissipative Ising case. The details related to the spectrum and the generic Ginibre-like features of this realization are shown in Fig.~\ref{fig:RL-sd-4-spectrum}. Unless stated otherwise, we use the $\mathtt{seed}=4$ realization to illustrate the numerical results in this section. The initial states are chosen according to Eq.~\eqref{eq:init_purity} for $\chi=1,2,$ and $4$. The purity of the initial state decreases under Lindbladian evolution. At late times, the system approaches its unique steady state $\rho_{ss}$, leading to saturation of its purity to $\tr{(\rho_{ss}^2)}$. In general, when the jump operators are not Hermitian, purity evolution is non-monotonic \cite{lidar2006purityMonotonicProof}. We observe two key features that we expand on below: First, the time evolution of the purity is independent of the specific initial state within the chosen ensemble of entangled states. Second, the early time decay of the purity is approximately captured by the decoherence rate shown using dashed curves in Fig.~\ref{fig:RL-10-purity}(a). 

%===============================================================
\begin{figure}
	\includegraphics{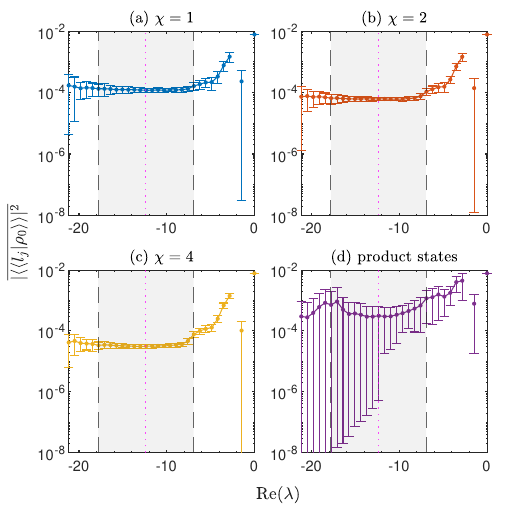}
	\caption{The overlap of initial states with the left eigenoperators in the random Lindblad model is plotted as a function of the real part of eigenvalue. The data is shown for parameters $N=7,J=1,\gamma_1=\gamma_2=0.25,$ and $\mathtt{seed=4}$. We show the coarse-grained values of the overlaps by averaging over both the eigenmodes with a fixed value of Re$(\lambda)$, and ensemble of initial states (see Eq.~\eqref{eq:vlrho-avg}) The data is shown for $100$ randomly chosen initial states according to Eq.~\eqref{eq:init_purity} with (a) $\chi=1$, (b) $\chi=2$ and (c) $\chi=4$. (d) translationally invariant Haar random product states defined in Eq.~\eqref{eq:product-state}. The coarse-graining window is set to $\Delta=0.68\approx\sigma_X/4$. The error bars show the standard deviations of the coarse-grained values for given $x_0$ over these initial states. The error bars are capped at the smallest value when the standard deviation is larger than the average value. The shaded region indicates the bulk portion of the spectrum defined by $|\text{Re}(\lambda)-\overline{X}|\le2\sigma_X$ (see Eq.~\cref{eq:xbar,eq:sigmaXY}).}
	\label{fig:RL-10-vlRho-disc} 
\end{figure}
%===============================================================

The initial state independence in the early time dynamics of the purity shown in Fig.~\ref{fig:RL-10-purity}(a) is a result of universality in the spectral overlaps of generic initial states with the Lindblad eigenoperators. Using Eq.~\eqref{eq:dyn}, the explicit expression of the time evolution of the purity can be obtained as
\eq{ \label{eq:purity-full-expr} P(t) = \sum_{j,k} e^{(\lambda_j^*+\lambda_k)t} \ \frac{\sbra \rho_0|l_j\sket \sbra r_j|r_k\sket\sbra l_k|\rho_0\sket}{ \sbra r_j|l_j\sket \sbra l_k | r_k \sket } .}
The dependence on the initial state $\rho_0$ is primarily controlled by its overlap $\sbra l_k|\rho_0\sket$ with the left eigenoperators. Therefore, we begin by analyzing the variation of this term %towards Eq.~\eqref{eq:purity-full-expr}. 
across different randomly chosen initial states.  
We first focus on the absolute value, which is sensitive to the initial density matrix's distribution across operators of different sizes.   % note that $\sbra l_k|\rho_0\sket$ are in general complex numbers. However, since 
Subsequently, we verify that the phases of these complex numbers are uncorrelated, which prohibits additional magnification of any fluctuations across different initial states. Specifically, we consider the coarse-grained overlaps defined by 
\eq{ \label{eq:vlrho-avg} \overline{[|\sbra l |\rho_0\sket|^2]_{\Delta}}(x_0) : = \sum_{a=1}^{N_r} \ \sum_{j:|\text{Re}(\lambda_j)-x_0|\le \Delta/2 }\frac{|\sbra l_j|\rho_0^{(a)}\sket |^2}{N_r n_\Delta},}
where the first averaging is done over a coarse-graining window of size $\Delta$, which in total contains $n_\Delta$ eigenmodes, and followed by averaging over $N_r$ randomly chosen initial states from the given ensemble. 

In Fig.~\ref{fig:RL-10-vlRho-disc}(a-c), we show the spectral overlaps (\ref{eq:vlrho-avg}) as a function of the real part of their eigenvalues for different ensembles of initially pure and mixed states defined in Eq.~\eqref{eq:init_purity}.
These values of spectral overlaps averaged over the initial states in the given ensemble are approximately constant in the bulk of the spectrum. Moreover, the fluctuations across the different states within this ensemble of random states remain small, as witnessed by the relatively small error bars. This corroborates our earlier expectation that a generic initial state $\rho_0$, being a non-local operator, will overlap with bulk eigenmodes of the spectrum, which themselves are non-local operators as shown in section~\ref{sec:eigenvecs}. The high IPR of bulk eigenmodes over non-local operators means that a typical high-weight operator is effectively a random vector in the Lindbladian eigenbasis, leading to the observed uniform coarse-grained spectral overlaps. The contribution of the overlaps
to the purity of the state defined in Eq.~\eqref{eq:purity-full-expr} is additionally modified by the overlap of left and right eigenoperators $1/\sbra r_j|l_j\sket$. In the appendix (see Fig.~\ref{fig:app-vlrho-alpha}), we show that the modified overlaps are still independent of the initial state. This uniform nature of spectral overlaps results in the initial state-independent early time dynamics of the purity observed in Fig.~\ref{fig:RL-10-purity}. 

Finally, we analyze the contribution from the phases of complex coefficients in Eq.~\eqref{eq:purity-full-expr} towards the decoherence rate. We separate the coefficients into two pieces. The initial state-dependent part is captured by $w_{j,k} := \frac{\sbra \rho_0|l_j\sket \sbra l_k|\rho_0\sket}{ \sbra r_j|l_j\sket \sbra l_k | r_k \sket } $, and the right eigenoperator overlaps are given by $z_{j,k}=\sbra r_j|r_k\sket$. We estimate the degree of correlation among the phases by analyzing their distribution at a fixed value of the index $j$. If these distributions are independent of $j$, then the phases can be treated as essentially being uncorrelated. Specifically, we define the average and the standard deviation of the phase difference
\eq{\label{eq:phase-mean-std}\eqsp{\mu_j &= \frac{1}{d_L}\sum_{k=0}^{d_L-1} e^{i(\arg{[w_{j,k}]} - \arg{[z_{j,k}]})}  \\
\Sigma_j^2 &= \frac{1}{d_L}\sum_{k=0}^{d_L-1} |e^{i(\arg{[w_{j,k}]} - \arg{[z_{j,k}]})}-\mu_j|^2 },}
where $d_L$ is the total number of eigenmodes. Fig~\ref{fig:app-vlrho-phase} in the appendix shows that both statistics vary little as a function of $j$. Since the phases are uncorrelated, the absolute values of the overlaps are sufficient to capture the dominant contributions to Eq.~\eqref{eq:purity-full-expr}.

We contrast this picture of decoherence of generic entangled states with that of the random product states defined by
\eq{\label{eq:product-state}\rho_{\mathrm{prod}}=\otimes_{i=1}^N |\phi\rangle_i\langle\phi|_i,}
where $\phi_i$ is a random single-qubit pure state sampled from the Haar random ensemble. In Fig.~\ref{fig:RL-10-purity}(a), the purple curve shows the time evolution of the purity of $100$ independently sampled product states from this ensemble. We observe markedly higher fluctuations in the initial decoherence rate across these initial states.  These larger fluctuations are explained by the higher variation in overlaps with bulk eigenvectors, and higher concentration of overlaps on states outside the bulk of the spectrum, shown in Fig.~\ref{fig:RL-10-vlRho-disc}(d).  Compared to the panels (a--c), we observe a higher degree of fluctuations in the value of the overlaps $\sbra l_j|\rho\sket$ across different initial states.  

A heuristic explanation of this difference is as follows. Using Eq.~\eqref{eq:decoherence-jump}, we see that the decoherence rate is given by the covariance of local jump operators evaluated in the initial state $\rho_0$. If $\rho_0$ is a generic entangled state, then it looks like an infinite temperature state with respect to such local operators. Consequently, the local expectation values in such a generic state, at least for large system sizes, approximately become independent of the specific choice of the state. However, for a product state in Eq.~\eqref{eq:product-state}, the expectation values of the jump operators will depend on the details of the chosen state from the ensemble. Thus, while all of the product states have identical operator size distribution (same as the $|0\rangle^{{\otimes}N}$ state), their IPR over the basis operators varies considerably with the choice of initial state. As a result, if the eigenoperators have additional structure apart from the leading size-dependence, the overlaps of the states from the product ensemble will also show significant variations. The dynamics of the product states can potentially probe these more refined properties of the eigenoperators. We leave this interesting question for future investigations.

Given that the early time dynamics of the purity show universality, we estimate the rate of decoherence during this time window. The early time dynamics can be approximated by the leading order term in the Taylor expansion of $P(t)$ as
\eq{\label{eq:purity-series} P(t)\approx P_0 + 2t\sbra \rho_0|\mathcal{L}|\rho_0\sket + \mathcal{O}(t^2), }
where $P_0$ is the initial purity, and we have used the fact that $\sbra\rho_0|\mathcal{L}|\rho_0\sket=\sbra\rho_0|\mathcal{L}^\dagger|\rho_0\sket$. Comparing this expression to an exponentially decaying ansatz $P_0 e^{-D t}$, we can define the decoherence rate \cite{lidar1998dfs} of state $\rho_0$ under Lindblad dynamics as 
\eq{\label{eq:decoherence} D_{\rho_0}:=-\frac{2}{P_0}\sbra\rho_0|\mathcal{L}|\rho_0\sket.}
Explicitly writing the decoherence rate in terms of the jump operators results in
\eq{ \label{eq:decoherence-jump} D_{\rho_0} = \frac{4}{P_0}\sum_{a} \big[\tr{(\rho_0^2 L^\dagger_a L_a)} - \tr{(\rho_0 L^\dagger_a \rho_0 L_a) } \big],} 
where the sum runs over all jump operators. In the models considered here, each site is acted on by a finite number of jump operators. Hence, the decoherence rate increases linearly with the total number of sites $N$ in the system. This is typically the case for many-body Lindbladians with local dissipation \cite{campo2019extreme}. We estimate the initial decay rate of the purity by fitting the early-time dynamics with an exponential curve $P_0e^{-b t}$. In Fig.~\ref{fig:RL-10-purity}(b), we observe that the fit parameter matches well with the decoherence rate $D_{\rho_0}$, and the agreement between $b$ and the expected decoherence rate $D_{\rho_0}$ gets better with the increasing system size. We also observe that universality in spectral overlaps leads to smaller spread of decoherence rates with increasing system size, as seen by the decrease in the scatter of data points for $N=7$, compared to $N=6$.

\begin{figure}
    \includegraphics{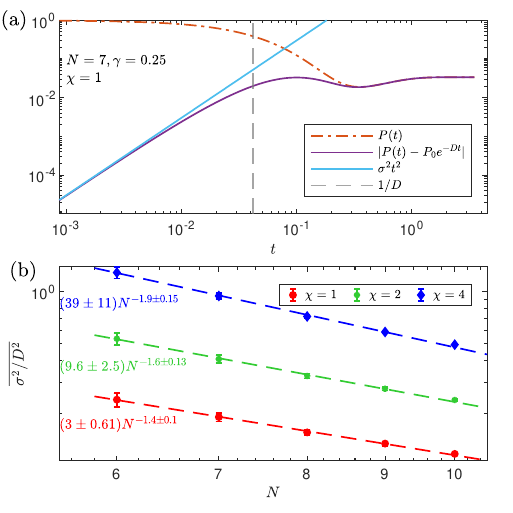}
    \caption{Scalings of initial state deviations in random Lindblad model $(J=1,\gamma_1=\gamma_2=0.25,\mathtt{seed=4})$: (a) The deviation of purity and exponential ansatz for a specific pure initial state in $N=7$ system size. (b) The approximate $\mathcal{O}(t^2)$ deviation $\delta = \sigma^2/D^2$ at characteristic time $\frac{1}{D}$ (depicted using a vertical dashed line in panel (a))as a function of system size $N$. The datapoints show the average over $500$ initial states from different ensembles parameterized by $\chi$, and the errorbars show the standard deviation over these initial states. The dashed lines show the power-law fits to the data.} 
    \label{fig:RL-10-purity_moments} 
\end{figure}

Truncating the infinite series in Eq.~\eqref{eq:purity-series} at a finite order is generally justified if the subsequent terms beyond that order are smaller in magnitude. The full series expression for the purity is given by 
\eq{ \label{eq:purity-inf-seris}P(t)= \sum_{m=0}^\infty \sum_{n=0}^\infty \frac{\sbra\rho_0|{\mathcal{L}^\dagger}^m \mathcal{L}^n|\rho_0\sket}{m!n!}t^{m+n}. }
For the models with local dissipation and interactions considered here, the Lindblad superoperators $\mathcal{L}$ and $\mathcal{L}^\dagger$ involve sums of $\mathcal{O}(N)$ local superoperators. This implies that the terms of form $\sbra\rho_0|{\mathcal{L}^\dagger}^m \mathcal{L}^n|\rho_0\sket$ will be upper-bounded by $(\kappa \gamma N)^{m+n}\sbra\rho_0|\rho_0\sket$ where $\kappa$ is an $\mathcal{O}(1)$ constant that depends on the microscopic details of the Lindbladian. Using this, we can estimate the magnitude of $j^{th}$ term $a_jt^j$ in Eq.~\eqref{eq:purity-inf-seris} as 
\eq{P_0t^j(\kappa N)^j \sum_{r=0}^j \frac{1}{r!(j-r)!} = \frac{P_0(2\kappa N )^j}{j!} t^j .}
The series remains convergent if $a_{j+1}t^{j+1}/(a_jt^j) =2\kappa N t/(j+1)<1$, meaning that the successive terms in the series become smaller and the series can be safely truncated at finite order. For example, after truncating the series at first order $j=1$, it remains convergent up to time of order $\tau\sim\frac{1}{\kappa N}$. While this timescale shrinks with the increasing system size, we have already observed that it is sufficiently long to capture the early time dynamics of the purity, which decays with rate $D_{\rho_0}\propto N$. 

To further justify this approximation, we show the difference between the exponential ansatz and the true evolution defined by $\delta(t):=|P(t)-P_0 e^{-Dt}|$ in Fig.~\ref{fig:RL-10-purity_moments}(a). We observe that during early times, this difference is approximated by the leading order term in the Taylor expansion given by 
$\delta(t) = \sigma^2 t^2 + \mathcal{O}(t^3)$, where 
\eq{\sigma^2 =  \sbra\rho_0 |\mathcal{L}^\dagger\mathcal{L} + \mathcal{L}^2|\rho_0\sket  - \frac{2\sbra\rho_0 |\mathcal{L}|\rho_0\sket^2}{P_0}.}
Since this expression can be computed without diagonalizing the Lindbladian, we can numerically access it for higher system sizes. We are interested in how the deviation $\delta(t)$ scales with the increasing system size $N$. While both terms in this expression scale as $N^2$, we numerically observe that this leading dependence is canceled out, resulting in a slower increase with $N$. In Fig.~\ref{fig:RL-10-purity_moments}(b), we show that the deviation at the characteristic decoherence time $t=1/D$ given by $\delta(t=\frac{1}{D})\approx\sigma^2/D^2$ decreases with increasing system size $N$. The sub-extensive nature of fluctuation observed here is in a similar spirit to its many-body counterpart in closed quantum systems, where the average energy of an arbitrary initial state grows $\sim N$, but the fluctuations of the energy only grow as $\sqrt{N}$. For the large system sizes, the average energy alone is sufficient to describe the values of observables in thermalizing systems \cite{rigol2008thermalization}.

Our results can be viewed as an analog of dynamical typicality~\cite{bartsch2009typicality} as applied to the early-time dynamics of open quantum systems.  At times $ t > \frac{1}{ \kappa N}$ our results from the bulk of the spectrum do not imply typicality; however, upon examining the variance of overlaps of random initial states with $|l_1 \sket,$ ref. \cite{bao2026lindbladTypicality} finds evidence of typicality in this regime as well.   

%% ======================== Non linear correlator ==================
\begin{figure}
    \includegraphics{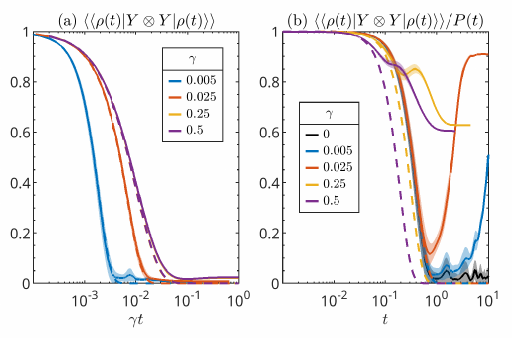}
    \caption{R\'enyi-2 correlator $\mathcal{R}_Y(t):=\frac{1}{P(t)}\sbra \rho(t) |\sigma^y_4\otimes {\sigma^y_4}^*|\rho(t)\sket$: The data is shown for the random Lindblad model ($\mathtt{seed=4}$) with $N=7$ sites. Hamiltonian interaction strength $J=1$, and varying dissipation rate $\gamma_1=\gamma_2=\gamma$. The solid lines depict the average value computed over $100$ random initial states (see Eq.~\eqref{eq:renyi-init}), and the shaded patch shows the standard deviation. (a) The numerator of the R\'enyi-2 correlator as a function of time. The X axis is rescaled by the dissipation strength. Notice that the curves for $\gamma\ge0.25$ approximately collapse onto each other, indicating that the $1/\gamma$ is the relevant timescale. The dashed lines correspond to the estimate of early-time decay given in $e^{-Dt-\Sigma t^2}$ (see Eqs.~\eqref{eq:decoherence}, and \eqref{eq:sigma-strange}) (b) Full R\'enyi-2 correlator $\mathcal{R}_Y$ as a function of time. The dashed curves show the estimated Gaussian decay $e^{-\Sigma t^2}$. The early-time evolution in the weak-dissipation limit closely follows the Hamiltonian-only dynamics (shown by the black-colored curve). } 
    \label{fig:RL10-strange} 
\end{figure}
%% ===================================================================

\subsection{R\'enyi-2 correlator}

A more general non-linear correlation function can be obtained by allowing for a local superoperator $\hat{A}\neq \mathbb{I}$ in Eq.~\eqref{eq:general-nonlinear}. A particularly useful correlation function is the R\'enyi-2 correlator, which can probe non-trivial features of mixed states such as spontaneous breaking of strong symmetry to a weak symmetry \cite{lessa2025strongtoweak,sala2024spontaneous}, critical systems with decoherence \cite{leeCenke2023decoherenceCriticality}, and properties of individual trajectories in measurement-induced transitions \cite{bao2021symmetry}. Recently, it has also been shown to be experimentally accessible on a trapped-ion quantum computing platform \cite{hsieh2025probingRenyi}. 

We define the R\'enyi-2 correlation function at time $t$ as 
\eq{ \label{eq:full-renyi}\mathcal{R}_A(t) = \frac{\sbra \rho(t)| A\otimes A^*|\rho(t)\sket}{\sbra\rho(t)|\rho(t)\sket}, }
where $A$ is some local unitary operator, and $\rho(t)$ is the state of the system. Recall that, according the superoperator notation in Eq.~\eqref{eq:opvec}, $A\otimes A^*|\rho\sket \longleftrightarrow A\rho A^\dagger$. In this work, we will focus on systems initialized in a pure state $\rho_0$, which satisfies 
\eq{\label{eq:renyi-init}\mathcal{A}|\rho_0\sket:=A\otimes A^*|\rho_0\sket=|\rho_0\sket \ .} 

To understand the early-time behavior of this correlation function, we Taylor expand Eq.~\eqref{eq:full-renyi} in powers of $t$. The linear in time term is given by 
\eq{ \mathcal{O}(t): \ \sbra{\rho_0} |\mathcal{A} \mathcal{L} + \mathcal{L}^\dagger \mathcal{A} - \sbra\rho_0|\mathcal{A}|\rho_0\sket( \mathcal{L} + \mathcal{L}^\dagger) |\rho_0\sket ,}
where $\mathcal{L}$ is the Lindblad superoperator. Using the condition from Eq.~\eqref{eq:renyi-init}, we observe that this $\mathcal{O}(t)$ term vanishes. Instead, the leading time dependence is governed by $\mathcal{O}(t^2)$ term given by 
\eq{ \label{eq:sigma-strange}\mathcal{O}(t^2): \quad -\Sigma := \sbra \rho_0 | \mathcal{L}^\dagger \mathcal{A} \mathcal{L} - \mathcal{L}^\dagger  \mathcal{L} |\rho_0\sket,}
where we have again used Eq.~\eqref{eq:renyi-init} to further simplify the expression. Unlike the example of the purity studied earlier, where early-time decoherence only depends on the dissipative terms (proportional to $\gamma$), the leading order time dependence of the R\'enyi-2 correlator gets contributions from both the Hamiltonian and the dissipative part of the Lindbladian, leading to more complex behavior of the relevant timescales as we show below.

In Fig.~\ref{fig:RL10-strange}, we show the evolution of the R\'enyi-2 correlator as a function of time for varying dissipation strengths $\gamma$, and a fixed value of the Hamiltonian interaction parameter $J=1$. Specifically, we set the local operator $\hat{A}$ to a Pauli operator $\sigma^y_4$ located on the central site in an $N=7$ sized system. We consider the initial state ensemble of randomly chosen entangled pure states defined according to Eq.~\eqref{eq:init_purity} with $\chi=1$. We further project the central site onto $+1$ eigenstate of $\sigma^y$ operator by transforming $\rho_0 \rightarrow \frac{1+\sigma^y_4}{2}\rho_0\frac{1+\sigma^y_4}{2}$, followed by appropriate normalization. This ensures that the initial state satisfies Eq.~\eqref{eq:renyi-init}.

First, in Fig.~\ref{fig:RL10-strange}(a), we plot the numerator \mbox{$\sbra \rho(t)|\sigma^y\otimes{\sigma^y}^*|\rho(t)\sket$} of $\mathcal{R}_Y$ as a function of $\gamma t$. When the dissipation strength is large enough, the curves collapse on each other. This implies that, similar to purity decay, the relevant timescale for early time evolution in this regime is proportional to $1/\gamma$. However, for smaller values of $\gamma$, the decay happens at a much faster rate. While the first order term in $t$ is same for both the numerator of $\mathcal{R}_Y$ and the purity, the second order term in the former case is of the form $\sim (J^2 + J\gamma+\gamma^2)t^2$, whereas the purity, which is unaffected by Hamiltonian dynamics, and has a second-order term of the form $( J\gamma  + \gamma^2)t^2$. For $\gamma\ll J$, the contributions from the $J^2$ term are more relevant in this case. Apart from these considerations, we observe that the $\sbra \rho(t)|\sigma^y\otimes{\sigma^y}^*|\rho(t)\sket$ is indeed independent of the specific choice of initial states within the ensemble considered here.

The time evolution of the full R\'enyi-2 correlator $\mathcal{R}_Y$ is shown in Fig.~\ref{fig:RL10-strange}(b), where we observe that its early-time decay can be approximately predicted in terms of the Gaussian ansatz $e^{-\Sigma t^2}$, where $\Sigma$ is the leading order term evaluated in Eq.~\eqref{eq:sigma-strange} with $\mathcal{A} = \sigma^y_4\otimes{\sigma^y_4}^*$. Here as well, we observe initial state independent universal decay for short times --- but long enough for purity to decay to small values. The late time value of $\mathcal{R}_Y$ depends on dissipation strength as it is more sensitive to changes in the steady state purity. 

To conclude, in this section, we have shown that the non-linear functions of the density operator, such as the purity and R\'enyi-2 correlator, involve contributions from the bulk part of the Lindblad spectrum. This leads to universal early-time decoherence in generic Lindblad dynamics, which is independent of the specific choice of initial state in a particular ensemble.
%% =======================================================================
\section{Open system dynamics of operators}
\label{sec:operator_dynamics}

In section~\ref{sec:eigenvecs}, we observed that the Pauli weights of the eigenoperators of our local Lindbladian have a strong dependence on the corresponding eigenvalue. Here, we show how this size dependence, which follows from the locality of the Lindbladian, dictates the nature of operator growth in generic open quantum systems. We begin by discussing well-known results on the operator growth in both closed and open quantum systems. Then we present the results on the time evolution of the operator size distribution and its normalization in the presence of decoherence. This is followed by the inspection of the operator scrambling due to open dynamics. At the end, we analyze the relevant timescales for the operator dynamics and compare them to the time over which the generic eigenmodes in the bulk of the spectrum play an important role. 

\subsection{Background on operator spreading}

Operator growth and scrambling under unitary dynamics have been extensively studied in a wide range of many-body quantum systems, from random quantum circuits to conformal field theories to chaotic black holes \cite{roberts2015otocCFT,maldacena2016bound,bohrdt2017scrambling,keyserlingk2018otoc,nahum2018randomunitary,khemani2018conserved,parker2019universalOGH}(see Ref.~\cite{fisher2023circuitreview} for a review). In recent years, several studies have also focused on the analysis of operator dynamics in open quantum systems \cite{schuster2023operator,shirai2024accelerated,zanardi2021,zhangHunagChen2019,yoshimuraSa2025dynamics,liu2024lindbladSykOperatorGrowth,bhattacharya2022operator,mori2024liouvilliangap,jacoby2025spectralGaps,zhang2025ruellePollicot}. Here, we briefly review the aspects of operator dynamics in closed systems relevant to this work, followed by an analysis of how the generic features of the Lindblad spectrum and eigenvectors described above shape the analogs in open quantum systems.

Let us consider a generic non-integrable Hamiltonian $H$, which controls the dynamics of an isolated closed quantum system. We are interested in the time evolution of an initial operator $A$ in the Heisenberg picture. According to Eq.~\eqref{eq:op-dyn}, this is given by 
\eq{\label{eq:unitary-heisen} A(t) = e^{\mathcal{L}_U^\dagger t}(A)  }
where $\mathcal{L}_U^\dagger:=i[H,.]$ is the adjoint Liouville superoperator. The operator at any given time can be expanded in terms of the Pauli basis operators according to Eq.~\eqref{eq:pauli-expansion}. Using this, the operator dynamics is understood in terms of the time evolution of the expansion coefficients
\eq{ \label{eq:op-expand}w_{\mathbf{m}}(t) = \sbra F_{\mathbf{m}}|A(t)\sket.} 
We notice that the unitary dynamics satisfies $\mathcal{L}_U^\dagger = -\mathcal{L}_U$, which implies that the normalization of the distribution of these coefficients, given by 
\eq{\label{eq:op-norm-unitary} \sbra A(t)|A(t)\sket = \sbra A|e^{\mathcal{L}_Ut} e^{\mathcal{L}^\dagger_Ut}|A\sket =  \sbra A|A\sket , }
remains conserved during time evolution. 

To analyze the spatial profile of the time-evolving operator, we consider an initial operator $A$ localized on a single site. The operator $A(t)$ at time $t$ can be obtained by explicitly expanding the expression in Eq.~\eqref{eq:unitary-heisen}, resulting in an infinite series given by
\eq{A(t) = A + it[H,A] + \frac{(it)^2}{2}[H,[H,A]] + \ldots  \label{Eq:TaylorHam}}
If the Hamiltonian $H$ only contains local interactions, the size of operators in the nested commutator at $k^{th}$ order in this expansion can grow at most linearly with $k$ \cite{abanin2015floquetPrethermal,mori2016floquetBound}. Meanwhile, the coefficient of the $k^{\text{th}}$ order term in Eq.~\eqref{Eq:TaylorHam} is grows approximately as $t^k/k! \approx (t/k)^k$, suggesting that the coefficients $w_{\mathbf{k}}(t)$ remain vanishingly small until times of order $k$. This intuitive description of operator growth can be quantified by modeling the dynamics of local interactions using random quantum circuits with local unitary gates. In this model, it was shown that the finite velocity for the propagation of information \cite{lieb1972finite} leads to ballistic growth of the length of the operators along with further diffusive broadening \cite{nahum2018randomunitary,keyserlingk2018otoc}. As a result of this operator growth, the generic Hamiltonian dynamics leads to the spreading of the operator size distribution defined in Eq.~\eqref{eq:size-dist} onto large size sectors with increasing time. Furthermore, due to the scrambling nature of local gates, the resulting operator is made up of a large number of Pauli strings, leading to growth of operator entanglement \cite{jonay2018operatorentangle}. The presence of conservation laws can further enrich this generic dynamics with additional slow modes corresponding to conserved densities \cite{khemani2018conserved,rakovszky2018conserved}.

When the quantum system becomes open, several features of operator growth get modified due to dissipation and decoherence. A variety of probes have been developed to illustrate the interplay of dissipative processes and unitary interactions on the operator dynamics, including modified out-of-time ordered correlators \cite{schuster2023operator,swingle2018scramblingmeasurements,zhangHunagChen2019,martinez2023opertorvariance,zanardi2021}, Krylov space complexity \cite{bhattacharya2022operator,liu2023krylovopen,srivatsa2024growthhypothesis,bhattacharya2023biLanczos,bhattacharjee2024perspective,carolan2024operator}, and dynamics of information-theoretic measures in random unitary circuits with decoherence \cite{leeJiang2024cmi,weinstein2023radiativeRUC,gopalakrishnan2024cmi,touil2021entropysinks}. Here, we will focus on the decay of the operator norm and evolution of the operator size distribution to illustrate the effect of open dynamics and related features in the spectrum of the Lindbladian.

The operator dynamics in open systems differs from that in closed systems in two qualitative ways. First, the operator norm is not conserved. As we discuss below, this leads to operator shrinking, since the decay rate of the norm of a Pauli string is proportional to its size. The impact of this correlation between the operator size and the decay rate was analyzed in ref.~\cite{schuster2023operator} to illustrate the generic operator growth in a variety of systems in the weak dissipation limit. Second, our Lindblad dynamics also includes processes that can directly convert a non-identity Pauli string to an identity operator, which are strictly absent in the unitary dynamics. These processes lead to a growth of the operator's trace, and can be modeled by layers of unitary gates interleaved by swap gates with ancilla. Ref.~\cite{weinstein2023radiativeRUC} showed that in such a model, when the density of the swaps is smaller than a critical value, the operator keeps growing, resembling the unitary dynamics. Only when the density of swaps is above the critical value does the system enter a phase where the entire operator is eventually swapped with the environment. In the following, we discuss the combined effect of these two processes on the operator dynamics in generic Lindblad systems for a wide range of dissipation strengths.

\begin{figure}
    \includegraphics{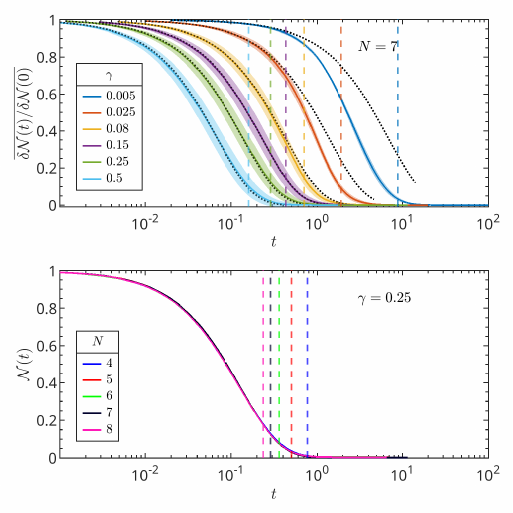}
    \caption{Operator norm $\mathcal{N}(t):=\sbra A(t)|A(t)\sket$: (a) The time evolution of the relative deviation of operator norm, in terms of $\delta\mathcal{N}(t) :=\mathcal{N}(t)-\mathcal{N}(\infty)$, averaged over initial operators is plotted as a function of time for random Lindblad model ($\mathtt{seed=4}$) with $7$ sites and Hamiltonian interaction strength $J=1$. The data is shown for varying values of dissipation rates $\gamma_1=\gamma_2=\gamma$. The solid lines show the average over 10 independent single-site initial operators chosen according to Eq.~\eqref{eq:rand-init-ops}. The shaded region shows the standard deviation around this average value. The dotted lines depict the prediction $e^{-\overline{D}t}$ based on the first-order term in the Taylor expansion. (b) The operator norm $\mathcal{N}(t)$ is plotted for a single initial operator (see Eq.~\eqref{eq:figs-init-op}). The data is shown for multiple system sizes and a fixed value of $\gamma=0.25$.  The vertical dashed lines in both panels show the timescale $\tau_F$ defined in Eq.~\eqref{eq:tau-freeze} up to which the bulk modes contribute towards the dynamics of the operator.\label{fig:RL10-op-norm}}      
\end{figure}

\begin{figure}
    \includegraphics{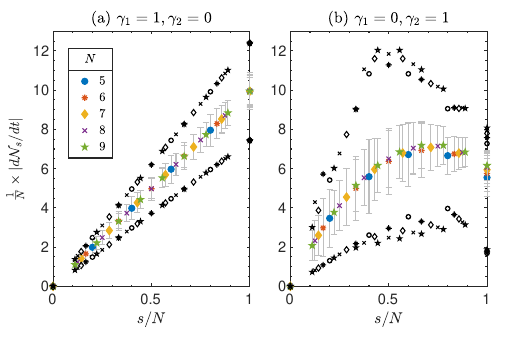}
    \caption{The average rate of change of norm for Pauli basis operator of size $s$ as a function of their size $s$. The average rate is defined as $\frac{d}{dt}\mathcal{N}_s=-\frac{2}{b_N(s)}\sum_\mathbf{m} \sbra F_{\mathbf{m}}|\mathcal{L}^\dagger|F_{\mathbf{m}}\sket\delta_{\mathtt{S}[\mathbf{m}],s}$, where $b_N(s)$ is tht total number of size $s$ basis operators. (a) The contribution from single-site jump operators, obtained by setting $\gamma_2 = 0$ (b) The contribution from two-site jump operators, obtained by setting $\gamma_1 = 0$. Each solid data point is obtained by averaging over basis operators $F_{\mathbf{m}}$ of size $s$, and the error bars show the standard deviation. The black colored open symbols show the smallest and the largest rate for a given basis size $s$. Both the x and y axes are rescaled by system size $N$. The data is shown for the random Lindblad model labeled by $\mathtt{seed=4}$. } 
    \label{fig:dndt-vs-s} 
\end{figure}
\subsection{Operator norm}
The operator norm $\mathcal{N}=\sbra A|A\sket$ is no longer a conserved quantity under open system dynamics. It evolves according to the Lindblad equation~\eqref{eq:heisen} as 
\eq{\label{eq:dndt} \frac{d}{dt}\mathcal{N} = 2\sbra A(t)|\mathcal{L}^\dagger|A(t)\sket ,}
where $\mathcal{L}^\dagger$ is the adjoint Lindblad superoperator that governs the dynamics in the Heisenberg picture. We numerically evaluate the norm of the operator initially located at a single site in the middle of the spin chain. Specifically, we consider an initial operator of the form
\eq{\label{eq:rand-init-ops} A =\sum_{\alpha={x,y,z}} r_\alpha \times \begin{cases*}
\hat{\sigma}^{\alpha}_{{(N+1)}/2}, \  & \text{if odd  } N \\
\hat{\sigma}^{\alpha}_{N/2} + \hat{\sigma}^{\alpha}_{{(N+2)}/2}, \  & \text{if even  } N
\end{cases*}  ,} 
where $r_\alpha$ are real numbers sampled from normal distribution, and $\sigma^{\alpha}_{j}$ are the Pauli matrices acting on the $j^{\text{th}}$ site. The resulting operator is then normalized such that $\sbra A|A\sket = 1$. For the numerical results shown in this section, the initial operator is 
\begin{equation}
    \label{eq:figs-init-op}
    \sqrt{2^N}A = \begin{cases*}
    \begin{aligned}
        &0.459\sigma^x_{\frac{N+1}{2}} + 0.681\sigma^y_{\frac{N+1}{2}} + 0.571\sigma^z_{\frac{N+1}{2}}, \ \text{if odd  } N \\
&\frac{0.459}{\sqrt{2}}(\sigma^x_{\frac{N}{2}}+\sigma^x_{\frac{N+2}{2}}) + \frac{0.681}{\sqrt{2}}(\sigma^y_{\frac{N}{2}}+\sigma^y_{\frac{N+2}{2}}) \\ & + \frac{0.571}{\sqrt{2}}(\sigma^z_{\frac{N}{2}}+\sigma^z_{\frac{N+2}{2}}), \ \text{if even  } N
    \end{aligned}
\end{cases*}
\end{equation}
The numerical results in this section are shown for $\mathtt{seed}=4$ realization of the random Lindblad model (see Fig.~\ref{fig:RL-sd-4-spectrum} for the details of its spectrum). The rest of the realizations, and the dissipative Ising model, show qualitatively similar features to those described below. However, initial transient features differ since short-time dynamics of initially local operators is expected to depend on the specific initial operator.

In Fig.~\ref{fig:RL10-op-norm}(a), we show the time evolution of the operator norm in a system evolving according to the random Lindblad model. For a finite value of dissipation $\gamma$, it initially decays exponentially with a rate given by $D=-2\sbra A|\mathcal{L}^\dagger|A\sket$ (indicated by black dotted lines in Fig.~\ref{fig:RL10-op-norm}(a)). At a very late time, when the system attains its steady state distribution $\rho_{ss}$, the operator evolves to $\tr{(A\rho_{ss})}\mathbb{I}$ as per Eq.~\eqref{eq:op-dyn}. Consequently, the operator norm approaches $\mathcal{N}_{\infty} = [\tr{(A\rho_{ss})}]^2$ at long time.

Further insight into the dynamics of the operator norm can be obtained by substituting the Lindblad equation~\eqref{eq:heisen} into Eq.~\eqref{eq:dndt} to obtain
\eq{\label{eq:dndt-2} \frac{d}{dt}\mathcal{N}(t) = 4\sum_{a}\tr{(A(t)\ [L^\dagger_a,A(t)]\ L_a)} ,}
where $L_a$ are the jump operators describing the open dynamics. The contribution from each of the jump operators $L_a$ to this rate vanishes unless the time-evolved operator $A(t)$ has support in the region occupied by that jump operator. If the jump operators are strictly single-site operators, then the rate is expected to be proportional to the average size of the operator $A(t)$, leading to 
\eq{ \label{eq:norm-size-relation} \frac{d}{dt}\mathcal{N}(t) \propto -c \ \mathcal{S}(t) \mathcal{N}(t). }
where $\mathcal{S}(A(t))$ is the size of the time evolved operator (see Eq.~\eqref{eq:avg-size}). The proportionality constant depends on the jump operators, i.e., $c\sim ||L^\dagger L||\sim \mathcal{O}(\gamma)$. This relation holds exactly if the dissipative part is generated by the fully depolarizing channel \cite{schuster2023operator}, since this channel decoheres all single-site Pauli operators with an identical rate. In Fig.~\ref{fig:dndt-vs-s}(a), we show the contribution from single-site jump operators in our model to the decay rate of the norm of a basis operator as a function of its size. We observe that while the rate averaged over basis operators of size $s$ increases linearly with $s$, there are fluctuations about this average value depending on the Pauli-content of the basis string. 

Additionally, however, our model includes jump operators that act on two nearest-neighbor sites. In this case, apart from the size, the spatial distribution of non-identity Pauli operators in the basis string also becomes relevant. In Fig.~\ref{fig:dndt-vs-s}(b), we show the contribution to the norm decay rate from the two-site jump operators. The average decay rate of the basis operators  $F_{\mathbf{m}}$ generally increases with the size of $m$, up to $m \sim N$.
However, the error bars in Fig.~\ref{fig:dndt-vs-s} reflect the fact that the specific type of single-site Pauli operators present in $F_{\mathbf{m}}$ also play a role in determining the value of its decay rate. This distinction between different Pauli types means that, unlike the fully depolarizing model, Eq.~\eqref{eq:norm-size-relation} only approximately holds true for generic Lindbladians, with deviations from this trend becoming more pronounced as system size increases. Moreover, when only two-site dissipation is present, we observe that there are some basis operators of size $s=N$ for which the operator norm decays at a rate comparable to that of single-site operators. This corroborates our earlier observation of unusually large eigenoperators at small values $|\text{Re}(\lambda)|$ in the large $\gamma_2$ parameter regime, discussed in section~\ref{sec:unusual-evec}. This in turn means that the late time operator dynamics in this regime is dominated by this slowly decaying operator, as discussed in section~\ref{sec:unusual-op-dyn}.

The overall impact of the relation between the operator norm decay rate and its size is shown in Fig.~\ref{fig:RL10-op-norm}(b). Here, we observe that the decoherence for an initially single-site operator is independent of the system size, as evidenced by the collapse of curves for early time dynamics. We note that the $\mathcal{O}(1)$ timescale of operator decay is significantly longer than the decoherence time for initial states, which happens over $t\sim\frac{1}{N}$ (see Eq.~\eqref{eq:decoherence-jump}).

Finally, we comment on how this size-dependence of the rate of change of norm emerges from the eigenoperator size distribution studied in section~\ref{sec:eigenvecs}. This rate can be expressed using the spectral decomposition of the Lindbladian as
\eq{ \label{eq:op-norm-spectrum} \frac{d}{dt}\sbra F_{\mathbf{m}} | F_{\mathbf{m}} \sket = 2\sum_{j=1}^{d_L} \lambda_j^*\frac{\sbra F_{\mathbf{m}} | l_j \sket\sbra r_j|F_{\mathbf{m}} \sket}{\sbra r_j|l_j\sket}.}

When $F_{\mathbf{m}}$ is a single-site basis operator, its overlap with right eigenoperators $\sbra r_j|F_{\mathbf{m}}\sket$ exponentially decreases with increasing value of $|\text{Re}(\lambda)|$ (see eg, $s=1$ curve in Fig.~\ref{fig:secAvg}(a)). As a result, only $|\lambda|\sim\mathcal{O}(1)$ terms in Eq.~\eqref{eq:op-norm-spectrum} contribute to this expression, leading to $\mathcal{O}(1)$ decay rate for the norm of single-site basis operators. Whereas, when $F_{\mathbf{m}}$ is a size $s\sim N$ operator, its overlap with both the left and the right eigenoperators is approximately constant in the bulk part of the spectrum. This suggests that the decay rate of the large-size operators is roughly proportional to the average over all eigenvalues. The center of the spectrum, defined in Eq.~\eqref{eq:xbar}, scales as $|\overline{X}|\sim N$. As a result, the norm of size $s\sim N$ basis operators also decay with a rate that approximately grows as $\sim N$ with the increasing system size. This simple argument qualitatively captures the relation between the observed norm decay rate and the size distribution of the eigenoperators.

%%% ======================== operator size plots ====================================
\begin{figure*}
    \includegraphics{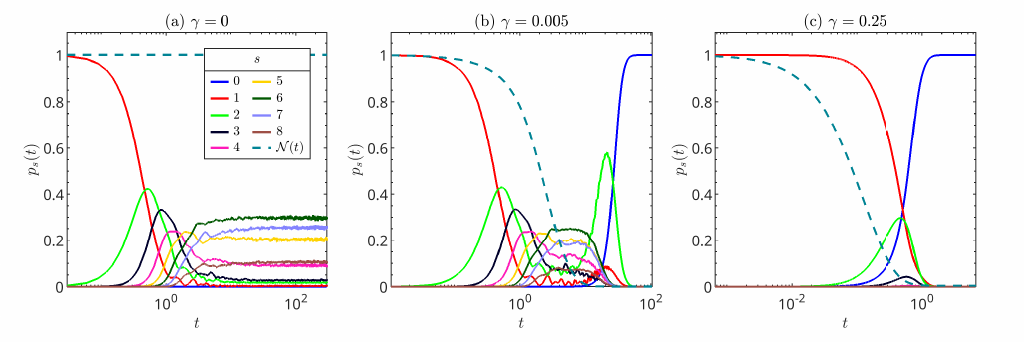}
    \caption{Operator size distribution: The operator size distribution of the time evolving operator $A(t)$ is plotted as a function of time for the random Lindblad model ($\mathtt{seed=4}$). The data is shown for the varying values of dissipation rates $\gamma_1=\gamma_2=\gamma$: (a) $\gamma=0$, (b)$\gamma=0.005$, and (c)$\gamma=0.25$, while the strength of the Hamiltonian interactions is set to $J=1$. In all three panels, a single initial operator is chosen according to Eq.~\eqref{eq:rand-init-ops}. The different colors show the size distribution on the basis operators of size $s$. The teal colored dashed line depicts the operator norm $\mathcal{N}(t)$. } 
    \label{fig:RL10-opDist-vs-t} 
\end{figure*}

\begin{figure}
    \includegraphics{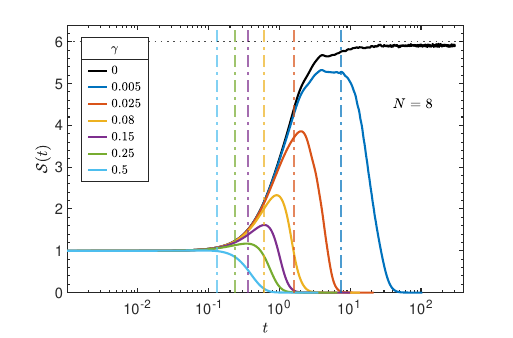}
    \caption{The average size of the time evolved operator $\mathcal{S}(t)$ defined in Eq.~\eqref{eq:avg-size} is plotted as a function of time for the random Lindblad model ($\mathtt{seed=4}$) with Hamiltonian strength $J=1$, and varying values of dissipation rates $\gamma_1=\gamma_2=\gamma$. The initial single-site operator is chosen as Eq.~\eqref{eq:figs-init-op}. The vertical dashed lines show the freezing timescale $\tau_F$, for every value of dissipation rate, defined in Eq.~\eqref{eq:tau-freeze}, after which the bulk eigenmodes have negligible contribution to the dynamics. For purely Hamiltonian dynamics $(\gamma=0)$, the operator grows to have size approximately equal to $3N/4$, corresponding to proportionately sampling all basis operators. With the increasing dissipation rate, the largest attained size shrinks, eventually reaching the steady state of zero size in accordance with the data shown in Fig.~\ref{fig:RL10-opDist-vs-t}.} 
    \label{fig:RL10-tot-len} 
\end{figure}
%% ==============================================================================
\subsection{Operator size distribution}

We now analyze how this time-evolved operator is distributed across the Pauli basis operators of different sizes using the operator size distribution $p_s(A)$ defined in Eq.~\eqref{eq:size-dist}. The denominator in this expression renormalizes the distribution by accounting for the changing norm $\mathcal{N}(t)$. The renormalization procedure partially captures the effect of the open system dynamics described in the previous subsection. However, there are additional modifications to the size distribution due to the decoherence that are not captured by this simple renormalization \cite{zhangHunagChen2019}.

Notably, the trace of an operator is not conserved under generic Lindblad evolution; Instead, it evolves according to 
\eq{ \frac{d}{dt} \tr{A(t)} = 2\sum_a\tr{\left([L_a,L_a^\dagger]A(t)\right)}\propto \sbra \mathbb{I}|\mathcal{L}^\dagger|A(t)\sket. \label{Eq:TrDecay}}
If $A$ is a short operator, the trace increases as Lindblad dynamics converts $A$ into an identity operator on all sites \cite{weinstein2023radiativeRUC}; this occurs at a rate proportional to $\gamma$.

For reference, we first show the time evolution of the distribution $p_s(t)$ for Hamiltonian dynamics with $\gamma=0$ in Fig.~\ref{fig:RL10-opDist-vs-t}(a). The operator is initially located on a single site. As time progresses, the interactions spread the operator to larger basis operators, which is visible in the increased value of $p_{s}$ for $s\ge2$. Eventually, at later times, the operator can grow to span the entire system. The unitary nature of the dynamics ensures that the initially traceless operator remains traceless at all times, as evidenced by $p_{s=0}(t)=0$.

In Fig.~\ref{fig:RL10-opDist-vs-t}(b), we contrast the time evolution generated by the Hamiltonian with a system undergoing open dynamics due to a small but finite dissipation $\gamma=0.005$. 
In this case, the dominant Hamiltonian interactions lead to rapid operator growth, which outpaces the direct down-conversion of operators to the identity described by Eq. (\ref{Eq:TrDecay}). As a result, in the weak dissipation regime, the effect of dissipation during the early time dynamics is predominantly apparent in the overall operator norm decay.
Since the operator norm is only affected by the dissipation, it changes slowly in this weak dissipation regime, and the renormalized size distribution $p_s$ aligns closely with that seen under unitary time evolution: the operator grows to have support on larger basis operators before the dissipation has time to significantly increase its trace by converting it into the identity operator. However, the operator norm of the larger basis strings decays with a comparably faster rate according to Eq.~\eqref{eq:dndt}. This results in a diminishing contribution of large basis operators relative to shorter strings, eg, $s=1,2$ in Fig.~\ref{fig:RL10-opDist-vs-t}(b), in comparison to the unitary case at late times. Dissipation then converts these short operators to $\mathbb{I}$, which is indeed the steady state operator of the time evolution in the Heisenberg picture.

When the dissipation becomes large, as shown in Fig.~\ref{fig:RL10-opDist-vs-t}(c) for $\gamma=0.25$, the operator only grows up to a few short-sized basis operators. In this parameter regime, both the decay rate of the operator norm and the rate of conversion into the identity become larger than the typical operator growth rate. As a result, the operators generically do not develop support on large basis operators. While we have shown results for a specific realization of the random Lindblad model in Fig.~\ref{fig:RL10-opDist-vs-t} ($\mathtt{seed}=4$), the late-time shrinking of operators is generically observed whenever the strength of single-site dissipation is not significantly smaller than the two-site dissipation. However, we also observe late-time operator growth for a special class of initial operators when $\gamma_2\gg\gamma_1$, which we discuss in section~\ref{sec:unusual-op-dyn}.

The collective effect of the operator size distribution observed above can be captured using the average size of the operator defined by
\eq{ \mathcal{S}(t) = \sum_{s_0=0}^N s_0 p_{s_0}(A(t)) ,}
where $N$ is the total number of sites. As shown in Ref.~\cite{schuster2023operator}, the size of an operator is equal to the degree of decoherence under fully depolarizing channel. As a result an equivalent expression of the average size is given by
\eq{ \label{eq:avg-size} \mathcal{S}(t) = \frac{1}{4}\sum_{i=1}^N \sum_{\alpha=0,x,y,z}\bigg[ 1-\frac{ \tr{(A(t)^\dagger\hat{\sigma}^\alpha_i A(t)\hat{\sigma}^\alpha_i)}}{\mathcal{N}(t)} \bigg] ,} 
where $\sigma^0_{i}\equiv \mathbb{I}_i$. The second term in this expression can be interpreted as a normalized out-of-time correlator (OTOC) of the time-evolved operator averaged over all single-site Pauli operators $\hat{\sigma}^\alpha$ \cite{schuster2023operator}. In Fig.~\ref{fig:RL10-tot-len}, we show the time evolution of the $\mathcal{S}$ for an initially single-site operator. In the weak dissipation regime, the growth of the average size follows the $\gamma=0$ curve for early times. Eventually, it shrinks to zero size as the contribution of $p_{s=0}$ starts to increase. In the strong dissipation regime, the average size does not grow much before it begins to decrease due to the increasing value of $p_{s=0}$.

\begin{figure}
    \includegraphics{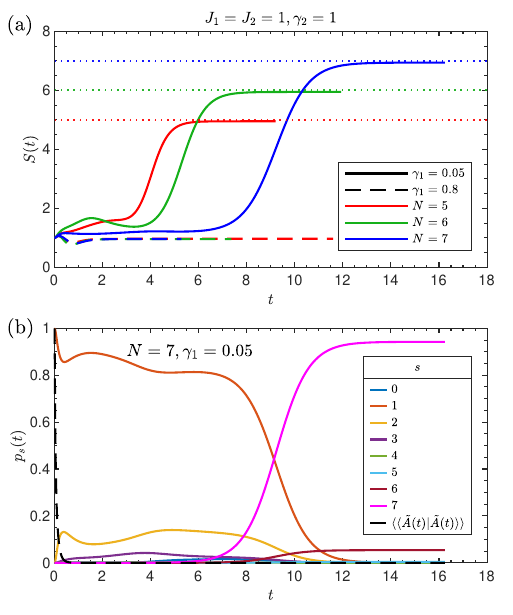}
    \caption{Unusual operator growth in large $\gamma_2$ limit ($J=\gamma_2=1,\mathtt{seed=4}$): (a) The average size $S(t)$  of the time-evolved operator (see Eq.~\eqref{eq:avg-size}) is shown for varying system sizes $N=5,6,7$. The initial operator, chosen according to Eq.~\eqref{eq:noSS-init}, is of form $\tilde{A}:=A - \tr{(\rho_{ss}A)}\mathbb{I}$. The dotted lines show the maximum possible size (i.e $s=N$). The solid lines correspond to regime $\gamma_2\gg\gamma_1$, for which the operator grows up to the maximum possible size ($\gamma_1=0.05$). The dashed lines represent a more generic case, where the operator does not grow due to a larger value of single-site dissipation $\gamma_1=0.8$.  (b) The size distribution $p_s(t)$ (see Eq.~\eqref{eq:size-dist}) is plotted as a function of time for $N=7$ system size (blue curve in panel (a) $\gamma_1=0.05$). The black dashed line shows the operator norm $\sbra \tilde{A}(t)|\tilde{A}(t)\sket$ (see Appendix~\ref{app:unusual-op}, for additional discussion of anomalous operator growth).} 
    \label{fig:unusual-growth} 
\end{figure}

\subsection{Anomalous operator growth}
\label{sec:unusual-op-dyn}

We conclude the discussion of the operator size distribution by pointing out the unusual operator growth observed for a range of parameters in the local random Lindblad model. We begin by expressing the time evolution of the operator in terms of its eigenmodes. According to Eq.~\eqref{eq:op-dyn}, the contribution from the bulk eigenmodes is exponentially suppressed at late times, and the operator evolves to
\eq{A(t) \overset{t\gg1}{\approx} c_0 |l_0\sket + e^{-\lambda_1 t} c_1 |l_1\sket, }
where the coefficient $c_j := \alpha_j^*\sbra r_j|A_0\sket$ captures the overlap of the initial operator with the corresponding right eigenoperator, and we assume $\lambda_1\in \mathbb{R}$. Since $l_0$ is always the identity operator due to the trace-preserving property of the dynamics, it has size equal to zero. As a result, the size of this late-time operator can be estimated using Eq.~\eqref{eq:avg-size} as 
\eq{\label{eq:late-op-dyn}S[A(t)] \approx \frac{ |c_1|^2 e^{2\lambda_1 t} \mathcal{S}[|l_1\sket]}{ |c_0|^2 + c_0c_1 e^{\lambda_1 t}\sbra l_0|l_1\sket + |c_1|^2e^{2\lambda_1 t} }, }
where we have used the normalization $\sbra l_j|l_j\sket = 1$ to simplify the expression. 

In section~\ref{sec:unusual-evec}, we observed that when $\gamma_2\gg\gamma_1$, the left eigenoperator ($\hat{l}_1$) corresponding to the slowest decaying eigenmode ($\lambda_1$) has size approximately equal to system size $N$. Typically, this would result in a portion of an operator being converted into this large operator at late times. However, this contribution also depends on the coefficient $c_1$, which accounts for the overlap of the right eigenoperator $\hat{r}_1$ with the initially local operator $A_0$. In the examples we have found (see appendix \ref{app:unusual-op}), $\hat{r}_1$ has much smaller overlap with short operators. As a result, $|c_1|$ is atypically small, and the contribution of $\hat{l}_1$ is suppressed compared to the bulk eigenmodes during the transient dynamics. This means that a local operator will grow to these large sizes only at relatively late times after the majority of other eigenmodes have frozen out. However, at these late times the operator size is dominated by the contribution $c_0\propto\tr{\left(A \rho_{ss}\right)}$ of the steady state left eigenoperator $\mathbb{I}$ (see denominator of the expression in Eq.~\eqref{eq:late-op-dyn}). Consequently, anomalous operator growth is typically hard to see for generic initial conditions.

One way to observe the dynamical impact of the unusually large left eigenoperator is to study the time evolution of operators that have comparably small overlap with the steady state $\rho_{ss}$. Here, we focus on a situation where this contribution is strictly zero by defining an initial operator as 
\eq{ \label{eq:noSS-init} \tilde{A} = A - \tr{(A \rho_{ss})} \times \mathbb{I},}
where $A$ is a single-site operator located at the center of the system, chosen according Eq.~\eqref{eq:rand-init-ops}  (the data in the figures is shown for the specific choice given in Eq.~\eqref{eq:figs-init-op}). In Fig.~\ref{fig:unusual-growth}(a), we show the time evolution of the average size $\mathcal{S}(t)$ of such an operator in the regime where $\gamma_1=0.05$ is much smaller than $\gamma_2=1$. Up to the numerically accessible system sizes, we observe that the average size of the operator saturates to a large value proportional to the system size. This is in stark contrast with a more generic case (see Fig.~\ref{fig:RL10-tot-len}), where operators eventually shrink as they develop weight on the identity operator. We stress that the subtraction of the steady state contribution in Eq.~\eqref{eq:noSS-init} is not the sole contributor towards this difference. Indeed, if we consider a parameter regime where the left eigenoperator is of short length ($\gamma_1=0.8$), the subtracted operator $\tilde{A}$ still shrinks to a small size independent of the system size $N$ at late times, as shown using dashed lines in Fig.~\ref{fig:unusual-growth}. 

In Fig.~\ref{fig:unusual-growth}(b), we show the size distribution $p_s$ of the time-evolved operator across different sizes $s$ of basis operators. As expected, initially, the single site-operator gets converted into operators of increasing sizes. Meanwhile, the operator norm (black dashed line) decays with a rate that is faster than the slowest decaying eigenmode. Since the left eigenoperators in the bulk are in general traceful (see Fig.~\ref{fig:app-unusual-secavg}), the time-evolving operator will develop some weight in the $p_{s=0}$ sector due to processes that convert short operators into $\mathbb{I}$. However, unlike in generic models, the presence of an unusually large eigenmode $l_1$ which decays with a small rate $-\lambda_1$ means that its relative contribution compared to the rest of the basis operators dominates, which leads to $p_{s=N}\approx 1$ at late times. The eigenoperator $\hat{l}_1$ is primarily composed of those size $s=N$ basis operators whose norm decay is smaller compared to the average decay rate of typical $s=N$ basis operators, as shown by the open symbols in Fig.~\ref{fig:dndt-vs-s}(b). As a result, this set of large operators dominates the dynamics at late times, similar to how the short operators contribute the most in the generic case. The inclusion of single-site dissipation $\gamma_1$ eventually removes this anomalous effect. Finally, the operator growth observed here is markedly different from the operator spreading in the Hamiltonian case shown in Fig.~\ref{fig:RL10-opDist-vs-t}(a), where the operators of all sizes $s>0$ are present at late times: the $\gamma_2$ dominated operator growth primarily involves operators of size $s=N$.

%%============================================================================

\begin{figure}
    \includegraphics{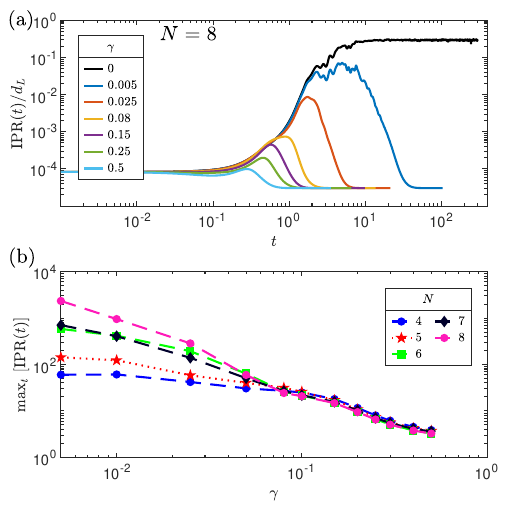}
    \caption{Inverse Participation Ratio (IPR): (a) The IPR of the time evolved operator, defined in Eq.~\eqref{eq:op-full_IPR}, is plotted as a function of time $t$. The data is shown for the $N=8$ site random Lindblad model ($\mathtt{seed=4}$) with Hamiltonian $J=1$, and varying values of dissipation rates $\gamma_1=\gamma_2=\gamma$, and the initially single-site operator defined in Eq.~\eqref{eq:figs-init-op}. The y-axis is rescaled by the total dimension of the operator space $d_L$. (b) The largest value of the IPR attained during the time evolution is plotted as a function of the dissipation rate for varying system size $N=4,5,6,7$, and $8$. } 
    \label{fig:RL10-IPR-vs-t} 
\end{figure}

\subsection{Operator scrambling}

Finally, we characterize how, given the size of the time-evolving operator, it is distributed over different Pauli basis operators. Following the treatment of the eigenoperators in section~\ref{sec:evec-ipr}, this can be quantified using the inverse participation ratio (IPR) defined as
\eq{ \label{eq:op-full_IPR} \text{IPR}(t) = \frac{\mathcal{N}(t)^2}{\sum_{\mathbf{m}=1}^{d_L}|\sbra F_{\mathbf{m}}|A(t)\sket|^4 } .}
In Fig.~\ref{fig:RL10-IPR-vs-t}(a), we show the time evolution of IPR for varying values of dissipation strengths $\gamma$. For the model with $\gamma=0$, the non-integrable Hamiltonian scrambles the initial operator over a large number of Pauli operators, leading to a large value of IPR at late time, which increases exponentially with increasing system size $N$. In the presence of dissipation, the operator is initially scrambled as its size increases, but eventually, the IPR vanishes as the weight gets transferred onto the identity operator. In Fig.~\ref{fig:RL10-IPR-vs-t}(b), we show the maximum value of the IPR attained during time evolution as a function of the dissipation strength.  In the presence of dissipation, the well-scrambled bulk eigenmodes rapidly decay and the operator grows for only a finite amount of time, both of which limit the degree of scrambling. When the dissipation is weak, the smaller system sizes saturate to an IPR on the order of the system size before hitting this limit. However, for larger values of $\gamma$, operator growth is cut off at values well below the system size, and the maximum value of the IPR becomes independent of the system size.

\begin{figure}
    \includegraphics{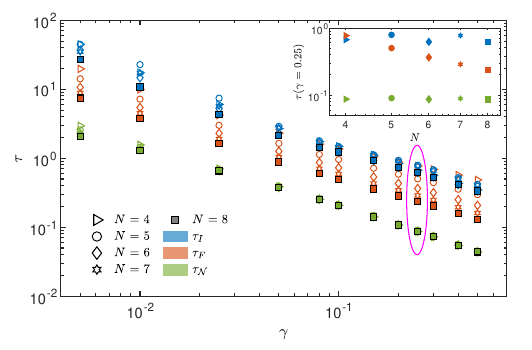}
    \caption{The relevant timescales for the operator dynamics are plotted as a function of dissipation rate $\gamma_1=\gamma_2=\gamma$ for multiple system sizes. The data is shown for the random Lindblad model ($\mathtt{seed}=4$) with $J=1$, and initial operator given in Eq.~\eqref{eq:figs-init-op}. $\tau_I$ (blue color) is the time for the weight on the identity component $p_{s=0}$ to reach $1/2$. $\tau_{\mathcal{N}}$ (green color) is the time required for the operator norm $\mathcal{N}(t)$ to decay below $1/2$. $\tau_F$ (orange color) is the freezing time defined in Eq.~\eqref{eq:tau-freeze} beyond which the contribution of the eigenmodes in the bulk of the spectrum becomes negligible. Different system sizes $N=4-8$ are indicated by different shaped symbols. (Inset) The time-scales are plotted as a function of system size $N$ for a fixed value of dissipation $\gamma=0.25$ (encircled in the main plot). The color scheme is identical to the main plot. } 
    \label{fig:RL10-op-timescales} 
\end{figure}

\subsection{Contribution from generic eigenmodes} 
Now that we have determined the generic dynamics of the operators, we come to the question of the influence of Ginibre-like spectral properties on these. From section~\ref{sec:spectrum}, we recall that the eigenmodes that are located in the bulk of the spectrum defined by $|\text{Re}(\lambda)-\overline{X}|\le2\sigma_X$ show generic Ginibre-like level repulsion. Here $\overline{X}$ and $\sigma_X$ are the center and the standard deviation of the spectrum along the real axis, respectively (see.~Eqs.~\ref{eq:xbar}, and ~\ref{eq:sigmaXY}). Each of these eigenmodes decays with a rate proportional to $|\text{Re}(\lambda)|$. Moreover, we recall that local operators generally have low overlaps with right eigenoperators of the eigenmodes in this bulk part of the spectrum (see Fig.~\ref{fig:secAvg}(a) and Fig.~\ref{fig:RL-secAvg}). Consequently, we expect that the dynamics of local operators will have limited contributions due to the bulk eigenmodes.

More specifically, we define a \emph{freezing} time $\tau_F$ as
\eq{\label{eq:tau-freeze} \tau_{F} =2\ \frac{1}{|\ \overline{X} + 2\sigma_X\ |},}
which approximates the time after which the eigenmodes outside of the bulk ($\text{Re}(\lambda)>-|\overline{X}|+2\sigma_X$) of the spectrum effectively describe the dynamics. In Fig.~\ref{fig:RL10-op-timescales}, we compare $\tau_F$ with the time $\tau_{\mathcal{N}}$ for operator normalization to reach $1/2$ and time $\tau_I$ for the weight on the identity operator $(p_{s=0})$ to become $1/2$. Since the location of spectral bulk approximately scales as $|\overline{X}|\approx N$ and $\sigma_X\approx\sqrt{N}$ (see Fig.~\ref{fig:ee4}(b)), the freezing time window $\tau_F$ becomes shorter and shorter with increasing system size. However, the timescales for the operator dynamics $\tau_{\mathcal{N}}$ and $\tau_I$ remain finite. While $\tau_F$ is larger than $\tau_\mathcal{N}$ for the numerically accessible system sizes that are explored here, we expect that $\tau_F$ will eventually become the shortest timescale. Hence, the contribution of generic eigenmodes is frozen on much shorter time scales and has a negligible contribution towards the operator dynamics for larger system sizes.

%% =========================================================
\section{Conclusions}
\label{sec:conclusions}

In this paper, we have investigated how Ginibre-like bulk statistics of two classes of local Lindbladian models influence the resulting Lindbladian dynamics.  We have shown that in all of these models, the size of both left and right eigenoperators is strongly correlated with their decay rate $|\text{Re}(\lambda_j)|$ --- a feature that is not seen in non-local Lindbladians with Ginibre-like spectra. We have also verified that eigenoperators in the bulk of the spectrum are essentially random within each size sector. These two basic observations yield considerable insight into the time evolution of various physical quantities. The non-local character of the bulk eigenoperators means that their dynamics is observed predominantly in non-linear correlation functions of the density matrix, which is dominated by bulk eigenmodes up to times of order $1/N$.  This early-time dynamics is universal, showing almost no variation across random ensembles of entangled initial states, reflecting the random nature of bulk eigenmodes.

The strong correlation between eigenoperator size and decay rate for both left and right eigenoperators implies that the growth and scrambling of local operators is dominated by eigenmodes far outside the bulk of the spectrum in the thermodynamic limit. Thus, as dissipation increases at fixed $N$ (or equivalently, as $N$ increases at fixed dissipation rate), our models rapidly enter a regime where operators neither grow to a substantial size nor become highly scrambled even over short Pauli strings. We have shown how this suppression of operator growth and scrambling at late times arises in two physically distinct measures of operator decoherence: the rate of decay of the operator norm, and the rate of increase of the operator trace.  We also have observed that in some models with finite system sizes, the size dependence of operators outside of the bulk of the spectrum is significantly altered when dissipation is dominated by strictly two-site terms, allowing certain operators on the order of the system size to be anomalously long-lived.  

Our work raises several intriguing questions about the dynamics of dissipative quantum many-body systems. First, we have shown that a coarse description of the eigenoperator properties in terms of their size distribution and IPR allows us to correctly predict the dynamics of random classes of entangled states.  However, in section~\ref{sec:state_dynamics}, we also observed that ensembles of random product states have a markedly different behavior than those of generic random pure states. It would be interesting to study further how this difference manifests within the eigenoperators and whether it has additional physical consequences. A promising avenue is to study the entanglement structure of eigenoperators local Lindbladians using existing metrics for operator entanglement, with some developments already made for Ginibre ensembles \cite{cipolloni2023ginibreEvecEntanglement}. Second, we have observed that interacting dissipation can lead to different long-time behavior than single-site dissipation.  The precise mechanism for this difference would be interesting to investigate, as would the question of whether this phenomenon can find useful applications in real dissipative systems. In this work, we have focused on Lindblad models with bulk dissipation. It will be interesting to explore the effect of generic boundary dissipation on the operator size distributions and operator growth.  

More broadly, there are also interesting questions in connecting our work to other approaches to RMT in Lindbladian dynamics.  For example, we have focused on the static signature of level repulsion, measured in terms of the complex spacing ratio. Other metrics, such as the dissipative spectral form factor \cite{li2021dsff}, have been recently shown to effectively diagnose chaotic behavior for non-local models with Ginibre statistics \cite{liProsenChan2024MultipleModelDsff,Shivam2024}.  It would be interesting to explore such diagnostics for local Lindbladians.  Similarly, we have focused on Lindbladians with Ginibre-like spectra; it would also be interesting to investigate whether additional universal dynamical signatures can be obtained for ensembles of Lindbladians in the more restricted symmetry classes introduced by Refs. \cite{altland2021fermionLindbladSymmetry,lieu2020tenFold,sa2023symmetry,kawabata2023symmetryLindblad}.

\begin{acknowledgements} 
The authors thank Sarang Gopalakrishnan for insightful discussions. FJB and SC acknowledge the support of NSF DMR-2313858. SC acknowledges the financial support from the Robert O. Pepin Fellowship for summer 2025. FJB is grateful for the support of the Tang family professorship.  The authors acknowledge the computational resources provided by the Minnesota Supercomputing Institute (MSI) at the University of Minnesota.
\end{acknowledgements}

\section*{Data Availability Statement}
The data that support the findings of this article are publicly available on Zenodo at \cite{zenodoRMT}.
%\clearpage

%%%========================================================
\appendix
%%%========================================================
%% change the labeling style for the appendix
\setcounter{equation}{0} 
\setcounter{figure}{0} 
\setcounter{table}{0} 
\setcounter{section}{0} 
\renewcommand{\theequation}{A\arabic{equation}} 
\renewcommand{\thetable}{T\arabic{table}} 
\renewcommand\thefigure{A\arabic{figure}} 
\renewcommand{\theHtable}{Appendix.\thetable} 
\renewcommand{\theHfigure}{Appendix.\thefigure}

%%%%%%%%%%%%%%%%%%%%%%%% Tables  %%%%%%%%%%%%%%%%%%%%%%%%

%%%% ----------------- HH-1 table ---------------------------
\begin{table}
    \centering
    \begin{tabular}{|c|c|}
         $\mathtt{seed}$& $Q_{3\times 1}$\\
         \hline
         1& $\begin{pmatrix}
            -0.3222 & 0.7884 & 0.9287
        \end{pmatrix}$\\ 
         2& $\begin{pmatrix}
             -0.0726 &   0.0193  & -0.3934
        \end{pmatrix}$\\
         3& $\begin{pmatrix}
             0.9457 &  -0.3435 &  -1.1318
        \end{pmatrix}$\\
         4& $\begin{pmatrix}
             -0.4152 &  -0.7234  &  0.0929
        \end{pmatrix}$\\
         17& $\begin{pmatrix}
             0.1164 &  -1.0986  &  0.76994
        \end{pmatrix}$\\
         20& $\begin{pmatrix}
             -0.0546 &  -0.2478 &   0.0285
        \end{pmatrix}$\\
        \hline
    \end{tabular}
    \caption{Single-site Hamiltonian terms in random Lindblad model (see Eq~\eqref{eq:RL3-ham})}
    \label{tab:h1}
\end{table}

%%%% ----------------- HH-2 table ---------------------------
\begin{table}
    \centering
    \begin{tabular}{|c|c|}
         $\mathtt{seed}$ & $R_{3\times 3}$\\
         \hline
         1& $\begin{pmatrix}
            -0.6490 &  -1.1096 &  -0.5587\\
            1.1812  & -0.8456  &  0.1784\\
           -0.7585 &  -0.5727  & -0.1969
        \end{pmatrix}$\\ 
         2& $\begin{pmatrix}
            -0.1242 &  -0.1960 &  -1.1289\\
           -2.5415  & -0.1962  &  0.1942\\
            0.2772  & -0.3057 &  -0.6071             
        \end{pmatrix}$\\
         3& $\begin{pmatrix}
             0.0685  &  0.0359  & -0.4112\\
            0.9512   & 1.1221  & -0.6418\\
           -0.3448   & 0.3237  & -0.6615
        \end{pmatrix}$\\
         4& $\begin{pmatrix}
             2.3459  &  0.7440 &   1.0007\\
            0.0893  &  0.6762  & -1.8874\\
            2.2103 &  -0.4959  & -1.2499
        \end{pmatrix}$\\
         17& $\begin{pmatrix}
             -0.3951  & -1.8820 &   0.2868\\
            0.1406   & 0.7965   & 0.2686\\
           -1.5172  &  0.2413  & -2.1682
        \end{pmatrix}$\\
         20& $\begin{pmatrix}
             0.4335  &  1.1127  & -0.4518\\
            2.2726  & -1.6630   & 0.0203\\
            2.2105  &  1.0826  &  0.7218
        \end{pmatrix}$\\
        \hline
    \end{tabular}
    \caption{Nearest neighbor Hamiltonian term in random Lindblad model (see Eq~\eqref{eq:RL3-ham})}
    \label{tab:h2}
\end{table}

%%%% ----------------- gamma-1 table ---------------------------
\begin{table}
    \centering
    \begin{tabular}{|c|c|}
         $\mathtt{seed}$ & $K_{3\times 1}$\\
         \hline
         1& $\begin{pmatrix}
            0.5864 - 1.5094i \\ -0.8519 + 0.8759i \\  0.8003 - 0.2428i
        \end{pmatrix}$\\ 
         2& $\begin{pmatrix}
             -0.8284 - 1.1221i \\  0.5358 + 0.0460i \\  0.1095 - 1.2386i
        \end{pmatrix}$\\
         3& $\begin{pmatrix}
             -0.2253 + 0.3087i \\ -0.8299 - 0.5701i \\ -0.1262 + 0.5388i
        \end{pmatrix}$\\
         4& $\begin{pmatrix}
             -0.2327 + 0.9440i \\  0.1599 + 1.6672i \\ -1.0078 - 0.9105i
        \end{pmatrix}$\\
         17& $\begin{pmatrix}
             -0.1562 + 0.3709i  \\ 1.7974 + 0.7810i \\ -1.4667 - 1.3599i
        \end{pmatrix}$\\
         20& $\begin{pmatrix}
             -1.3543 + 0.8499i \\  -0.9625 + 1.6579i  \\ 0.8736 + 0.6706i
        \end{pmatrix}$\\
        \hline
    \end{tabular}
    \caption{Single-site jump operator term in random Lindblad model (see Eq~\eqref{eq:RL3-lind})}
    \label{tab:g1}
\end{table}

%%%% ----------------- gamma-2 table ---------------------------

\begin{table*}
    \centering
    \begin{tabular}{|c|c|}
         $\mathtt{seed}$ & $D_{3\times 3}$\\
         \hline
         1& $\begin{pmatrix}
            0.1668 - 1.8651i &  1.1752 + 1.4022i &  0.6037 + 1.2708i\\
          -1.9654 - 1.0511i &  2.0292 - 1.3677i  & 1.7813 + 0.0660i\\
          -1.2701 - 0.4174i & -0.2752 - 0.2925i &  1.7737 + 0.4513i
        \end{pmatrix}$\\ 
         2& $\begin{pmatrix}
             0.6382 - 1.1398i &  0.6610 - 1.6300i & -1.0163 - 0.0846i\\
           1.1452 + 0.3198i & -2.5455 - 0.9364i & -0.1156 - 0.8837i\\
          -0.0159 - 0.5715i &  0.0125 - 0.2786i & -0.7763 + 0.5118i
        \end{pmatrix}$\\
         3& $\begin{pmatrix}
             0.4669 + 0.1826i & -0.6617 - 1.3769i  & 0.6549 + 0.8313i\\
              -1.9741 + 1.2400i & -0.2770 - 0.4378i & -0.1366 + 0.4271i\\
               0.3630 - 0.4302i & -0.5213 + 0.3819i & -1.0028 + 1.7635i
        \end{pmatrix}$\\
         4& $\begin{pmatrix}
            0.4261 - 0.9377i & -1.6404 - 1.5713i &  0.4931 + 0.1329i\\
          -1.7079 - 0.5498i  & 1.9991 + 1.0665i  & 1.5809 + 1.9775i\\
          -0.2570 + 0.1921i & -0.0522 - 0.4822i &  1.0607 + 0.0485i
        \end{pmatrix}$\\
         17& $\begin{pmatrix}
             0.3476 - 0.8426i & -0.0437 + 0.8957i & -0.1180 + 1.4216i\\
           0.0843 - 1.2889i & -0.0934 - 0.3961i &  0.7153 + 0.2375i\\
           0.2321 - 1.3960i &  0.1947 - 0.9414i & -0.2241 - 0.5402i
        \end{pmatrix}$\\        
         20& $\begin{pmatrix}
             -2.3623 + 0.2897i &  0.8103 - 0.0050i & -0.6235 - 0.4999i\\
          -0.7797 + 1.6267i &  2.9868 + 0.6006i &  0.7898 - 1.9554i\\
          -0.5591 - 0.0911i &  0.3307 + 0.6567i & -0.0221 + 1.0997i
        \end{pmatrix}$\\
        \hline
    \end{tabular}
    \caption{Two-site jump operator term in random Lindblad model (see Eq~\eqref{eq:RL3-lind})}
    \label{tab:g2}
\end{table*}

%%%%%%%%%%%%%%%%%%%%%%%%%%%%%%%%%%%%%%%%%%%%%%%%%%%%%%%%%%%%%%

\begin{figure}
	\includegraphics{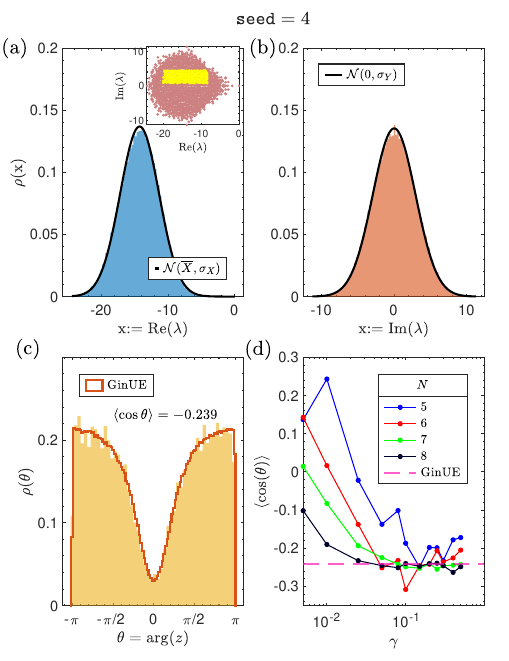}
	\caption{\label{fig:RL-sd-4-spectrum} The eigenvalue spectrum of the realization $\mathtt{seed=4}$ of the random Lindblad model used in section~\ref{sec:state_dynamics} and \ref{sec:operator_dynamics} for studying the dynamics ($N=8,J=1,\gamma=0.25$): (a)The density of eigenvalues as a function of the real part of the eigenvalues $\lambda$. The black-colored line shows the Gaussian distribution with mean $\overline{X}$ and variance $\sigma^2_X$ (see~\cref{eq:xbar,eq:sigmaXY}). (b) The analogous density plot is shown as a function of the imaginary part of the eigenvalues. The inset in panel (a) shows the full complex spectrum (red). The eigenvalues from the bulk of the spectrum used for CSR computation are colored in yellow. (c) The marginal distribution of $\theta=\text{arg}(z)$, where $z$ is the CSR defined in Eq.~\eqref{eq:zz}. The orange curve shows the corresponding distribution for the complex Ginibre matrix. (d) The average of $\cos(\theta)$ is shown as a function of the dissipation strength $\gamma_1=\gamma_2=\gamma$ for varying system size values $N$. The pink dashed line indicates the result for the Ginibre ensemble \cite{sa2020csr}. } 
\end{figure}

\begin{figure}
    \includegraphics{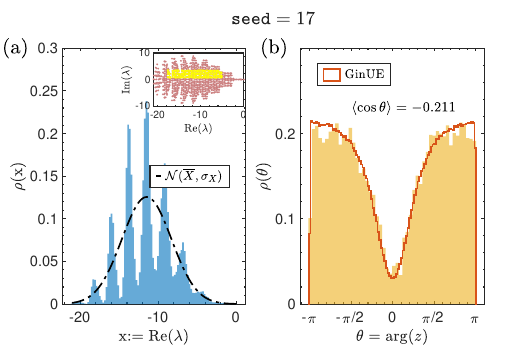}
    \caption{\label{fig:RL-fragment-spectrum} The eigenvalue spectrum of the realization  $\mathtt{seed=17}$ of the random Lindblad model, which exhibits a fragmented spectrum ($N=8,J=1,\gamma=0.25$): (a) The inset shows the eigenvalue spectrum in the complex plane. The main panel shows the density $\rho_X$ of the eigenvalues as a function of their real part. The black dashed curve represents the Gaussian probability distribution function with mean $\overline{X}$ and variance $\sigma^2_X$ (see~\cref{eq:xbar,eq:sigmaXY}) (b) The histogram shows the marginal angular distribution of the CSR (corresponding to the eigenvalues in the bulk shown using yellow color in the inset of panel (a)). The orange curve depicts the distribution obtained for the Ginibre matrices. } 
\end{figure}

\begin{figure}
	\includegraphics{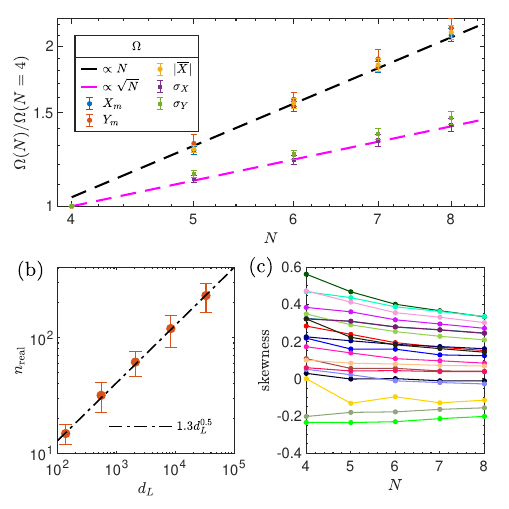}
	\caption{\label{fig:RL-spectrum-shape} The spectrum of random Lindblad model across multiple realizations: The results are shown for Hamiltonian strength $J=1$, and dissipation rate $\gamma_1=\gamma_2=0.25$. (a) The spectral shape in terms of the largest absolute real part ($X_m$), the largest imaginary value ($Y_m$), center of the spectrum ($\overline{X}$), spread in real ($\sigma_X$) and the imaginary direction ($\sigma_Y$) is shown as a function of the total number of sites $N$ (see~\cref{eq:xm-ym,eq:xbar,eq:sigmaXY} for detailed expressions). The trendlines show that the center and the extremal boundaries are proportional to $N$, whereas the standard deviations approximately grow as $\sqrt{N}$. (b) The number of exactly real eigenvalues is plotted as a function of the dimension of the operator space $d_L\approx 2^{2N-1}.$ In (a-b), the data points are obtained by averaging over 20 independent realizations, and the error bars show the standard deviation across these realizations. (c) The skewness $\tilde{\mu}_3$ of the real part of eigenvalues (see Eq.~\eqref{eq:skew}) as a function of the system size $N$ for 20 realizations of the model. } 
\end{figure}

\begin{figure}
	\includegraphics{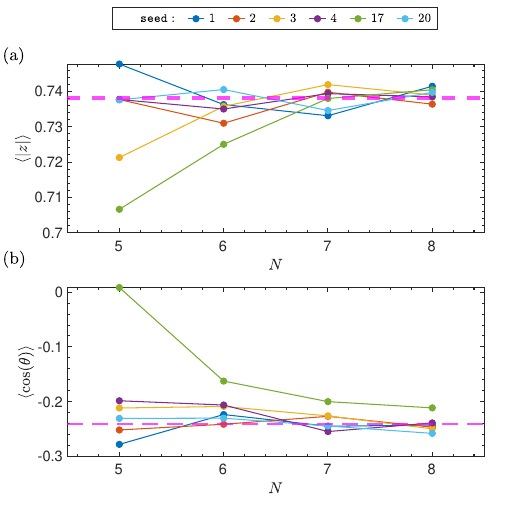}
	\caption{\label{fig:RL-csr} Statistics of complex spacing ratio (CSR) $z=|z|e^{i\theta}$ (see Eq.~\eqref{eq:zz}) for the random Lindblad model $(J=1,\gamma_1=\gamma_2=0.25)$: (a) The average of absolute value $|z|$, (b) the average of $\cos{\theta}$ is plotted as a function of increasing system sizes $N$. The data is shown for the realizations that appear in the main text, and they can be identified by the associated label $\mathtt{seed}$. We only consider the eigenvalues in the window $|\text{Re}(\lambda)-\overline{X}|\le2\sigma_X,$ and $\sigma_Y/4\le\text{Im}(\lambda)\le3\sigma_Y/2$, to avoid the eigenmodes near the boundary of the spectrum while computing the CSR statistics. The dashed magenta-colored horizontal lines indicate the respective values for the Ginibre ensemble \cite{sa2020csr}. } 
\end{figure}
%------------- alpha ------------------

\begin{figure}
	\includegraphics{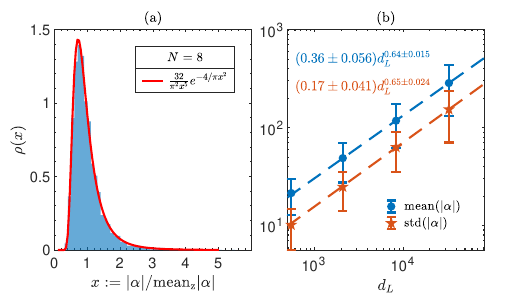}
    	\caption{\label{fig:RL-alpha} Statistics of eigenvector overlaps $\alpha_j:=1/\sbra l_j|r_j\sket$ for the random Lindblad model. (a) Probability distribution of the absolute value $|\alpha|/\text{mean}_z(|\alpha|)$ is shown for $\mathtt{seed=4}$ realization of the random model. (b) The average and the standard deviation of $|\alpha|$ are shown as a function of the dimension of the operator space $d_L$. The solid data points and error bars represent the average and the standard deviation over 20 independent realizations, respectively. The dashed lines are a fit to the function $y=a x^b$. The eigenvalue window to gather the data is separately chosen for each realization as described in the caption for Fig.~\ref{fig:alpha}.} 
\end{figure}

\section{Spectrum of random Lindblad model}
\label{app:RL-spectrum}

In this appendix, we present numerical results related to the eigenvalue spectrum of the random Lindblad model defined by Eqs.~\eqref{eq:RL3-ham} and~\eqref{eq:RL3-lind} to shed further light on which features of the eigenvalue spectrum are universal across a range of local Lindblad models. The numerical values of the model parameters for all of the independently sampled realizations are tablulated in \cref{tab:h1,tab:h2,tab:g1,tab:g2}.

We begin with the discussion of the eigenvalue spectrum of the realization of the random Lindblad model (labeled as $\mathtt{seed=4}$), which is used in section ~\ref{sec:state_dynamics} and ~\ref{sec:operator_dynamics} to illustrate generic dynamics. In Fig.~\ref{fig:RL-sd-4-spectrum}(a--c), we show data for $N=8$ system size and $J=1,\gamma_1=\gamma_2=0.25$. The density of the eigenvalues as a function of their real part, shown in Fig.~\ref{fig:RL-sd-4-spectrum}(a), is well approximated by a Gaussian distribution whose average is given by the center of the spectrum $\overline{X}$, and standard deviation along the real axis by $\sigma_X$. Similar to the dissipative Ising model, there is an asymmetry along the real axis as seen by a finite skewness of $\tilde{\mu}_3 =-0.0112 $ (see Eq.~\eqref{eq:skew}). The analogous comparison of the density of the eigenvalues along the imaginary axis is shown in Fig.~\ref{fig:RL-sd-4-spectrum}(b). We characterize the level repulsion in the eigenvalue spectrum by analyzing the statistics of the complex spacing ratios (see Eq.~\eqref{eq:zz}) in the bulk part of the spectrum indicated in the inset of Fig.~\ref{fig:RL-sd-4-spectrum}(a). In Fig.~\ref{fig:RL-sd-4-spectrum}(c), we observe that the angular distribution $\rho(\theta)$ of the CSR $z=|z|e^{i\theta}$ matches closely with that of the Ginibre random matrices. Furthermore, in Fig.~\ref{fig:RL-sd-4-spectrum}(d), we show the average value of $\langle\cos(\theta)\rangle$ as a function of the dissipation strength $\gamma$. For all values of $\gamma$, $\langle \cos \theta \rangle$ approaches the Ginibre value as the system size increases, with the trend being monotonic for most values of $\gamma$. The approach towards the Ginibre value with increasing $N$ is slower for smaller values of $\gamma$, since the Hamiltonian interactions are more dominant in the weak dissipation limit.

While $\mathtt{seed=4}$ realization exhibits a unimodal spectrum centered around $\overline{X}$, this behavior is not observed for all models that show Ginibre-like features in the CSR. In Fig.~\ref{fig:RL-fragment-spectrum}(a), we show the spectrum of another realization of the random Lindblad model labeled by $\mathtt{seed=17}$, which has a stripe-like fragmented structure. The model parameters are same as before: $J=1,\gamma_1=\gamma_2=0.25$. In this case, we see that the eigenvalues are clustered around a few well-separated points. However, most of the eigenvalues are still within the $\overline{X}\pm 2\sigma_X$ region of the complex plane. While eigenvalues separate into different clusters, they still exhibit generic local level repulsion as shown in Fig.~\ref{fig:RL-fragment-spectrum}(b). Here we see that the marginal angular distribution of CSR $z=|z|e^{i\theta}$ qualitatively agrees with the generic curve for the Ginibre random matrices. However, the agreement is weaker compared to $\mathtt{seed}=4$ model as seen in Fig.~\ref{fig:RL-csr}. Even though the value of the parameter $\gamma_1$ is equal to $\gamma_2$, the operator norm of the sampled single-site jump operator $L_1$ is much higher compared to the norm of the two-site jump operator $L_2$ for $\mathtt{seed}=17$ (for this sample, $\sbra L_1|L_1\sket\approx 10 \sbra L_2|L_2\sket$). This suggests that the fragmentation in the spectrum arises due to strong size dependence of the single-site jump operators (similar to the non-interacting model studied in section~\ref{sec:eigenvecs}), but the remaining randomly oriented terms of finite strength mix the eigenmodes near each fragment, leading to near-generic level repulsion.

In Fig.~\ref{fig:RL-spectrum-shape}, we show data for additional independently sampled realizations of this random Lindblad model. We provide additional evidence for the scaling of the spectral features as a function of the system size $N$ in Fig.~\ref{fig:RL-spectrum-shape}(a). We observe that the extent of the spectrum in both real and imaginary directions, parameterized using $X_m$ and $Y_m$, respectively, is proportional to the system size $N$. Similarly, the center of the spectrum scales as $\overline{X}\sim N$. Moreover, the spread of the spectrum around its center, computed using the standard deviations $\sigma_X$ and $\sigma_Y$, increases as $\sqrt{N}$. Both of these scalings are recovered for multiple independent realizations of the random Lindblad model as evidenced by small error bars in Fig.~\ref{fig:RL-spectrum-shape}(a). Along with the results for the dissipative Ising model in Fig.~\ref{fig:ee4}(b), this shows that the Lindbladians with local and bounded terms have spectra with these scalings, in accordance with expectations of extensivity and the central limit theorem. Importantly, this means that most of the eigenmodes in the bulk of the spectrum of the local Lindbladian will have characteristically shorter lifetimes compared to eigenmodes near the steady state, in the thermodynamic limit. Fig.~\ref{fig:RL-spectrum-shape}(b) shows that the number of exactly real eigenvalues increases as $\sqrt{d_L}$, as expected for the real Ginibre matrices. Finally, Fig.~\ref{fig:RL-spectrum-shape}(c), we show the skewness of the spectrum along the real axis, which in general remains non-zero even for large system sizes. This is also expected generically, as (unlike for Hamiltonian systems) there is no physical reason for the fast-decaying eigenmodes at large values $|\text{Re}(\lambda)|$ to resemble those at small $|\text{Re}(\lambda)|$. To complete the analysis of the eigenvalues, we show the CSR statistics for $\langle | z | \rangle$ and  $\langle \cos(\theta) \rangle$ in Fig.~\ref{fig:RL-csr}, for all of the realizations of the random Lindblad model that are presented in the main text.

\begin{figure}
    \includegraphics{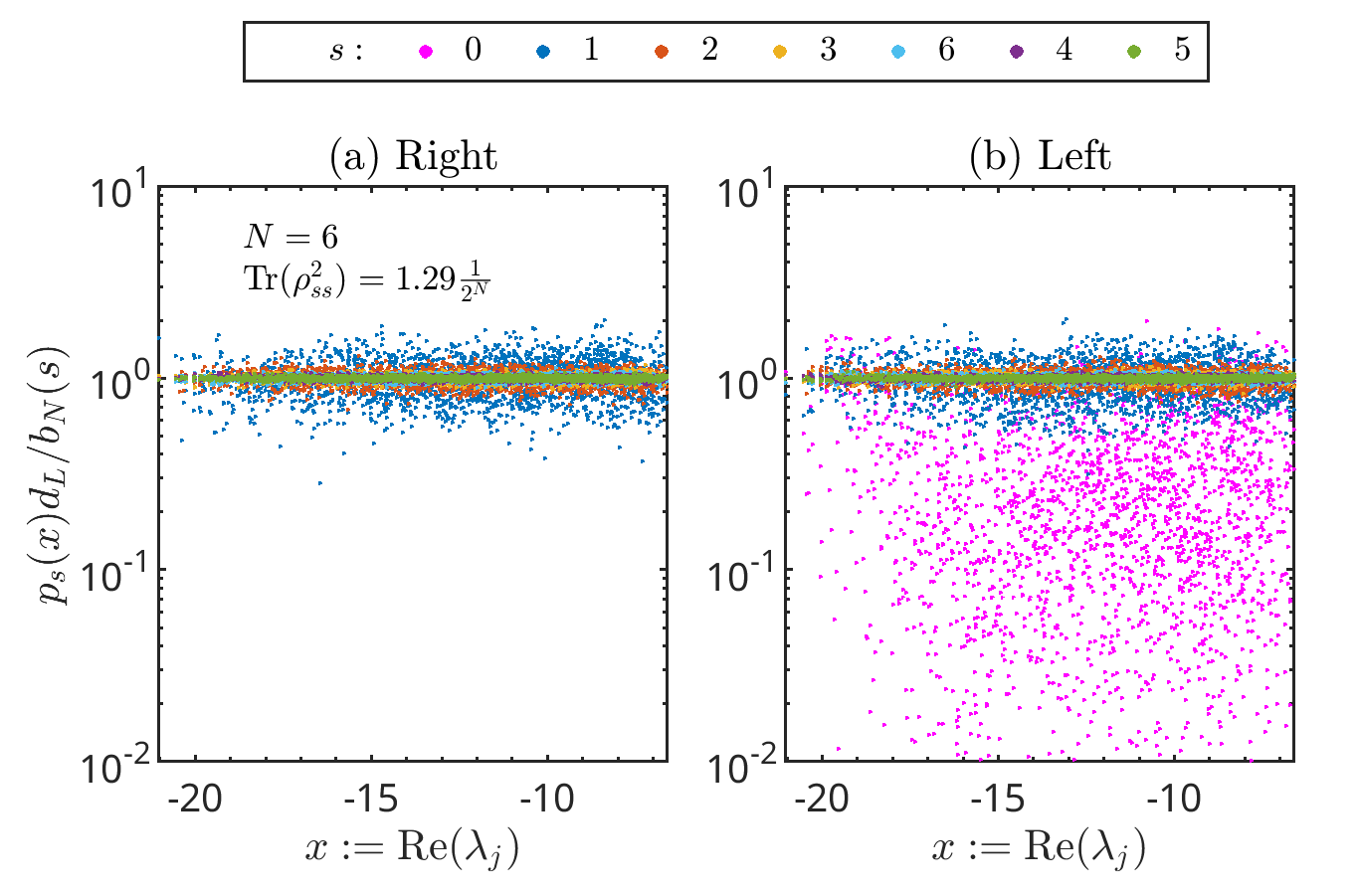}
    \caption{The size distribution of the (a) right and (b) left eigenoperators for the non-local Lindblad model in Eq.~\eqref{eq:md-nonLoc} is plotted as a function of the real part of eigenvalue $\lambda$. The steady state eigenmode $\lambda=0$ is not shown here. The y-axis is rescaled by the operator space dimension $d_L=4^N$ and the number of size $s$ basis operators $b_N(s)=3^s\binom{N}{s}$. The collapse indicates that the eigenoperators are uniformly distributed as expected in Eq.~\eqref{eq:nonLoc-size-uniform}. }  
    \label{fig:nonLoc-size} 
\end{figure}

\begin{figure}
    \includegraphics{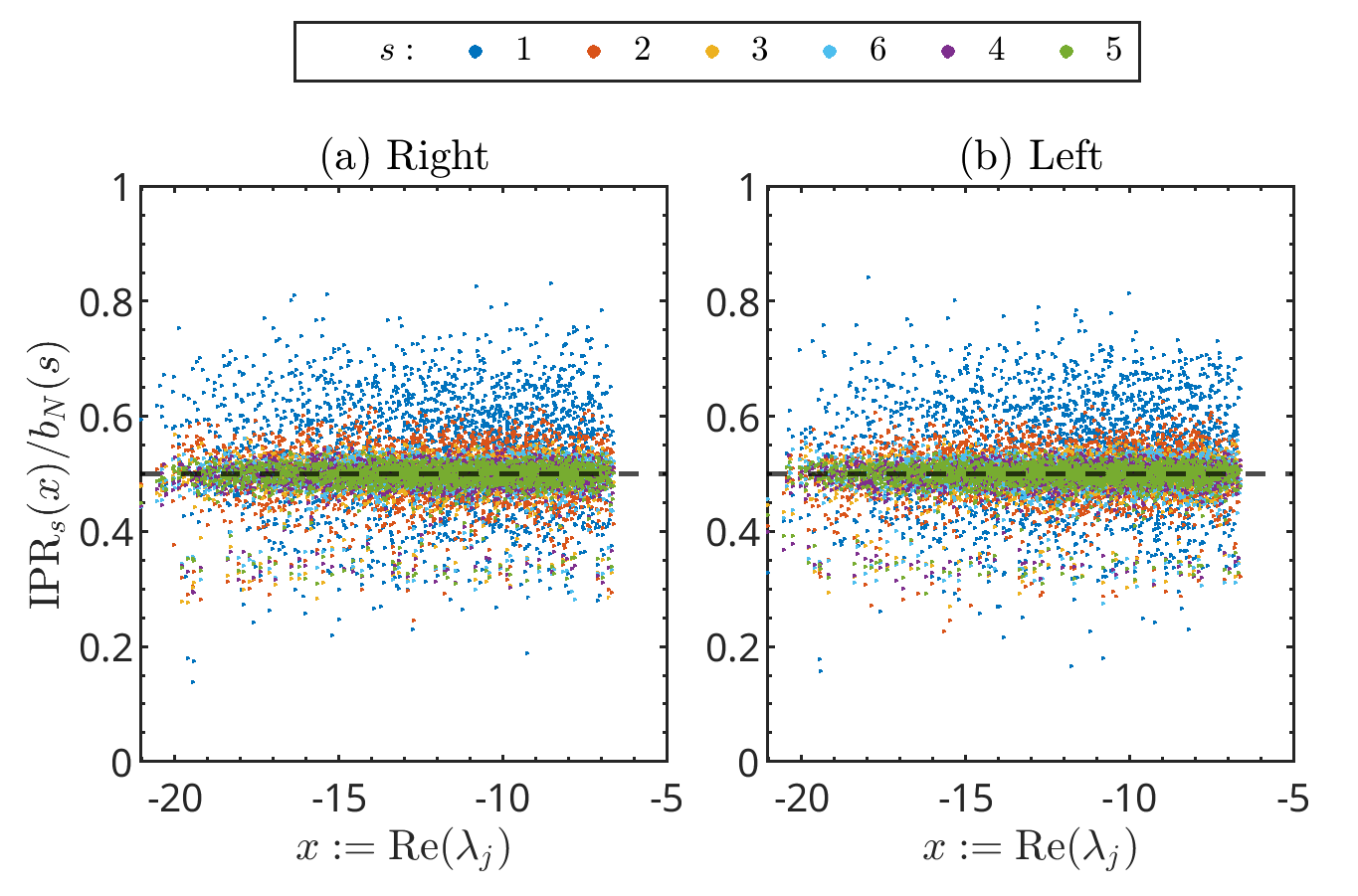}
    \caption{The IPR over the basis operators of size $s$ of the (a) right and (b) left eigenoperators for the non-local Lindblad model in Eq.~\eqref{eq:md-nonLoc} is plotted as a function of the real part of eigenvalue $\lambda$. The steady state eigenmode $\lambda=0$ is not shown here. The collapse indicates that the IPR saturates to $b_N(s)/2$. Here $b_N(s)=3^s\binom{N}{s}$ is the total number of size-s basis operators in full operator space. } 
    \label{fig:nonLoc-IPR} 
\end{figure}

\begin{figure}
    \includegraphics{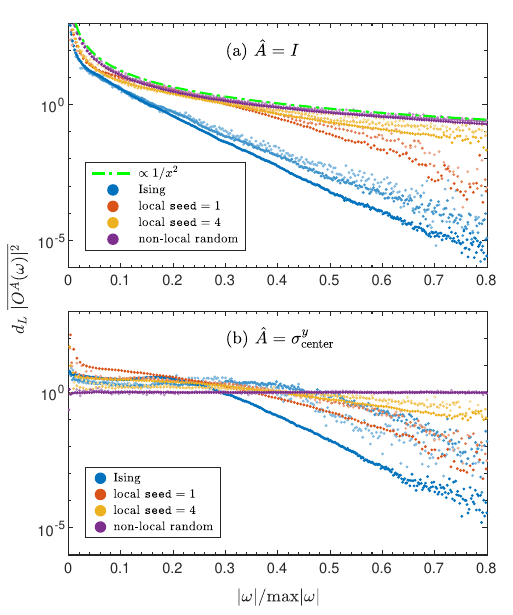}
    \caption{ The overlap of the right eigenoperators with superoperator (a) $\mathcal{A} = \mathbb{I}\otimes\mathbb{I}$ (b) $\mathcal{A} = \sigma^y_m\otimes{\sigma^y_m}^*$ is shown as a function of coarse-grained eigenvalue difference $|\omega|$. The $\sigma^y_m$ acts on site $(N+1)/2$ for odd $N$ and on a pair of sites $N/2,(N+2)/2$ for even values of $N$. The data is coarse-grained over a window of size $\delta_\omega=0.1$ (see Eq.~\eqref{eq:sup-op-avg} for details of the averaging procedure). The y-axis is rescaled by the dimension of the operator space $d_L$. The data is shown for four different models: (blue) dissipative Ising model with $\gamma=0.8$ (yellow and orange) two independent realizations of the local random Lindblad model with $\gamma_1=\gamma_2=0.25$ (purple) non-local random Lindblad model. The light to darker shades indicate increasing system sizes $N=5,6,7$.}
	\label{fig:app-combined-sup_vs_ee}
\end{figure}
Finally, we analyze the statistics of the overlap of left and right eigenoperators defined as $\alpha_j:=1/\sbra l_j|r_j\sket$. In Fig.~\ref{fig:RL-alpha}, we show the distribution of the $\alpha$ rescaled by its average value $\langle |\alpha|\rangle$. As for the dissipative Ising model (see Fig.~\ref{fig:alpha}), it matches the known result for the Ginibre ensemble. This provides further evidence for the universality of the local correlations among eigenoperators. In Fig.~\ref{fig:RL-alpha}(b), we show the average value $\langle |\alpha|\rangle$ and its standard deviation as a function of increasing dimension of the operator space $d_L$. The fit suggests that they increase as $\approx d_L^{0.65}$, similar to the Ising model. We note that the uncertainty in the fit parameters is comparatively smaller compared to the Ising model due to averaging over multiple realizations. However, data for larger system sizes would be needed to get more accurate fit for individual realizations.

To summarize, we have confirmed that a range of independently sampled realizations of the random local Lindblad model exhibit generic Ginibre-like eigenvalue correlations. While the overall shape of the spectrum significantly depends on the specific choice of the model, the generic level statistics is observed for sufficiently strong dissipation and large values of system size.

\section{Non-local Lindblad model}
\label{app:non-local}
In this appendix, we discuss the structure of eigenoperators of Lindblad models with non-local interactions and dissipation. We focus on the random Lindbladians that exhibit generic eigenvalue repulsion and hence obey Ginibre-like statistical properties for their eigenvalues. However, since they can couple sites far apart from each other, the correspondence between the size of the eigenoperator and their eigenvalue established in section~\ref{sec:eigenvecs} for local models is lost. This analysis will help further contrast the features induced in generic open systems due to locality.

We consider a system with $N$ sites, each of which hosts a spin-$1/2$ degree of freedom similar to the models considered in the main text. Let $A$ and $B$ be random square matrices of size $2^N\times 2^N$. The entries of these matrices are complex numbers $x_{jk}+i y_{jk}$ whose real and imaginary parts are independent random variables sampled from the standard normal distribution with zero mean and unit variance. The Hamiltonian and the dissipative part of the Lindbladian are then chosen as
\eq{ \label{eq:md-nonLoc} \eqsp{H &= \frac{N}{4 \ 2^{N/2}}(A +  A^\dagger),\\
L_\alpha &= \frac{1}{\sqrt{2 \ 2^N}} B , \qquad \alpha = 1,2,\ldots N.}}
The pre-factors ensure that $||H|| \sim N, ||L_\alpha||\sim 1$, so that the strength of the unitary and the dissipative parts are comparable to each other.

In Fig.~\ref{fig:nonLoc-size}, we show the size distribution of the eigenoperators for this model. As expected, the size distribution is agnostic of the eigenvalue (or equivalently, decay rate) of the eigenmode. Expecting a uniform distribution then implies that it will be given by 
\eq{\label{eq:nonLoc-size-uniform} p(s) = \frac{b_N(s)}{d_L}}
where $d_L=4^N$ is the total dimension of the operator space and $b_N(s)=3^s\binom{N}{s}$ is the total number of basis operators of size $s$ in an $N$ site system. We note that the $s=0$ curve for the left eigenoperator in Fig.~\ref{fig:nonLoc-size}(b) is unusually suppressed. This can be explained by noticing that even in this random all-to-all interacting model, the steady state right eigenoperator ($\hat{r}_0$) affects the left eigenoperators. The purity of the steady state for this particular realization of the model is $\tr{(\rho_{ss}^2)}\approx 1.29\frac{1}{2^N}$. Indeed, this means that the steady state has considerable weight on the identity basis operator ($s=0$), compared to the remaining right eigenoperators. To satisfy the bi-orthogonality condition
\eq{\sbra l_j |r_0\sket = 0 \quad \text{for  } j=1,2,\ldots 4^N-1 \ , }
The left eigenoperators need to have much smaller contributions from the $\mathbb{I}$ operator, leading to a suppressed value of $p_{s=0}$.

Finally, we also analyze the inverse participation ratio of the eigenoperators over different basis operators. The average value of the IPR saturates to a finite value of $\frac{1}{2}b_N(s)$ for all of the basis sizes (see Fig.~\ref{fig:nonLoc-IPR}). This confirms that the eigenoperators are indeed highly scrambled (similar to the eigenvectors of Ginibre matrices) across all Pauli operators. We observe larger fluctuations around this average value for small values of $s$, since there are comparably fewer basis operators, leading to larger statistical fluctuations.

\section{Superoperator overlaps}
\label{app:sup-ops}

In this appendix, we analyze the \emph{matrix elements} of local superoperators evaluated in the eigenbasis of the Lindbladian. In a recent work \cite{ferrari2025openStripeETH}, these overlaps were shown to be smooth functions of their eigenvalues for generic Lindbladians.  These matrix elements control the dynamics of non-linear correlation functions discussed in Eq.~\eqref{eq:general-nonlinear}. Specifically, we will consider the overlaps defined by 
\eq{ O^A_{jk} : = \sbra r_j |A\otimes A^* |r_k\sket ,}
where $r_j(r_k)$ is the right eigenoperator of the Lindbladian with eigenvalues $\lambda_j (\lambda_k)$, and $A$ is a local operator. 

In the following, we focus on the eigenvalue dependence of $O^I$ and $O^Y$, which appear in the expression of the purity and R\'enyi-2 correlator, respectively (see Eq.~\eqref{eq:purity}, and Eq.~\eqref{eq:renyi-init}). We are interested in finding whether the overlap $O_{jk}$ remains finite when the eigenvalues corresponding to $r_j$ and $r_k$ are of order $\sim N$, which would situate these eigenmodes in the bulk part of the spectrum that exhibits generic Ginibre statistics.  We will focus on the functional dependence of these overlaps on the eigenvalue differences at a fixed average eigenvalue, using a coarse-grained description defined as
\eq{ \label{eq:sup-op-avg}|O^A(\omega)|^2 = \frac{1}{\widetilde{n}_\omega}\sum\nolimits_{j\neq k}^{''} |\sbra r_j | A \otimes A^*|r_k\sket|^2 .}
Let us unpack the averaging procedure in detail: Here, $\sum^{''}_{j\neq k}$ is performed over those pairs of eigenmodes which have their average eigenvalue constrained by \mbox{$|\text{Re}\frac{(\lambda_j+\lambda_k)}{2} - \overline{X}|\le\frac{\sigma_X}{3}$} and \mbox{$|\text{Im}\frac{(\lambda_j+\lambda_k)}{2} |\le\frac{\sigma_Y}{3}$}, and eigenvalue difference centered around $\omega$ as $||\lambda_j-\lambda_k| -\omega|\le \frac{\delta_\omega}{2}$. The parameters $\overline{X}$ and $\sigma_{X(Y)}$ represent the center and the spread of the spectrum along the real (imaginary) axis, respectively (see~\cref{eq:xbar,eq:sigmaXY}). $\delta_\omega>0$ is the size of the discretization window. Finally, $\widetilde{n}_\omega$ is the total number of eigenvalue pairs that satisfy these constraints and participate in the summation $\sum^{''}_{j\neq k}$ for a given value of $\omega$.

Let us begin by analyzing the case where $A=\mathbb{I}$ corresponding to the overlap $\sbra r_j|r_k\sket$, which in general is non-zero since the eigenvectors are not orthogonal to each other. We recall that the eigenoperators are normalized such that $\sbra r_j|r_j\sket=1$. If $|r_j\sket,|r_k\sket$ are independent and identically distributed random vectors, then, according to Eq.~\eqref{eq:rr-scaling}, we expect $|\sbra r_j|r_k\sket|^2\sim\frac{1}{d_L}$. However, even for random Ginibre matrices, it is known that the eigenoperators are not truly uncorrelated random operators; instead, for Ginibre matrices, the overlap depends on the eigenvalue difference as $d_L |\sbra r_j|r_k\sket|^2 \sim 1/|\omega|^2$\cite{fyodorov2018statistics,bourgade2020distribution}.
In Fig.~\ref{fig:app-combined-sup_vs_ee}(a), we show the associated matrix elements $O^I(\omega)$ as a function of the absolute value of eigenvalue difference $|\omega|$ for a variety of models considered here. For the non-local Lindblad model in Eq.~\eqref{eq:md-nonLoc}, we recover both the anticipated scaling with $d_L$, as observed by the system-size collapse for the purple curves in Fig.~\ref{fig:app-combined-sup_vs_ee}(a), and a good agreement with the theoretical prediction of $1/ |\omega|^2$. The overlaps for the local Lindblad models deviate considerably from both predictions, showing an exponential dependence on $|\omega|$ at larger complex eigenvalue differences, and slightly more variation over different system sizes. 
We conjecture that, owing to the locality, the eigenoperators that are far apart in the complex plane will have considerably different size distribution, which should lead to higher suppression of the overlaps for large values of $|\omega|$ compared to the non-local Lindblad model.

Next, we consider the overlap between a pair of eigenoperators upon insertion of the local superoperator $\mathcal{Y}={\sigma}^y\otimes {{\sigma}^y}^*$ acting at the center of the spin-chain; the results are shown in Fig.~\ref{fig:app-combined-sup_vs_ee}(b).   
For eigenoperators of non-local Lindblad operators, the local change induced by this superoperator removes the Ginibre-like correlations characterized by the $1/|\omega|^2$ dependence, and turns the pair of eigenoperators into essentially uncorrelated random vectors. In this case, we recover the prediction in Eq.~\eqref{eq:rr-scaling}, where the overlap becomes independent of $\omega$ and equal to $1/d_L$ (purple curve in Fig.~\ref{fig:app-combined-sup_vs_ee}(b)). In the local Lindblad model, however, this local rotation does not lead to an $|O^Y(\omega)|^2$ that is independent of $|\omega|$: instead, the exponential fall-off of matrix element overlaps with $|\omega|$ persists, similar to the behavior of off-diagonal matrix elements of local operators in Hamiltonians satisfying ETH. We conjecture that the reason for the two phenomena is similar: For local models, the eigenvalue specifies the density of excitations in the system. When the eigenvalues corresponding to a pair of eigenmodes differ significantly, the density of excitations in the two eigenoperators is different, and thus their local structure is expected to differ also. This leads to higher suppression of overlaps when the eigenvalue difference $|\omega|$ is large.

Moreover, we observe additional features emerge in $\sbra r_j|\mathcal{Y}|r_k\sket$ at small values of $|\omega|$ in models with local terms, similar to off-diagonal matrix elements in the case of generic local Hamiltonians \cite{rigolETHreview}. We leave the detailed study of this dependence and its implications on the observable dynamics for future work.

In summary, we observe that the overlaps of the superoperators with eigenoperators corresponding to pairs of nearby eigenmodes from the bulk of the spectrum are not heavily suppressed even when we restrict the interactions to be local. This, in turn, means that these generic bulk eigenmodes meaningfully contribute to the dynamics of the system at early times, when we consider the non-linear correlation functions that are controlled by such superoperator overlaps.

\begin{figure}
	\includegraphics{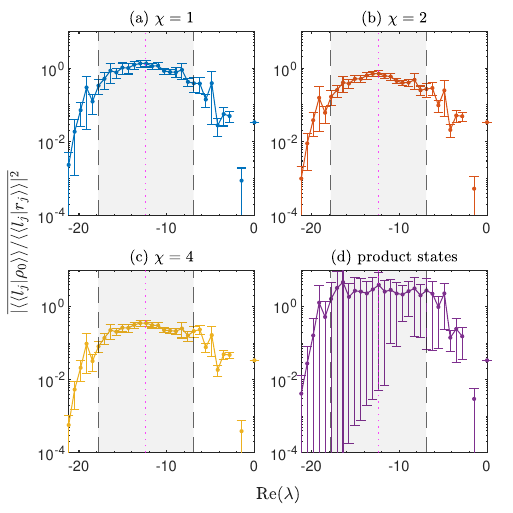}
    \caption{The overlaps of initial state $\rho_0$ with left eigenoperators defined in Eq.~\eqref{eq:vlrho-with-alpha}. The data is shown for
parameters $N=7,J=1,\gamma_1=\gamma_2=0.25$, and $\mathtt{seed} = 4$. The initial states are chosen from the ensemble defined in Eq.\eqref{eq:init_purity} (a) $\chi=1$ (b)$\chi=2$ (c)$\chi=4$ and (d) Haar random product states. (See captions of Fig.~\ref{fig:RL-10-vlRho-disc} for additional details).   }
    \label{fig:app-vlrho-alpha} 
\end{figure}

\begin{figure}
    \includegraphics{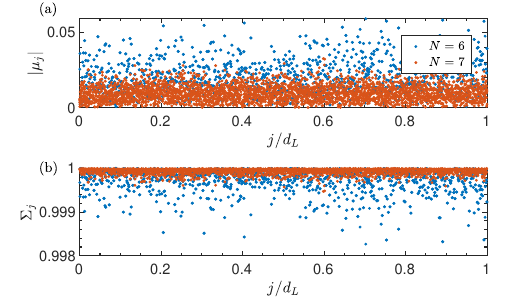}
    \caption{The statistics of the complex phases of the initial state overlaps (see Eq.~\eqref{eq:app-phase-mean-std}) (a) The mean value $\mu_j$ and (b) the standard deviation $\Sigma_j$ as a function of the index $j$ that labels the eigenmodes. The x-axis is rescaled by the total number of eigenmodes $d_L$. The data is shown for
parameters $J=1,\gamma_1=\gamma_2=0.25$, $\mathtt{seed} = 4$ and 100 independently chosen random states according to Eq.~\eqref{eq:init_purity} with $\chi=1$. Two different system sizes $N=6$(blue) and $N=7$(orange) are shown, where the data approaches the uncorrelated values with increasing value of $N$. } 
    \label{fig:app-vlrho-phase} 
\end{figure}

%%%%%%%%%%%%%%%%%%%%%%%%%%%%%%%%%%%%%%%%%%%%%%%%%%

\section{Initial state spectral overlaps}
\label{app:initial-ovlp}

In this section, we provide additional numerical results on the overlaps of the density matrix $\rho_0$ and the eigenoperators $|r_j\sket$ and $|l_j\sket$.

In the main text, we observed that the absolute values of overlaps $|\sbra l_j|\rho_0\sket$ are independent of the initial states chosen from the random ensembles. However, the expression of the purity in Eq.~\eqref{eq:purity-full-expr} has an additional term $\frac{1}{\sbra r_j|l_j\sket}$ that also depends on the left eigenoperators. Specifically, we define the modified coarse-grained overlap 
\eq{ \label{eq:vlrho-with-alpha} \eqsp{\overline{[\alpha|\sbra l |\rho_0\sket|^2]_{\Delta}}(x_0) : =& \\ \sum_{a=1}^{N_r} \ \sum_{j:|\text{Re}(\lambda_j)-x_0|\le \Delta/2 }&\frac{1}{N_r n_\Delta}\frac{|\sbra l_j|\rho_0^{(a)}\sket |^2}{|\sbra r_j|l_j\sket|^2}},}
where, similar to the main text, the first averaging is done over a coarse-graining window of size $\Delta$, which in total contains $n_\Delta$ eigenmodes, and followed by averaging over $N_r$ randomly chosen initial states from the given ensemble. In Fig.~\ref{fig:app-vlrho-alpha}, we plot these overlaps as a function of the real part of the eigenvalue. They share similar qualitative features to the overlaps plotted in the main text Fig.~\ref{fig:RL-10-vlRho-disc}. The overlaps are approximately constant in the bulk part of the spectrum. Additionally, smaller error bars indicate fewer variations across different initial states chosen from the random ensemble. As expected, the data for the product states in Fig.~\ref{fig:app-vlrho-alpha} (d) show markedly higher variations.

Next, we analyze the correlations between phases of complex numbers $\sbra r_j | r_k\sket$ and $\alpha_j^*\sbra \rho_0|l_j\sket\sbra l_k|\rho_0\sket\alpha_k$. Here, $\alpha_j=1/\sbra l_j|r_j\sket$. The correlations are captured in terms of the mean $\mu_j$ and standard deviation $\Sigma_j$ defined in Eq.~\eqref{eq:phase-mean-std}, which we reproduce here:
\eq{\label{eq:app-phase-mean-std}\eqsp{\mu_j &= \frac{1}{d_L}\sum_{k=0}^{d_L-1} e^{i(\arg{[w_{j,k}]} - \arg{[z_{j,k}]})}  \\
\Sigma_j^2 &= \frac{1}{d_L}\sum_{k=0}^{d_L-1} |e^{i(\arg{[w_{j,k}]} - \arg{[z_{j,k}]})}-\mu_j|^2 },}
In Fig~\ref{fig:app-vlrho-phase}, we plot the statistics for a single state $\rho_0$ randomly chosen from the ensemble of entangled pure states. Both $\mu_j$ and $\Sigma_j$ are approximately independent of the index $j$, and their variation decreases with increasing system size. This observation agrees with the phases being uncorrelated with each other. For uncorrelated phases, $\mu_j$ is expected to be a small number since it corresponds to the addition of unit vectors with randomly distributed arguments. If $\mu_j\approx 0$, then according to Eq.~\eqref{eq:app-phase-mean-std}, the standard deviation $\Sigma_j$ becomes close to $1$.

In summary, we indeed confirm that the independence of overlaps $|\sbra l|\rho_0\sket|$ on the specific state $\rho_0$ is a good indicator of the state-independent decoherence rate. We numerically verify that the phases of the complex coefficients and the details of the eigenoperator overlaps do not qualitatively alter this dependence.

%%%%%%%%%%%%%%%%%%%%%%%%%%%%%%%%%%%%%%%%%%%%%%%%%%

\begin{figure}
	\includegraphics{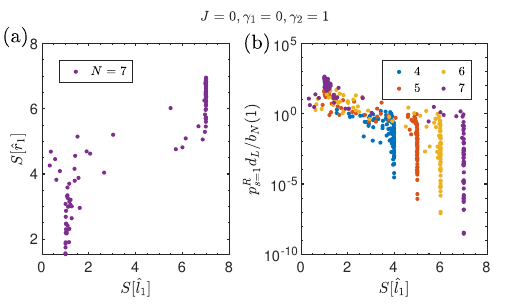}
    \caption{(a) The size of the right eigenoperator as a function of the size of the left eigenoperator for $100$ randomly sampled realizations. (b) The size distribution of the right eigenoperator on size $s=1$ basis operators as a function of the total size of the left eigenoperator. The y-axis is scaled by the number of single-site basis operators $b_N(s=1)$ and the total dimension of the operator space $d_L$. } 
    \label{fig:app-unusual-right} 
\end{figure}

\begin{figure}
	\includegraphics{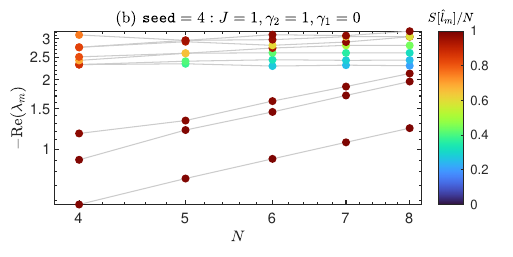}\\	
    \includegraphics{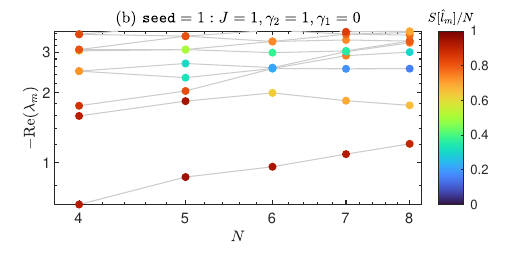}
    \caption{First $10$ eigenmodes of random Lindblad model ordered according to their $|\text{Re}(\lambda)|$ are plotted as a function of the system size $N$. The steady state eigenmode ($\lambda=0$), which has size $s=0$ is not shown. Each data point is colored according to the size of the corresponding left eigenoperator normalized by the system size, i.e.,$S[\hat{l}_m]/N$. The data is shown for (a) $\mathtt{seed}=4,\gamma_1=0$ and (b) $\mathtt{seed}=1,\gamma_1=0$. The axes are represented on a log-log scale. In both panels, the eigenvalues of left eigenoperators with size $\sim \mathcal{O}(N)$ (red colored dots) increases with increasing values of $N$. } 
    \label{fig:unusual-ee-vs-N} 
\end{figure}

\section{Anomalous eigenoperators: additional details}
\label{app:unusual-op}
\begin{figure*}
    \includegraphics{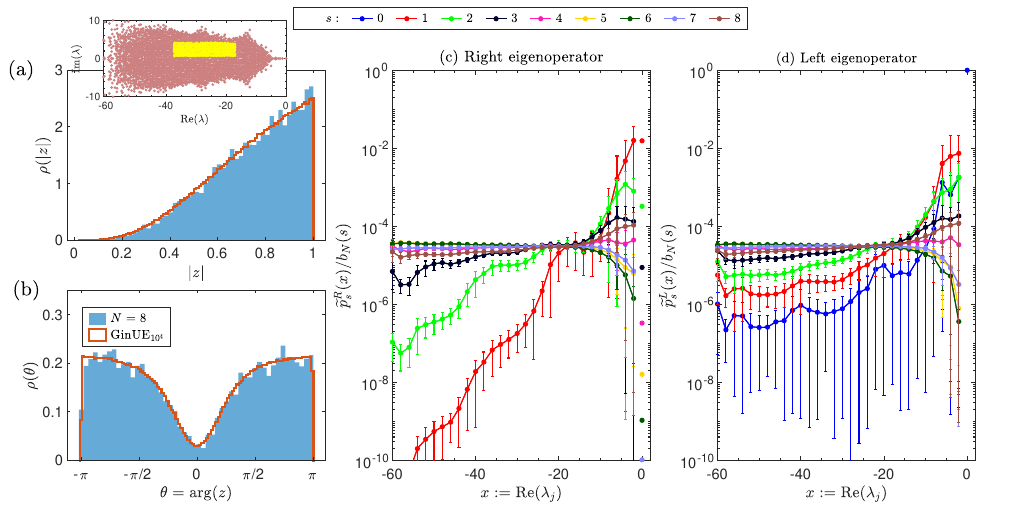}
    \caption{Random Lindblad model realization $\mathtt{seed}=4,J=1,\gamma_1=0,\gamma_2=1$: (a) The average of the absolute value of CSR ($|z|$) and (b) the average of $\cos{\theta}$, where $\theta$ is the argument of CSR. The orange curves show the respective results for the Ginibre matrix. The inset in panel (a) shows the complex eigenvalue spectrum (red), and window used for computing CSR (yellow). The discretized operator size distribution (see Eq.~\eqref{eq:ps_tilde}, with $\Delta=2$) of the (c) right and (d) left eigenoperators of the random Lindblad model. This is the $\gamma_1=0$ limit of the model shown in Fig.~\ref{fig:unusual-growth}, which exhibits large operator growth at late times. Note that the $\hat{r}_1,\hat{l}_1$ will have the largest weight in the $s=8$ sector, which is not apparent because of the rescaling of the y-axis by $b_N(s)$ (number of size $s$ basis operators). The error bars are defined similarly to Fig.~\ref{fig:secAvg}.} 
    \label{fig:app-unusual-secavg} 
\end{figure*}
In this section, we provide additional details related to the parameter regime $\gamma_2\gg\gamma_1$ where the anomalous operator growth is observed at late times.

The initially local operator is converted into the left eigenoperator $\hat{l}_1$ corresponding to the eigenvalue with the smallest absolute real part at very late times. However, the corresponding coefficient depends on the overlap
\eq{ \frac{\sbra r_1|A_0\sket}{\sbra r_1|l_1\sket} ,}
where $A_0$ is the initial operator. If the operator $r_1$ is also of size $N$, then this overlap will be small, which may lead to a very long timescale for this eigenmode to dominate the dynamics even in the presence of a finite Lindblad gap. Indeed, in a local model, this must be the case: an initially small operator cannot grow into an operator on the order of the system size in a system-size independent time.  In Fig.~\ref{fig:app-unusual-right}(a), we compare the size $S$ of the right and the left eigenoperators for $100$ independent realizations of the model in the two-site only dissipation limit $(\gamma_1=J=0)$. We observe that if the $S[\hat{l}_1]$ is of $\mathcal{O}(N)$, the corresponding right eigenoperator also has a large size. In Fig.~\ref{fig:app-unusual-right}(b), we compare the weight of the right eigenoperator on the basis operators of size $s=1$ by analyzing $p_{s=1}$. For a random operator which is an equal mixture of all possible basis operators, we would expect $p^R_{s=1}\sim \frac{b_N(s=1)}{d_L}$. Here, $b_N(s=1)=3\lceil N/2\rceil$ is the number of reflection-symmetric single-site basis operators, and $d_L$ is the total dimension of the operator space. When the left eigenoperator has a large size, the weight $p^R_{s=1}$ of the right-eigenoperator on single-site bases is not larger than the expectation of a random vector, and is often much smaller. These observations agree with general constraints on the time evolution of a local operator, where the time for the operator to grow up to size $\sim N$ increases with the system size. We also analyze how the eigenvalue of this anamolous eigenmode changes as a function of system size $N$. In Fig.~\ref{fig:unusual-ee-vs-N}, we plot the first few eigenmodes of Lindbadian that are closest to the steady state for increasing values of $N$. The eigenvalues $|\text{Re}(\lambda)|$ of eigenmodes whose left eigenoperators have size of $\mathcal{O}(N)$ increase with $N$. However, the eigenmodes that have left eigenoperators with size $\mathcal{O}(1)$ approximately remain independent of system size. This trend suggests that the size of the slowest decaying left eigenoperator $\hat{l}_1$ will be of order $\mathcal{O}(1)$ in the thermodynamic limit.

Finally, we also show the properties of eigenmodes in the bulk of the spectrum for this unusual model in Fig.~\ref{fig:app-unusual-secavg} for $N=8$ system size. The statistics of CSR computed for the bulk eigenmodes agrees well with those of the Ginibre ensemble as seen in Fig.~\ref{fig:app-unusual-secavg}(a--b). The size distribution of the eigenoperators of this model is shown in Fig.~\ref{fig:app-unusual-secavg}(c--d). The weight of the right eigenoperators on $s=1$ basis operators is exponentially suppressed with increasing $|\text{Re}(\lambda)|$. Whereas the contribution of longer basis operators is fairly constant for eigenmodes in the bulk part of the spectrum. Comparing these observations with generic size-distributions shown in Fig.~\ref{fig:secAvg} and Fig.~\ref{fig:RL-secAvg}, we see that the eigenmodes in the bulk of this model also exhibit a size distribution that is characteristic of underlying local interactions. However, the size distribution of eigenoperators corresponding to eigenmodes with small values of $|\text{Re}(\lambda)|$ is highly atypical. Here we observe that the large basis operators with sizes $s=7,8$ have considerably higher contributions to these slowly decaying eigenmodes. This is in strict contrast with the generic case where $p_s$ corresponding to a large value of $s$ is highly suppressed in this part of the spectrum.

\clearpage %to force bibliography at the end, maybe remove it later..
\bibliography{ref.bib}

\begin{thebibliography}{139}%
\makeatletter
\providecommand \@ifxundefined [1]{%
 \@ifx{#1\undefined}
}%
\providecommand \@ifnum [1]{%
 \ifnum #1\expandafter \@firstoftwo
 \else \expandafter \@secondoftwo
 \fi
}%
\providecommand \@ifx [1]{%
 \ifx #1\expandafter \@firstoftwo
 \else \expandafter \@secondoftwo
 \fi
}%
\providecommand \natexlab [1]{#1}%
\providecommand \enquote  [1]{``#1''}%
\providecommand \bibnamefont  [1]{#1}%
\providecommand \bibfnamefont [1]{#1}%
\providecommand \citenamefont [1]{#1}%
\providecommand \href@noop [0]{\@secondoftwo}%
\providecommand \href [0]{\begingroup \@sanitize@url \@href}%
\providecommand \@href[1]{\@@startlink{#1}\@@href}%
\providecommand \@@href[1]{\endgroup#1\@@endlink}%
\providecommand \@sanitize@url [0]{\catcode `\\12\catcode `\$12\catcode
  `\&12\catcode `\#12\catcode `\^12\catcode `\_12\catcode `\%12\relax}%
\providecommand \@@startlink[1]{}%
\providecommand \@@endlink[0]{}%
\providecommand \url  [0]{\begingroup\@sanitize@url \@url }%
\providecommand \@url [1]{\endgroup\@href {#1}{\urlprefix }}%
\providecommand \urlprefix  [0]{URL }%
\providecommand \Eprint [0]{\href }%
\providecommand \doibase [0]{https://doi.org/}%
\providecommand \selectlanguage [0]{\@gobble}%
\providecommand \bibinfo  [0]{\@secondoftwo}%
\providecommand \bibfield  [0]{\@secondoftwo}%
\providecommand \translation [1]{[#1]}%
\providecommand \BibitemOpen [0]{}%
\providecommand \bibitemStop [0]{}%
\providecommand \bibitemNoStop [0]{.\EOS\space}%
\providecommand \EOS [0]{\spacefactor3000\relax}%
\providecommand \BibitemShut  [1]{\csname bibitem#1\endcsname}%
\let\auto@bib@innerbib\@empty
%</preamble>
\bibitem [{\citenamefont {Wigner}(1955)}]{wigner}%
  \BibitemOpen
  \bibfield  {author} {\bibinfo {author} {\bibfnamefont {E.~P.}\ \bibnamefont
  {Wigner}},\ }\bibfield  {title} {\bibinfo {title} {Characteristic vectors of
  bordered matrices with infinite dimensions},\ }\href
  {http://www.jstor.org/stable/1970079} {\bibfield  {journal} {\bibinfo
  {journal} {Annals of Mathematics}\ }\textbf {\bibinfo {volume} {62}},\
  \bibinfo {pages} {548} (\bibinfo {year} {1955})}\BibitemShut {NoStop}%
\bibitem [{\citenamefont {Bohigas}\ \emph {et~al.}(1984)\citenamefont
  {Bohigas}, \citenamefont {Giannoni},\ and\ \citenamefont
  {Schmit}}]{bohigas1984bgsconjecture}%
  \BibitemOpen
  \bibfield  {author} {\bibinfo {author} {\bibfnamefont {O.}~\bibnamefont
  {Bohigas}}, \bibinfo {author} {\bibfnamefont {M.~J.}\ \bibnamefont
  {Giannoni}},\ and\ \bibinfo {author} {\bibfnamefont {C.}~\bibnamefont
  {Schmit}},\ }\bibfield  {title} {\bibinfo {title} {Characterization of
  chaotic quantum spectra and universality of level fluctuation laws},\ }\href
  {https://doi.org/10.1103/PhysRevLett.52.1} {\bibfield  {journal} {\bibinfo
  {journal} {Phys. Rev. Lett.}\ }\textbf {\bibinfo {volume} {52}},\ \bibinfo
  {pages} {1} (\bibinfo {year} {1984})}\BibitemShut {NoStop}%
\bibitem [{\citenamefont {Deutsch}(1991)}]{deutsch1991eth}%
  \BibitemOpen
  \bibfield  {author} {\bibinfo {author} {\bibfnamefont {J.~M.}\ \bibnamefont
  {Deutsch}},\ }\bibfield  {title} {\bibinfo {title} {Quantum statistical
  mechanics in a closed system},\ }\href
  {https://doi.org/10.1103/PhysRevA.43.2046} {\bibfield  {journal} {\bibinfo
  {journal} {Phys. Rev. A}\ }\textbf {\bibinfo {volume} {43}},\ \bibinfo
  {pages} {2046} (\bibinfo {year} {1991})}\BibitemShut {NoStop}%
\bibitem [{\citenamefont {Srednicki}(1994)}]{srednicki1994eth}%
  \BibitemOpen
  \bibfield  {author} {\bibinfo {author} {\bibfnamefont {M.}~\bibnamefont
  {Srednicki}},\ }\bibfield  {title} {\bibinfo {title} {Chaos and quantum
  thermalization},\ }\href {https://doi.org/10.1103/PhysRevE.50.888} {\bibfield
   {journal} {\bibinfo  {journal} {Phys. Rev. E}\ }\textbf {\bibinfo {volume}
  {50}},\ \bibinfo {pages} {888} (\bibinfo {year} {1994})}\BibitemShut
  {NoStop}%
\bibitem [{\citenamefont {Srednicki}(1999)}]{srednicki1999}%
  \BibitemOpen
  \bibfield  {author} {\bibinfo {author} {\bibfnamefont {M.}~\bibnamefont
  {Srednicki}},\ }\bibfield  {title} {\bibinfo {title} {The approach to thermal
  equilibrium in quantized chaotic systems},\ }\href
  {https://doi.org/10.1088/0305-4470/32/7/007} {\bibfield  {journal} {\bibinfo
  {journal} {Journal of Physics A: Mathematical and General}\ }\textbf
  {\bibinfo {volume} {32}},\ \bibinfo {pages} {1163} (\bibinfo {year}
  {1999})},\ \Eprint {https://arxiv.org/abs/cond-mat/9809360}
  {arXiv:cond-mat/9809360} \BibitemShut {NoStop}%
\bibitem [{\citenamefont {Rigol}\ \emph {et~al.}(2008)\citenamefont {Rigol},
  \citenamefont {Dunjko},\ and\ \citenamefont
  {Olshanii}}]{rigol2008thermalization}%
  \BibitemOpen
  \bibfield  {author} {\bibinfo {author} {\bibfnamefont {M.}~\bibnamefont
  {Rigol}}, \bibinfo {author} {\bibfnamefont {V.}~\bibnamefont {Dunjko}},\ and\
  \bibinfo {author} {\bibfnamefont {M.}~\bibnamefont {Olshanii}},\ }\bibfield
  {title} {\bibinfo {title} {Thermalization and its mechanism for generic
  isolated quantum systems},\ }\href {https://doi.org/10.1038/nature06838}
  {\bibfield  {journal} {\bibinfo  {journal} {Nature}\ }\textbf {\bibinfo
  {volume} {452}},\ \bibinfo {pages} {854} (\bibinfo {year}
  {2008})}\BibitemShut {NoStop}%
\bibitem [{\citenamefont {D'Alessio}\ \emph {et~al.}(2016)\citenamefont
  {D'Alessio}, \citenamefont {Kafri}, \citenamefont {Polkovnikov},\ and\
  \citenamefont {Rigol}}]{rigolETHreview}%
  \BibitemOpen
  \bibfield  {author} {\bibinfo {author} {\bibfnamefont {L.}~\bibnamefont
  {D'Alessio}}, \bibinfo {author} {\bibfnamefont {Y.}~\bibnamefont {Kafri}},
  \bibinfo {author} {\bibfnamefont {A.}~\bibnamefont {Polkovnikov}},\ and\
  \bibinfo {author} {\bibfnamefont {M.}~\bibnamefont {Rigol}},\ }\bibfield
  {title} {\bibinfo {title} {From quantum chaos and eigenstate thermalization
  to statistical mechanics and thermodynamics},\ }\href
  {https://doi.org/10.1080/00018732.2016.1198134} {\bibfield  {journal}
  {\bibinfo  {journal} {Advances in Physics}\ }\textbf {\bibinfo {volume}
  {65}},\ \bibinfo {pages} {239} (\bibinfo {year} {2016})}\BibitemShut
  {NoStop}%
\bibitem [{\citenamefont {M\"uller}\ \emph {et~al.}(2004)\citenamefont
  {M\"uller}, \citenamefont {Heusler}, \citenamefont {Braun}, \citenamefont
  {Haake},\ and\ \citenamefont
  {Altland}}]{mullerAltland2004semiclassicalChaos}%
  \BibitemOpen
  \bibfield  {author} {\bibinfo {author} {\bibfnamefont {S.}~\bibnamefont
  {M\"uller}}, \bibinfo {author} {\bibfnamefont {S.}~\bibnamefont {Heusler}},
  \bibinfo {author} {\bibfnamefont {P.}~\bibnamefont {Braun}}, \bibinfo
  {author} {\bibfnamefont {F.}~\bibnamefont {Haake}},\ and\ \bibinfo {author}
  {\bibfnamefont {A.}~\bibnamefont {Altland}},\ }\bibfield  {title} {\bibinfo
  {title} {Semiclassical foundation of universality in quantum chaos},\ }\href
  {https://doi.org/10.1103/PhysRevLett.93.014103} {\bibfield  {journal}
  {\bibinfo  {journal} {Phys. Rev. Lett.}\ }\textbf {\bibinfo {volume} {93}},\
  \bibinfo {pages} {014103} (\bibinfo {year} {2004})}\BibitemShut {NoStop}%
\bibitem [{\citenamefont {Kos}\ \emph {et~al.}(2018)\citenamefont {Kos},
  \citenamefont {Ljubotina},\ and\ \citenamefont
  {Prosen}}]{kosProsen2018analyticRMT}%
  \BibitemOpen
  \bibfield  {author} {\bibinfo {author} {\bibfnamefont {P.}~\bibnamefont
  {Kos}}, \bibinfo {author} {\bibfnamefont {M.}~\bibnamefont {Ljubotina}},\
  and\ \bibinfo {author} {\bibfnamefont {T.~c.~v.}\ \bibnamefont {Prosen}},\
  }\bibfield  {title} {\bibinfo {title} {Many-body quantum chaos: Analytic
  connection to random matrix theory},\ }\href
  {https://doi.org/10.1103/PhysRevX.8.021062} {\bibfield  {journal} {\bibinfo
  {journal} {Phys. Rev. X}\ }\textbf {\bibinfo {volume} {8}},\ \bibinfo {pages}
  {021062} (\bibinfo {year} {2018})}\BibitemShut {NoStop}%
\bibitem [{\citenamefont {Abanin}\ \emph {et~al.}(2019)\citenamefont {Abanin},
  \citenamefont {Altman}, \citenamefont {Bloch},\ and\ \citenamefont
  {Serbyn}}]{abanin2019mblReview}%
  \BibitemOpen
  \bibfield  {author} {\bibinfo {author} {\bibfnamefont {D.~A.}\ \bibnamefont
  {Abanin}}, \bibinfo {author} {\bibfnamefont {E.}~\bibnamefont {Altman}},
  \bibinfo {author} {\bibfnamefont {I.}~\bibnamefont {Bloch}},\ and\ \bibinfo
  {author} {\bibfnamefont {M.}~\bibnamefont {Serbyn}},\ }\bibfield  {title}
  {\bibinfo {title} {Colloquium: Many-body localization, thermalization, and
  entanglement},\ }\href {https://doi.org/10.1103/RevModPhys.91.021001}
  {\bibfield  {journal} {\bibinfo  {journal} {Rev. Mod. Phys.}\ }\textbf
  {\bibinfo {volume} {91}},\ \bibinfo {pages} {021001} (\bibinfo {year}
  {2019})}\BibitemShut {NoStop}%
\bibitem [{\citenamefont {Basko}\ \emph {et~al.}(2006)\citenamefont {Basko},
  \citenamefont {Aleiner},\ and\ \citenamefont
  {Altshuler}}]{basko2006metalInsulator}%
  \BibitemOpen
  \bibfield  {author} {\bibinfo {author} {\bibfnamefont {D.~M.}\ \bibnamefont
  {Basko}}, \bibinfo {author} {\bibfnamefont {I.~L.}\ \bibnamefont {Aleiner}},\
  and\ \bibinfo {author} {\bibfnamefont {B.~L.}\ \bibnamefont {Altshuler}},\
  }\bibfield  {title} {\bibinfo {title} {Metal--insulator transition in a
  weakly interacting many-electron system with localized single-particle
  states},\ }\href {https://doi.org/https://doi.org/10.1016/j.aop.2005.11.014}
  {\bibfield  {journal} {\bibinfo  {journal} {Annals of physics}\ }\textbf
  {\bibinfo {volume} {321}},\ \bibinfo {pages} {1126} (\bibinfo {year}
  {2006})}\BibitemShut {NoStop}%
\bibitem [{\citenamefont {Nandkishore}\ and\ \citenamefont
  {Huse}(2015)}]{nandkishoreHuse2015mblReview}%
  \BibitemOpen
  \bibfield  {author} {\bibinfo {author} {\bibfnamefont {R.}~\bibnamefont
  {Nandkishore}}\ and\ \bibinfo {author} {\bibfnamefont {D.~A.}\ \bibnamefont
  {Huse}},\ }\bibfield  {title} {\bibinfo {title} {Many-body localization and
  thermalization in quantum statistical mechanics},\ }\href
  {https://doi.org/https://doi.org/10.1146/annurev-conmatphys-031214-014726}
  {\bibfield  {journal} {\bibinfo  {journal} {Annu. Rev. Condens. Matter
  Phys.}\ }\textbf {\bibinfo {volume} {6}},\ \bibinfo {pages} {15} (\bibinfo
  {year} {2015})}\BibitemShut {NoStop}%
\bibitem [{\citenamefont {Vidmar}\ and\ \citenamefont
  {Rigol}(2016)}]{vidmarRigol2016GGEintegrable}%
  \BibitemOpen
  \bibfield  {author} {\bibinfo {author} {\bibfnamefont {L.}~\bibnamefont
  {Vidmar}}\ and\ \bibinfo {author} {\bibfnamefont {M.}~\bibnamefont {Rigol}},\
  }\bibfield  {title} {\bibinfo {title} {Generalized gibbs ensemble in
  integrable lattice models},\ }\href
  {https://doi.org/https://doi.org/10.1088/1742-5468/2016/06/064007} {\bibfield
   {journal} {\bibinfo  {journal} {Journal of Statistical Mechanics: Theory and
  Experiment}\ }\textbf {\bibinfo {volume} {2016}},\ \bibinfo {pages} {064007}
  (\bibinfo {year} {2016})}\BibitemShut {NoStop}%
\bibitem [{\citenamefont {Turner}\ \emph {et~al.}(2018)\citenamefont {Turner},
  \citenamefont {Michailidis}, \citenamefont {Abanin}, \citenamefont {Serbyn},\
  and\ \citenamefont {Papi{\'c}}}]{turner2018manyBodyScars}%
  \BibitemOpen
  \bibfield  {author} {\bibinfo {author} {\bibfnamefont {C.~J.}\ \bibnamefont
  {Turner}}, \bibinfo {author} {\bibfnamefont {A.~A.}\ \bibnamefont
  {Michailidis}}, \bibinfo {author} {\bibfnamefont {D.~A.}\ \bibnamefont
  {Abanin}}, \bibinfo {author} {\bibfnamefont {M.}~\bibnamefont {Serbyn}},\
  and\ \bibinfo {author} {\bibfnamefont {Z.}~\bibnamefont {Papi{\'c}}},\
  }\bibfield  {title} {\bibinfo {title} {Weak ergodicity breaking from quantum
  many-body scars},\ }\href
  {https://doi.org/https://doi.org/10.1038/s41567-018-0137-5} {\bibfield
  {journal} {\bibinfo  {journal} {Nature Physics}\ }\textbf {\bibinfo {volume}
  {14}},\ \bibinfo {pages} {745} (\bibinfo {year} {2018})}\BibitemShut
  {NoStop}%
\bibitem [{\citenamefont {Moudgalya}\ \emph {et~al.}(2022)\citenamefont
  {Moudgalya}, \citenamefont {Bernevig},\ and\ \citenamefont
  {Regnault}}]{moudgalya2022scarsReview}%
  \BibitemOpen
  \bibfield  {author} {\bibinfo {author} {\bibfnamefont {S.}~\bibnamefont
  {Moudgalya}}, \bibinfo {author} {\bibfnamefont {B.~A.}\ \bibnamefont
  {Bernevig}},\ and\ \bibinfo {author} {\bibfnamefont {N.}~\bibnamefont
  {Regnault}},\ }\bibfield  {title} {\bibinfo {title} {Quantum many-body scars
  and hilbert space fragmentation: A review of exact results},\ }\href
  {https://doi.org/10.1088/1361-6633/ac73a0} {\bibfield  {journal} {\bibinfo
  {journal} {Reports on Progress in Physics}\ }\textbf {\bibinfo {volume}
  {85}},\ \bibinfo {pages} {086501} (\bibinfo {year} {2022})}\BibitemShut
  {NoStop}%
\bibitem [{\citenamefont {Chandran}\ \emph {et~al.}(2023)\citenamefont
  {Chandran}, \citenamefont {Iadecola}, \citenamefont {Khemani},\ and\
  \citenamefont {Moessner}}]{chandran2023scarsReview}%
  \BibitemOpen
  \bibfield  {author} {\bibinfo {author} {\bibfnamefont {A.}~\bibnamefont
  {Chandran}}, \bibinfo {author} {\bibfnamefont {T.}~\bibnamefont {Iadecola}},
  \bibinfo {author} {\bibfnamefont {V.}~\bibnamefont {Khemani}},\ and\ \bibinfo
  {author} {\bibfnamefont {R.}~\bibnamefont {Moessner}},\ }\bibfield  {title}
  {\bibinfo {title} {Quantum {{Many-Body Scars}}: {{A Quasiparticle
  Perspective}}},\ }\href
  {https://doi.org/10.1146/annurev-conmatphys-031620-101617} {\bibfield
  {journal} {\bibinfo  {journal} {Annual Review of Condensed Matter Physics}\
  }\textbf {\bibinfo {volume} {14}},\ \bibinfo {pages} {443} (\bibinfo {year}
  {2023})}\BibitemShut {NoStop}%
\bibitem [{\citenamefont {O'Dea}\ \emph {et~al.}(2020)\citenamefont {O'Dea},
  \citenamefont {Burnell}, \citenamefont {Chandran},\ and\ \citenamefont
  {Khemani}}]{odea2020scarsLieAlgebra}%
  \BibitemOpen
  \bibfield  {author} {\bibinfo {author} {\bibfnamefont {N.}~\bibnamefont
  {O'Dea}}, \bibinfo {author} {\bibfnamefont {F.}~\bibnamefont {Burnell}},
  \bibinfo {author} {\bibfnamefont {A.}~\bibnamefont {Chandran}},\ and\
  \bibinfo {author} {\bibfnamefont {V.}~\bibnamefont {Khemani}},\ }\bibfield
  {title} {\bibinfo {title} {From tunnels to towers: Quantum scars from lie
  algebras and $q$-deformed lie algebras},\ }\href
  {https://doi.org/10.1103/PhysRevResearch.2.043305} {\bibfield  {journal}
  {\bibinfo  {journal} {Phys. Rev. Res.}\ }\textbf {\bibinfo {volume} {2}},\
  \bibinfo {pages} {043305} (\bibinfo {year} {2020})}\BibitemShut {NoStop}%
\bibitem [{\citenamefont {Foss-Feig}\ \emph {et~al.}(2024)\citenamefont
  {Foss-Feig}, \citenamefont {Pagano}, \citenamefont {Potter},\ and\
  \citenamefont {Yao}}]{fossYao2024trappedIonReview}%
  \BibitemOpen
  \bibfield  {author} {\bibinfo {author} {\bibfnamefont {M.}~\bibnamefont
  {Foss-Feig}}, \bibinfo {author} {\bibfnamefont {G.}~\bibnamefont {Pagano}},
  \bibinfo {author} {\bibfnamefont {A.~C.}\ \bibnamefont {Potter}},\ and\
  \bibinfo {author} {\bibfnamefont {N.~Y.}\ \bibnamefont {Yao}},\ }\bibfield
  {title} {\bibinfo {title} {Progress in trapped-ion quantum simulation},\
  }\bibfield  {journal} {\bibinfo  {journal} {Annual Review of Condensed Matter
  Physics}\ }\textbf {\bibinfo {volume} {16}},\ \href
  {https://doi.org/https://doi.org/10.1146/annurev-conmatphys-032822-045619}
  {https://doi.org/10.1146/annurev-conmatphys-032822-045619} (\bibinfo {year}
  {2024})\BibitemShut {NoStop}%
\bibitem [{\citenamefont {Jiang}\ \emph {et~al.}(2025)\citenamefont {Jiang},
  \citenamefont {Deng}, \citenamefont {Fan}, \citenamefont {Li}, \citenamefont
  {Sun}, \citenamefont {Tan}, \citenamefont {Wang}, \citenamefont {Xue},
  \citenamefont {Yan}, \citenamefont {Yu} \emph
  {et~al.}}]{jiang2025superconductQubitAdvance}%
  \BibitemOpen
  \bibfield  {author} {\bibinfo {author} {\bibfnamefont {Y.-Y.}\ \bibnamefont
  {Jiang}}, \bibinfo {author} {\bibfnamefont {C.}~\bibnamefont {Deng}},
  \bibinfo {author} {\bibfnamefont {H.}~\bibnamefont {Fan}}, \bibinfo {author}
  {\bibfnamefont {B.-Y.}\ \bibnamefont {Li}}, \bibinfo {author} {\bibfnamefont
  {L.}~\bibnamefont {Sun}}, \bibinfo {author} {\bibfnamefont {X.-S.}\
  \bibnamefont {Tan}}, \bibinfo {author} {\bibfnamefont {W.}~\bibnamefont
  {Wang}}, \bibinfo {author} {\bibfnamefont {G.-M.}\ \bibnamefont {Xue}},
  \bibinfo {author} {\bibfnamefont {F.}~\bibnamefont {Yan}}, \bibinfo {author}
  {\bibfnamefont {H.-F.}\ \bibnamefont {Yu}}, \emph {et~al.},\ }\bibfield
  {title} {\bibinfo {title} {Advancements in superconducting quantum
  computing},\ }\href {https://doi.org/https://doi.org/10.1093/nsr/nwaf246}
  {\bibfield  {journal} {\bibinfo  {journal} {National Science Review}\
  }\textbf {\bibinfo {volume} {12}},\ \bibinfo {pages} {nwaf246} (\bibinfo
  {year} {2025})}\BibitemShut {NoStop}%
\bibitem [{\citenamefont {Cornish}\ \emph {et~al.}(2024)\citenamefont
  {Cornish}, \citenamefont {Tarbutt},\ and\ \citenamefont
  {Hazzard}}]{cornish2024ultraColdReview}%
  \BibitemOpen
  \bibfield  {author} {\bibinfo {author} {\bibfnamefont {S.~L.}\ \bibnamefont
  {Cornish}}, \bibinfo {author} {\bibfnamefont {M.~R.}\ \bibnamefont
  {Tarbutt}},\ and\ \bibinfo {author} {\bibfnamefont {K.~R.}\ \bibnamefont
  {Hazzard}},\ }\bibfield  {title} {\bibinfo {title} {Quantum computation and
  quantum simulation with ultracold molecules},\ }\href
  {https://doi.org/https://doi.org/10.1038/s41567-024-02453-9} {\bibfield
  {journal} {\bibinfo  {journal} {Nature Physics}\ }\textbf {\bibinfo {volume}
  {20}},\ \bibinfo {pages} {730} (\bibinfo {year} {2024})}\BibitemShut
  {NoStop}%
\bibitem [{\citenamefont {Lindblad}(1976)}]{lindblad1976}%
  \BibitemOpen
  \bibfield  {author} {\bibinfo {author} {\bibfnamefont {G.}~\bibnamefont
  {Lindblad}},\ }\bibfield  {title} {\bibinfo {title} {On the generators of
  quantum dynamical semigroups},\ }\href {https://doi.org/10.1007/BF01608499}
  {\bibfield  {journal} {\bibinfo  {journal} {Communications in Mathematical
  Physics}\ }\textbf {\bibinfo {volume} {48}},\ \bibinfo {pages} {119}
  (\bibinfo {year} {1976})}\BibitemShut {NoStop}%
\bibitem [{\citenamefont {Gorini}\ \emph {et~al.}(1976)\citenamefont {Gorini},
  \citenamefont {Kossakowski},\ and\ \citenamefont {Sudarshan}}]{gorini1976}%
  \BibitemOpen
  \bibfield  {author} {\bibinfo {author} {\bibfnamefont {V.}~\bibnamefont
  {Gorini}}, \bibinfo {author} {\bibfnamefont {A.}~\bibnamefont
  {Kossakowski}},\ and\ \bibinfo {author} {\bibfnamefont {E.~C.~G.}\
  \bibnamefont {Sudarshan}},\ }\bibfield  {title} {\bibinfo {title} {Completely
  positive dynamical semigroups of n‐level systems},\ }\href
  {https://doi.org/10.1063/1.522979} {\bibfield  {journal} {\bibinfo  {journal}
  {Journal of Mathematical Physics}\ }\textbf {\bibinfo {volume} {17}},\
  \bibinfo {pages} {821} (\bibinfo {year} {1976})}\BibitemShut {NoStop}%
\bibitem [{\citenamefont {Breuer}\ and\ \citenamefont
  {Petruccione}(2007)}]{breuer2002theory}%
  \BibitemOpen
  \bibfield  {author} {\bibinfo {author} {\bibfnamefont {H.-P.}\ \bibnamefont
  {Breuer}}\ and\ \bibinfo {author} {\bibfnamefont {F.}~\bibnamefont
  {Petruccione}},\ }\href
  {https://doi.org/10.1093/acprof:oso/9780199213900.001.0001} {\emph {\bibinfo
  {title} {The Theory of Open Quantum Systems}}}\ (\bibinfo  {publisher}
  {Oxford University Press},\ \bibinfo {year} {2007})\BibitemShut {NoStop}%
\bibitem [{\citenamefont {Prosen}\ and\ \citenamefont {\ifmmode \check{Z}\else
  \v{Z}\fi{}nidari\ifmmode~\check{c}\else
  \v{c}\fi{}}(2013)}]{prosen2013steadyStateLevelSpacing}%
  \BibitemOpen
  \bibfield  {author} {\bibinfo {author} {\bibfnamefont {T.~c.~v.}\
  \bibnamefont {Prosen}}\ and\ \bibinfo {author} {\bibfnamefont
  {M.}~\bibnamefont {\ifmmode \check{Z}\else
  \v{Z}\fi{}nidari\ifmmode~\check{c}\else \v{c}\fi{}}},\ }\bibfield  {title}
  {\bibinfo {title} {Eigenvalue statistics as an indicator of integrability of
  nonequilibrium density operators},\ }\href
  {https://doi.org/10.1103/PhysRevLett.111.124101} {\bibfield  {journal}
  {\bibinfo  {journal} {Phys. Rev. Lett.}\ }\textbf {\bibinfo {volume} {111}},\
  \bibinfo {pages} {124101} (\bibinfo {year} {2013})}\BibitemShut {NoStop}%
\bibitem [{\citenamefont {Richter}\ \emph {et~al.}(2025)\citenamefont
  {Richter}, \citenamefont {S{\'a}},\ and\ \citenamefont
  {Haque}}]{richterSa2025integrableNESS}%
  \BibitemOpen
  \bibfield  {author} {\bibinfo {author} {\bibfnamefont {J.}~\bibnamefont
  {Richter}}, \bibinfo {author} {\bibfnamefont {L.}~\bibnamefont {S{\'a}}},\
  and\ \bibinfo {author} {\bibfnamefont {M.}~\bibnamefont {Haque}},\ }\bibfield
   {title} {\bibinfo {title} {Integrability versus chaos in the steady state of
  many-body open quantum systems},\ }\href
  {https://doi.org/10.1103/PhysRevE.111.064103} {\bibfield  {journal} {\bibinfo
   {journal} {Physical Review E}\ }\textbf {\bibinfo {volume} {111}},\ \bibinfo
  {pages} {064103} (\bibinfo {year} {2025})}\BibitemShut {NoStop}%
\bibitem [{\citenamefont {Rufo}\ \emph {et~al.}(2025)\citenamefont {Rufo},
  \citenamefont {Rufo}, \citenamefont {Ribeiro},\ and\ \citenamefont
  {Chesi}}]{rufo2025semiclassical}%
  \BibitemOpen
  \bibfield  {author} {\bibinfo {author} {\bibfnamefont {G.}~\bibnamefont
  {Rufo}}, \bibinfo {author} {\bibfnamefont {S.}~\bibnamefont {Rufo}}, \bibinfo
  {author} {\bibfnamefont {P.}~\bibnamefont {Ribeiro}},\ and\ \bibinfo {author}
  {\bibfnamefont {S.}~\bibnamefont {Chesi}},\ }\href
  {https://doi.org/10.48550/arXiv.2506.14961} {\bibinfo {title} {Quantum and
  {Semi}-{Classical} {Signatures} of {Dissipative} {Chaos} in the {Steady}
  {State}}} (\bibinfo {year} {2025})\BibitemShut {NoStop}%
\bibitem [{\citenamefont {S{\'a}}\ \emph
  {et~al.}(2020{\natexlab{a}})\citenamefont {S{\'a}}, \citenamefont {Ribeiro},\
  and\ \citenamefont {Prosen}}]{sa2020spectral}%
  \BibitemOpen
  \bibfield  {author} {\bibinfo {author} {\bibfnamefont {L.}~\bibnamefont
  {S{\'a}}}, \bibinfo {author} {\bibfnamefont {P.}~\bibnamefont {Ribeiro}},\
  and\ \bibinfo {author} {\bibfnamefont {T.}~\bibnamefont {Prosen}},\
  }\bibfield  {title} {\bibinfo {title} {Spectral and steady-state properties
  of random {{Liouvillians}}},\ }\href
  {https://doi.org/10.1088/1751-8121/ab9337} {\bibfield  {journal} {\bibinfo
  {journal} {Journal of Physics A: Mathematical and Theoretical}\ }\textbf
  {\bibinfo {volume} {53}},\ \bibinfo {pages} {20} (\bibinfo {year}
  {2020}{\natexlab{a}})},\ \Eprint {https://arxiv.org/abs/1905.02155}
  {arXiv:1905.02155} \BibitemShut {NoStop}%
\bibitem [{\citenamefont {Grobe}\ \emph {et~al.}(1988)\citenamefont {Grobe},
  \citenamefont {Haake},\ and\ \citenamefont {Sommers}}]{grobe1988quantum}%
  \BibitemOpen
  \bibfield  {author} {\bibinfo {author} {\bibfnamefont {R.}~\bibnamefont
  {Grobe}}, \bibinfo {author} {\bibfnamefont {F.}~\bibnamefont {Haake}},\ and\
  \bibinfo {author} {\bibfnamefont {H.-J.}\ \bibnamefont {Sommers}},\
  }\bibfield  {title} {\bibinfo {title} {Quantum distinction of regular and
  chaotic dissipative motion},\ }\href
  {https://doi.org/10.1103/PhysRevLett.61.1899} {\bibfield  {journal} {\bibinfo
   {journal} {Phys. Rev. Lett.}\ }\textbf {\bibinfo {volume} {61}},\ \bibinfo
  {pages} {1899} (\bibinfo {year} {1988})}\BibitemShut {NoStop}%
\bibitem [{\citenamefont {Grobe}\ and\ \citenamefont
  {Haake}(1989)}]{grobe1989cubic}%
  \BibitemOpen
  \bibfield  {author} {\bibinfo {author} {\bibfnamefont {R.}~\bibnamefont
  {Grobe}}\ and\ \bibinfo {author} {\bibfnamefont {F.}~\bibnamefont {Haake}},\
  }\bibfield  {title} {\bibinfo {title} {Universality of cubic-level repulsion
  for dissipative quantum chaos},\ }\href
  {https://doi.org/10.1103/PhysRevLett.62.2893} {\bibfield  {journal} {\bibinfo
   {journal} {Phys. Rev. Lett.}\ }\textbf {\bibinfo {volume} {62}},\ \bibinfo
  {pages} {2893} (\bibinfo {year} {1989})}\BibitemShut {NoStop}%
\bibitem [{\citenamefont {Akemann}\ \emph {et~al.}(2019)\citenamefont
  {Akemann}, \citenamefont {Kieburg}, \citenamefont {Mielke},\ and\
  \citenamefont {Prosen}}]{akemann2019}%
  \BibitemOpen
  \bibfield  {author} {\bibinfo {author} {\bibfnamefont {G.}~\bibnamefont
  {Akemann}}, \bibinfo {author} {\bibfnamefont {M.}~\bibnamefont {Kieburg}},
  \bibinfo {author} {\bibfnamefont {A.}~\bibnamefont {Mielke}},\ and\ \bibinfo
  {author} {\bibfnamefont {T.~c.~v.}\ \bibnamefont {Prosen}},\ }\bibfield
  {title} {\bibinfo {title} {Universal signature from integrability to chaos in
  dissipative open quantum systems},\ }\href
  {https://doi.org/10.1103/PhysRevLett.123.254101} {\bibfield  {journal}
  {\bibinfo  {journal} {Phys. Rev. Lett.}\ }\textbf {\bibinfo {volume} {123}},\
  \bibinfo {pages} {254101} (\bibinfo {year} {2019})}\BibitemShut {NoStop}%
\bibitem [{\citenamefont {Altland}\ \emph {et~al.}(2021)\citenamefont
  {Altland}, \citenamefont {Fleischhauer},\ and\ \citenamefont
  {Diehl}}]{altland2021fermionLindbladSymmetry}%
  \BibitemOpen
  \bibfield  {author} {\bibinfo {author} {\bibfnamefont {A.}~\bibnamefont
  {Altland}}, \bibinfo {author} {\bibfnamefont {M.}~\bibnamefont
  {Fleischhauer}},\ and\ \bibinfo {author} {\bibfnamefont {S.}~\bibnamefont
  {Diehl}},\ }\bibfield  {title} {\bibinfo {title} {Symmetry {{Classes}} of
  {{Open Fermionic Quantum Matter}}},\ }\bibfield  {journal} {\bibinfo
  {journal} {Physical Review X}\ }\textbf {\bibinfo {volume} {11}},\ \href
  {https://doi.org/10.1103/PhysRevX.11.021037} {10.1103/PhysRevX.11.021037}
  (\bibinfo {year} {2021}),\ \Eprint {https://arxiv.org/abs/2007.10448}
  {arXiv:2007.10448} \BibitemShut {NoStop}%
\bibitem [{\citenamefont {Lieu}\ \emph {et~al.}(2020)\citenamefont {Lieu},
  \citenamefont {McGinley},\ and\ \citenamefont {Cooper}}]{lieu2020tenFold}%
  \BibitemOpen
  \bibfield  {author} {\bibinfo {author} {\bibfnamefont {S.}~\bibnamefont
  {Lieu}}, \bibinfo {author} {\bibfnamefont {M.}~\bibnamefont {McGinley}},\
  and\ \bibinfo {author} {\bibfnamefont {N.~R.}\ \bibnamefont {Cooper}},\
  }\bibfield  {title} {\bibinfo {title} {Tenfold {{Way}} for {{Quadratic
  Lindbladians}}},\ }\bibfield  {journal} {\bibinfo  {journal} {Physical Review
  Letters}\ }\textbf {\bibinfo {volume} {124}},\ \href
  {https://doi.org/10.1103/PhysRevLett.124.040401}
  {10.1103/PhysRevLett.124.040401} (\bibinfo {year} {2020}),\ \Eprint
  {https://arxiv.org/abs/1908.08834} {arXiv:1908.08834} \BibitemShut {NoStop}%
\bibitem [{\citenamefont {S\'a}\ \emph {et~al.}(2023)\citenamefont {S\'a},
  \citenamefont {Ribeiro},\ and\ \citenamefont {Prosen}}]{sa2023symmetry}%
  \BibitemOpen
  \bibfield  {author} {\bibinfo {author} {\bibfnamefont {L.}~\bibnamefont
  {S\'a}}, \bibinfo {author} {\bibfnamefont {P.}~\bibnamefont {Ribeiro}},\ and\
  \bibinfo {author} {\bibfnamefont {T.~c.~v.}\ \bibnamefont {Prosen}},\
  }\bibfield  {title} {\bibinfo {title} {Symmetry classification of many-body
  lindbladians: Tenfold way and beyond},\ }\href
  {https://doi.org/10.1103/PhysRevX.13.031019} {\bibfield  {journal} {\bibinfo
  {journal} {Phys. Rev. X}\ }\textbf {\bibinfo {volume} {13}},\ \bibinfo
  {pages} {031019} (\bibinfo {year} {2023})}\BibitemShut {NoStop}%
\bibitem [{\citenamefont {Kawabata}(2023)}]{kawabata2023symmetryLindblad}%
  \BibitemOpen
  \bibfield  {author} {\bibinfo {author} {\bibfnamefont {K.}~\bibnamefont
  {Kawabata}},\ }\bibfield  {title} {\bibinfo {title} {Symmetry of {{Open
  Quantum Systems}}: {{Classification}} of {{Dissipative Quantum Chaos}}},\
  }\bibfield  {journal} {\bibinfo  {journal} {PRX Quantum}\ }\textbf {\bibinfo
  {volume} {4}},\ \href {https://doi.org/10.1103/PRXQuantum.4.030328}
  {10.1103/PRXQuantum.4.030328} (\bibinfo {year} {2023})\BibitemShut {NoStop}%
\bibitem [{\citenamefont {Hamazaki}\ \emph {et~al.}(2020)\citenamefont
  {Hamazaki}, \citenamefont {Kawabata}, \citenamefont {Kura},\ and\
  \citenamefont {Ueda}}]{hamazaki2020nonhermitianUniversality}%
  \BibitemOpen
  \bibfield  {author} {\bibinfo {author} {\bibfnamefont {R.}~\bibnamefont
  {Hamazaki}}, \bibinfo {author} {\bibfnamefont {K.}~\bibnamefont {Kawabata}},
  \bibinfo {author} {\bibfnamefont {N.}~\bibnamefont {Kura}},\ and\ \bibinfo
  {author} {\bibfnamefont {M.}~\bibnamefont {Ueda}},\ }\bibfield  {title}
  {\bibinfo {title} {Universality classes of non-{{Hermitian}} random
  matrices},\ }\bibfield  {journal} {\bibinfo  {journal} {Physical Review
  Research}\ }\textbf {\bibinfo {volume} {2}},\ \href
  {https://doi.org/10.1103/PhysRevResearch.2.023286}
  {10.1103/PhysRevResearch.2.023286} (\bibinfo {year} {2020}),\ \Eprint
  {https://arxiv.org/abs/1904.13082} {arXiv:1904.13082} \BibitemShut {NoStop}%
\bibitem [{\citenamefont {Li}\ \emph {et~al.}(2021)\citenamefont {Li},
  \citenamefont {Prosen},\ and\ \citenamefont {Chan}}]{li2021dsff}%
  \BibitemOpen
  \bibfield  {author} {\bibinfo {author} {\bibfnamefont {J.}~\bibnamefont
  {Li}}, \bibinfo {author} {\bibfnamefont {T.~c.~v.}\ \bibnamefont {Prosen}},\
  and\ \bibinfo {author} {\bibfnamefont {A.}~\bibnamefont {Chan}},\ }\bibfield
  {title} {\bibinfo {title} {Spectral statistics of non-hermitian matrices and
  dissipative quantum chaos},\ }\href
  {https://doi.org/10.1103/PhysRevLett.127.170602} {\bibfield  {journal}
  {\bibinfo  {journal} {Phys. Rev. Lett.}\ }\textbf {\bibinfo {volume} {127}},\
  \bibinfo {pages} {170602} (\bibinfo {year} {2021})}\BibitemShut {NoStop}%
\bibitem [{\citenamefont {S{\'a}}\ \emph
  {et~al.}(2020{\natexlab{b}})\citenamefont {S{\'a}}, \citenamefont {Ribeiro},\
  and\ \citenamefont {Prosen}}]{sa2020csr}%
  \BibitemOpen
  \bibfield  {author} {\bibinfo {author} {\bibfnamefont {L.}~\bibnamefont
  {S{\'a}}}, \bibinfo {author} {\bibfnamefont {P.}~\bibnamefont {Ribeiro}},\
  and\ \bibinfo {author} {\bibfnamefont {T.}~\bibnamefont {Prosen}},\
  }\bibfield  {title} {\bibinfo {title} {Complex {{Spacing Ratios}}: {{A
  Signature}} of {{Dissipative Quantum Chaos}}},\ }\bibfield  {journal}
  {\bibinfo  {journal} {Physical Review X}\ }\textbf {\bibinfo {volume} {10}},\
  \href {https://doi.org/10.1103/PhysRevX.10.021019}
  {10.1103/PhysRevX.10.021019} (\bibinfo {year} {2020}{\natexlab{b}}),\ \Eprint
  {https://arxiv.org/abs/1910.12784} {arXiv:1910.12784} \BibitemShut {NoStop}%
\bibitem [{\citenamefont {S\'a}\ \emph {et~al.}(2020)\citenamefont {S\'a},
  \citenamefont {Ribeiro}, \citenamefont {Can},\ and\ \citenamefont
  {Prosen}}]{sa2020randomKraus}%
  \BibitemOpen
  \bibfield  {author} {\bibinfo {author} {\bibfnamefont {L.}~\bibnamefont
  {S\'a}}, \bibinfo {author} {\bibfnamefont {P.}~\bibnamefont {Ribeiro}},
  \bibinfo {author} {\bibfnamefont {T.}~\bibnamefont {Can}},\ and\ \bibinfo
  {author} {\bibfnamefont {T.~c.~v.}\ \bibnamefont {Prosen}},\ }\bibfield
  {title} {\bibinfo {title} {Spectral transitions and universal steady states
  in random kraus maps and circuits},\ }\href
  {https://doi.org/10.1103/PhysRevB.102.134310} {\bibfield  {journal} {\bibinfo
   {journal} {Phys. Rev. B}\ }\textbf {\bibinfo {volume} {102}},\ \bibinfo
  {pages} {134310} (\bibinfo {year} {2020})}\BibitemShut {NoStop}%
\bibitem [{\citenamefont {Peyruchat}\ \emph {et~al.}(2025)\citenamefont
  {Peyruchat}, \citenamefont {Minganti}, \citenamefont {Scigliuzzo},
  \citenamefont {Ferrari}, \citenamefont {Jouanny}, \citenamefont {Nori},
  \citenamefont {Savona},\ and\ \citenamefont
  {Scarlino}}]{peyruchat2025landau}%
  \BibitemOpen
  \bibfield  {author} {\bibinfo {author} {\bibfnamefont {L.}~\bibnamefont
  {Peyruchat}}, \bibinfo {author} {\bibfnamefont {F.}~\bibnamefont {Minganti}},
  \bibinfo {author} {\bibfnamefont {M.}~\bibnamefont {Scigliuzzo}}, \bibinfo
  {author} {\bibfnamefont {F.}~\bibnamefont {Ferrari}}, \bibinfo {author}
  {\bibfnamefont {V.}~\bibnamefont {Jouanny}}, \bibinfo {author} {\bibfnamefont
  {F.}~\bibnamefont {Nori}}, \bibinfo {author} {\bibfnamefont {V.}~\bibnamefont
  {Savona}},\ and\ \bibinfo {author} {\bibfnamefont {P.}~\bibnamefont
  {Scarlino}},\ }\bibfield  {title} {\bibinfo {title} {Landau--{{Zener}}
  without a qubit: Multiphoton sidebands interaction and signatures of
  dissipative quantum chaos},\ }\href
  {https://doi.org/10.1038/s41534-025-00984-4} {\bibfield  {journal} {\bibinfo
  {journal} {npj Quantum Information}\ }\textbf {\bibinfo {volume} {11}},\
  \bibinfo {pages} {62} (\bibinfo {year} {2025})}\BibitemShut {NoStop}%
\bibitem [{\citenamefont {Ferrari}\ \emph
  {et~al.}(2025{\natexlab{a}})\citenamefont {Ferrari}, \citenamefont {Gravina},
  \citenamefont {Eeltink}, \citenamefont {Scarlino}, \citenamefont {Savona},\
  and\ \citenamefont {Minganti}}]{ferrari2025trajectories}%
  \BibitemOpen
  \bibfield  {author} {\bibinfo {author} {\bibfnamefont {F.}~\bibnamefont
  {Ferrari}}, \bibinfo {author} {\bibfnamefont {L.}~\bibnamefont {Gravina}},
  \bibinfo {author} {\bibfnamefont {D.}~\bibnamefont {Eeltink}}, \bibinfo
  {author} {\bibfnamefont {P.}~\bibnamefont {Scarlino}}, \bibinfo {author}
  {\bibfnamefont {V.}~\bibnamefont {Savona}},\ and\ \bibinfo {author}
  {\bibfnamefont {F.}~\bibnamefont {Minganti}},\ }\bibfield  {title} {\bibinfo
  {title} {Dissipative quantum chaos unveiled by stochastic quantum
  trajectories},\ }\href {https://doi.org/10.1103/PhysRevResearch.7.013276}
  {\bibfield  {journal} {\bibinfo  {journal} {Physical Review Research}\
  }\textbf {\bibinfo {volume} {7}},\ \bibinfo {pages} {013276} (\bibinfo {year}
  {2025}{\natexlab{a}})}\BibitemShut {NoStop}%
\bibitem [{\citenamefont {Villaseñor}\ \emph {et~al.}(2024)\citenamefont
  {Villaseñor}, \citenamefont {Santos},\ and\ \citenamefont
  {Barberis-Blostein}}]{villasenor2024breakdownCSR}%
  \BibitemOpen
  \bibfield  {author} {\bibinfo {author} {\bibfnamefont {D.}~\bibnamefont
  {Villaseñor}}, \bibinfo {author} {\bibfnamefont {L.~F.}\ \bibnamefont
  {Santos}},\ and\ \bibinfo {author} {\bibfnamefont {P.}~\bibnamefont
  {Barberis-Blostein}},\ }\bibfield  {title} {\bibinfo {title} {Breakdown of
  the {Quantum} {Distinction} of {Regular} and {Chaotic} {Classical} {Dynamics}
  in {Dissipative} {Systems}},\ }\href
  {https://doi.org/10.1103/PhysRevLett.133.240404} {\bibfield  {journal}
  {\bibinfo  {journal} {Physical Review Letters}\ }\textbf {\bibinfo {volume}
  {133}},\ \bibinfo {pages} {240404} (\bibinfo {year} {2024})}\BibitemShut
  {NoStop}%
\bibitem [{\citenamefont {Villase\~nor}\ and\ \citenamefont
  {Barberis-Blostein}(2024)}]{villasenor2024opendicke}%
  \BibitemOpen
  \bibfield  {author} {\bibinfo {author} {\bibfnamefont {D.}~\bibnamefont
  {Villase\~nor}}\ and\ \bibinfo {author} {\bibfnamefont {P.}~\bibnamefont
  {Barberis-Blostein}},\ }\bibfield  {title} {\bibinfo {title} {Analysis of
  chaos and regularity in the open dicke model},\ }\href
  {https://doi.org/10.1103/PhysRevE.109.014206} {\bibfield  {journal} {\bibinfo
   {journal} {Phys. Rev. E}\ }\textbf {\bibinfo {volume} {109}},\ \bibinfo
  {pages} {014206} (\bibinfo {year} {2024})}\BibitemShut {NoStop}%
\bibitem [{\citenamefont {Mondal}\ \emph {et~al.}(2025)\citenamefont {Mondal},
  \citenamefont {Santos},\ and\ \citenamefont {Sinha}}]{mondal2025transient}%
  \BibitemOpen
  \bibfield  {author} {\bibinfo {author} {\bibfnamefont {D.}~\bibnamefont
  {Mondal}}, \bibinfo {author} {\bibfnamefont {L.~F.}\ \bibnamefont {Santos}},\
  and\ \bibinfo {author} {\bibfnamefont {S.}~\bibnamefont {Sinha}},\ }\href
  {https://arxiv.org/abs/2506.05475} {\bibinfo {title} {Transient and
  steady-state chaos in dissipative quantum systems}} (\bibinfo {year}
  {2025}),\ \Eprint {https://arxiv.org/abs/2506.05475} {arXiv:2506.05475
  [quant-ph]} \BibitemShut {NoStop}%
\bibitem [{\citenamefont {Álvaro Rubio-García}\ \emph
  {et~al.}(2022)\citenamefont {Álvaro Rubio-García}, \citenamefont {Molina},\
  and\ \citenamefont {Dukelsky}}]{rubio2022integrability}%
  \BibitemOpen
  \bibfield  {author} {\bibinfo {author} {\bibnamefont {Álvaro
  Rubio-García}}, \bibinfo {author} {\bibfnamefont {R.~A.}\ \bibnamefont
  {Molina}},\ and\ \bibinfo {author} {\bibfnamefont {J.}~\bibnamefont
  {Dukelsky}},\ }\bibfield  {title} {\bibinfo {title} {{From integrability to
  chaos in quantum Liouvillians}},\ }\href
  {https://doi.org/10.21468/SciPostPhysCore.5.2.026} {\bibfield  {journal}
  {\bibinfo  {journal} {SciPost Phys. Core}\ }\textbf {\bibinfo {volume} {5}},\
  \bibinfo {pages} {026} (\bibinfo {year} {2022})}\BibitemShut {NoStop}%
\bibitem [{\citenamefont {Almeida}\ \emph {et~al.}(2025)\citenamefont
  {Almeida}, \citenamefont {Ribeiro}, \citenamefont {Haque},\ and\
  \citenamefont {Sá}}]{almeida2025openETH}%
  \BibitemOpen
  \bibfield  {author} {\bibinfo {author} {\bibfnamefont {G.}~\bibnamefont
  {Almeida}}, \bibinfo {author} {\bibfnamefont {P.}~\bibnamefont {Ribeiro}},
  \bibinfo {author} {\bibfnamefont {M.}~\bibnamefont {Haque}},\ and\ \bibinfo
  {author} {\bibfnamefont {L.}~\bibnamefont {Sá}},\ }\href
  {https://arxiv.org/abs/2504.10261} {\bibinfo {title} {Universality,
  robustness, and limits of the eigenstate thermalization hypothesis in open
  quantum systems}} (\bibinfo {year} {2025}),\ \Eprint
  {https://arxiv.org/abs/2504.10261} {arXiv:2504.10261 [cond-mat.stat-mech]}
  \BibitemShut {NoStop}%
\bibitem [{\citenamefont {Ferrari}\ \emph
  {et~al.}(2025{\natexlab{b}})\citenamefont {Ferrari}, \citenamefont {Savona},\
  and\ \citenamefont {Minganti}}]{ferrari2025openStripeETH}%
  \BibitemOpen
  \bibfield  {author} {\bibinfo {author} {\bibfnamefont {F.}~\bibnamefont
  {Ferrari}}, \bibinfo {author} {\bibfnamefont {V.}~\bibnamefont {Savona}},\
  and\ \bibinfo {author} {\bibfnamefont {F.}~\bibnamefont {Minganti}},\ }\href
  {https://arxiv.org/abs/2505.18260} {\bibinfo {title} {Chaos and
  thermalization in open quantum systems}} (\bibinfo {year}
  {2025}{\natexlab{b}}),\ \Eprint {https://arxiv.org/abs/2505.18260}
  {arXiv:2505.18260 [quant-ph]} \BibitemShut {NoStop}%
\bibitem [{\citenamefont {Hamazaki}\ \emph {et~al.}(2019)\citenamefont
  {Hamazaki}, \citenamefont {Kawabata},\ and\ \citenamefont
  {Ueda}}]{hamazaki2019nonHermitianMBL}%
  \BibitemOpen
  \bibfield  {author} {\bibinfo {author} {\bibfnamefont {R.}~\bibnamefont
  {Hamazaki}}, \bibinfo {author} {\bibfnamefont {K.}~\bibnamefont {Kawabata}},\
  and\ \bibinfo {author} {\bibfnamefont {M.}~\bibnamefont {Ueda}},\ }\bibfield
  {title} {\bibinfo {title} {Non-hermitian many-body localization},\ }\href
  {https://doi.org/10.1103/PhysRevLett.123.090603} {\bibfield  {journal}
  {\bibinfo  {journal} {Phys. Rev. Lett.}\ }\textbf {\bibinfo {volume} {123}},\
  \bibinfo {pages} {090603} (\bibinfo {year} {2019})}\BibitemShut {NoStop}%
\bibitem [{\citenamefont {Tzortzakakis}\ \emph {et~al.}(2020)\citenamefont
  {Tzortzakakis}, \citenamefont {Makris},\ and\ \citenamefont
  {Economou}}]{tzortzakakis2020nonHermitianDisroder}%
  \BibitemOpen
  \bibfield  {author} {\bibinfo {author} {\bibfnamefont {A.~F.}\ \bibnamefont
  {Tzortzakakis}}, \bibinfo {author} {\bibfnamefont {K.~G.}\ \bibnamefont
  {Makris}},\ and\ \bibinfo {author} {\bibfnamefont {E.~N.}\ \bibnamefont
  {Economou}},\ }\bibfield  {title} {\bibinfo {title} {Non-hermitian disorder
  in two-dimensional optical lattices},\ }\href
  {https://doi.org/10.1103/PhysRevB.101.014202} {\bibfield  {journal} {\bibinfo
   {journal} {Phys. Rev. B}\ }\textbf {\bibinfo {volume} {101}},\ \bibinfo
  {pages} {014202} (\bibinfo {year} {2020})}\BibitemShut {NoStop}%
\bibitem [{\citenamefont {Huang}\ and\ \citenamefont
  {Shklovskii}(2020)}]{huang2020andersonNonHermitian}%
  \BibitemOpen
  \bibfield  {author} {\bibinfo {author} {\bibfnamefont {Y.}~\bibnamefont
  {Huang}}\ and\ \bibinfo {author} {\bibfnamefont {B.~I.}\ \bibnamefont
  {Shklovskii}},\ }\bibfield  {title} {\bibinfo {title} {Anderson transition in
  three-dimensional systems with non-hermitian disorder},\ }\href
  {https://doi.org/10.1103/PhysRevB.101.014204} {\bibfield  {journal} {\bibinfo
   {journal} {Phys. Rev. B}\ }\textbf {\bibinfo {volume} {101}},\ \bibinfo
  {pages} {014204} (\bibinfo {year} {2020})}\BibitemShut {NoStop}%
\bibitem [{\citenamefont {Li}\ \emph {et~al.}(2025)\citenamefont {Li},
  \citenamefont {Wang},\ and\ \citenamefont
  {Liu}}]{li2025opensystemBerryConjecture}%
  \BibitemOpen
  \bibfield  {author} {\bibinfo {author} {\bibfnamefont {Y.}~\bibnamefont
  {Li}}, \bibinfo {author} {\bibfnamefont {Y.}~\bibnamefont {Wang}},\ and\
  \bibinfo {author} {\bibfnamefont {Y.-C.}\ \bibnamefont {Liu}},\ }\href
  {https://doi.org/10.48550/arXiv.2509.14644} {\bibinfo {title} {Open-system
  analogy of {{Berry}} conjecture}} (\bibinfo {year} {2025}),\ \Eprint
  {https://arxiv.org/abs/2509.14644} {arXiv:2509.14644 [quant-ph]} \BibitemShut
  {NoStop}%
\bibitem [{\citenamefont {Yang}\ \emph {et~al.}(2024)\citenamefont {Yang},
  \citenamefont {Xu},\ and\ \citenamefont {del Campo}}]{campo2024purity}%
  \BibitemOpen
  \bibfield  {author} {\bibinfo {author} {\bibfnamefont {Y.}~\bibnamefont
  {Yang}}, \bibinfo {author} {\bibfnamefont {Z.}~\bibnamefont {Xu}},\ and\
  \bibinfo {author} {\bibfnamefont {A.}~\bibnamefont {del Campo}},\ }\bibfield
  {title} {\bibinfo {title} {Decoherence rate in random lindblad dynamics},\
  }\href {https://doi.org/10.1103/PhysRevResearch.6.023229} {\bibfield
  {journal} {\bibinfo  {journal} {Phys. Rev. Res.}\ }\textbf {\bibinfo {volume}
  {6}},\ \bibinfo {pages} {023229} (\bibinfo {year} {2024})}\BibitemShut
  {NoStop}%
\bibitem [{\citenamefont {Hamazaki}\ \emph {et~al.}(2022)\citenamefont
  {Hamazaki}, \citenamefont {Nakagawa}, \citenamefont {Haga},\ and\
  \citenamefont {Ueda}}]{hamazaki2022mbl}%
  \BibitemOpen
  \bibfield  {author} {\bibinfo {author} {\bibfnamefont {R.}~\bibnamefont
  {Hamazaki}}, \bibinfo {author} {\bibfnamefont {M.}~\bibnamefont {Nakagawa}},
  \bibinfo {author} {\bibfnamefont {T.}~\bibnamefont {Haga}},\ and\ \bibinfo
  {author} {\bibfnamefont {M.}~\bibnamefont {Ueda}},\ }\bibfield  {title}
  {\bibinfo {title} {Lindbladian many-body localization},\ }\Eprint
  {https://arxiv.org/abs/2206.02984} {arXiv:2206.02984 [cond-mat.dis-nn]}
  (\bibinfo {year} {2022})\BibitemShut {NoStop}%
\bibitem [{\citenamefont {Garc\'{\i}a-Garc\'{\i}a}\ \emph
  {et~al.}(2026)\citenamefont {Garc\'{\i}a-Garc\'{\i}a}, \citenamefont {Lu},
  \citenamefont {S\'a},\ and\ \citenamefont
  {Verbaarschot}}]{garciaVerbaarschot2025lindbladScars}%
  \BibitemOpen
  \bibfield  {author} {\bibinfo {author} {\bibfnamefont {A.~M.}\ \bibnamefont
  {Garc\'{\i}a-Garc\'{\i}a}}, \bibinfo {author} {\bibfnamefont
  {Z.}~\bibnamefont {Lu}}, \bibinfo {author} {\bibfnamefont {L.}~\bibnamefont
  {S\'a}},\ and\ \bibinfo {author} {\bibfnamefont {J.~J.~M.}\ \bibnamefont
  {Verbaarschot}},\ }\bibfield  {title} {\bibinfo {title} {Lindblad many-body
  scars},\ }\href {https://doi.org/10.1103/x15w-zchv} {\bibfield  {journal}
  {\bibinfo  {journal} {Phys. Rev. E}\ }\textbf {\bibinfo {volume} {113}},\
  \bibinfo {pages} {024116} (\bibinfo {year} {2026})}\BibitemShut {NoStop}%
\bibitem [{\citenamefont {Wang}\ \emph {et~al.}(2020)\citenamefont {Wang},
  \citenamefont {Piazza},\ and\ \citenamefont {Luitz}}]{wang2020hierarchy}%
  \BibitemOpen
  \bibfield  {author} {\bibinfo {author} {\bibfnamefont {K.}~\bibnamefont
  {Wang}}, \bibinfo {author} {\bibfnamefont {F.}~\bibnamefont {Piazza}},\ and\
  \bibinfo {author} {\bibfnamefont {D.~J.}\ \bibnamefont {Luitz}},\ }\bibfield
  {title} {\bibinfo {title} {Hierarchy of relaxation timescales in local random
  liouvillians},\ }\href {https://doi.org/10.1103/PhysRevLett.124.100604}
  {\bibfield  {journal} {\bibinfo  {journal} {Phys. Rev. Lett.}\ }\textbf
  {\bibinfo {volume} {124}},\ \bibinfo {pages} {100604} (\bibinfo {year}
  {2020})}\BibitemShut {NoStop}%
\bibitem [{\citenamefont {Schuster}\ and\ \citenamefont
  {Yao}(2023)}]{schuster2023operator}%
  \BibitemOpen
  \bibfield  {author} {\bibinfo {author} {\bibfnamefont {T.}~\bibnamefont
  {Schuster}}\ and\ \bibinfo {author} {\bibfnamefont {N.~Y.}\ \bibnamefont
  {Yao}},\ }\bibfield  {title} {\bibinfo {title} {Operator growth in open
  quantum systems},\ }\href {https://doi.org/10.1103/PhysRevLett.131.160402}
  {\bibfield  {journal} {\bibinfo  {journal} {Phys. Rev. Lett.}\ }\textbf
  {\bibinfo {volume} {131}},\ \bibinfo {pages} {160402} (\bibinfo {year}
  {2023})}\BibitemShut {NoStop}%
\bibitem [{\citenamefont {Weinstein}\ \emph {et~al.}(2023)\citenamefont
  {Weinstein}, \citenamefont {Kelly}, \citenamefont {Marino},\ and\
  \citenamefont {Altman}}]{weinstein2023radiativeRUC}%
  \BibitemOpen
  \bibfield  {author} {\bibinfo {author} {\bibfnamefont {Z.}~\bibnamefont
  {Weinstein}}, \bibinfo {author} {\bibfnamefont {S.~P.}\ \bibnamefont
  {Kelly}}, \bibinfo {author} {\bibfnamefont {J.}~\bibnamefont {Marino}},\ and\
  \bibinfo {author} {\bibfnamefont {E.}~\bibnamefont {Altman}},\ }\bibfield
  {title} {\bibinfo {title} {Scrambling transition in a radiative random
  unitary circuit},\ }\href {https://doi.org/10.1103/PhysRevLett.131.220404}
  {\bibfield  {journal} {\bibinfo  {journal} {Phys. Rev. Lett.}\ }\textbf
  {\bibinfo {volume} {131}},\ \bibinfo {pages} {220404} (\bibinfo {year}
  {2023})}\BibitemShut {NoStop}%
\bibitem [{\citenamefont {Zhang}\ \emph {et~al.}(2019)\citenamefont {Zhang},
  \citenamefont {Huang},\ and\ \citenamefont {Chen}}]{zhangHunagChen2019}%
  \BibitemOpen
  \bibfield  {author} {\bibinfo {author} {\bibfnamefont {Y.-L.}\ \bibnamefont
  {Zhang}}, \bibinfo {author} {\bibfnamefont {Y.}~\bibnamefont {Huang}},\ and\
  \bibinfo {author} {\bibfnamefont {X.}~\bibnamefont {Chen}},\ }\bibfield
  {title} {\bibinfo {title} {Information scrambling in chaotic systems with
  dissipation},\ }\href {https://doi.org/10.1103/PhysRevB.99.014303} {\bibfield
   {journal} {\bibinfo  {journal} {Phys. Rev. B}\ }\textbf {\bibinfo {volume}
  {99}},\ \bibinfo {pages} {014303} (\bibinfo {year} {2019})}\BibitemShut
  {NoStop}%
\bibitem [{\citenamefont {Xu}\ \emph {et~al.}(2019)\citenamefont {Xu},
  \citenamefont {García-Pintos}, \citenamefont {Chenu},\ and\ \citenamefont
  {del Campo}}]{campo2019extreme}%
  \BibitemOpen
  \bibfield  {author} {\bibinfo {author} {\bibfnamefont {Z.}~\bibnamefont
  {Xu}}, \bibinfo {author} {\bibfnamefont {L.~P.}\ \bibnamefont
  {García-Pintos}}, \bibinfo {author} {\bibfnamefont {A.}~\bibnamefont
  {Chenu}},\ and\ \bibinfo {author} {\bibfnamefont {A.}~\bibnamefont {del
  Campo}},\ }\bibfield  {title} {\bibinfo {title} {Extreme {Decoherence} and
  {Quantum} {Chaos}},\ }\href {https://doi.org/10.1103/PhysRevLett.122.014103}
  {\bibfield  {journal} {\bibinfo  {journal} {Physical Review Letters}\
  }\textbf {\bibinfo {volume} {122}},\ \bibinfo {pages} {014103} (\bibinfo
  {year} {2019})}\BibitemShut {NoStop}%
\bibitem [{\citenamefont {Nathan}\ and\ \citenamefont
  {Rudner}(2020)}]{rudner2020universalLindblad}%
  \BibitemOpen
  \bibfield  {author} {\bibinfo {author} {\bibfnamefont {F.}~\bibnamefont
  {Nathan}}\ and\ \bibinfo {author} {\bibfnamefont {M.~S.}\ \bibnamefont
  {Rudner}},\ }\bibfield  {title} {\bibinfo {title} {Universal lindblad
  equation for open quantum systems},\ }\href
  {https://doi.org/10.1103/PhysRevB.102.115109} {\bibfield  {journal} {\bibinfo
   {journal} {Phys. Rev. B}\ }\textbf {\bibinfo {volume} {102}},\ \bibinfo
  {pages} {115109} (\bibinfo {year} {2020})}\BibitemShut {NoStop}%
\bibitem [{\citenamefont {Shiraishi}\ \emph {et~al.}(2025)\citenamefont
  {Shiraishi}, \citenamefont {Nakagawa}, \citenamefont {Mori},\ and\
  \citenamefont {Ueda}}]{shiraishi2025manybodyLindblad}%
  \BibitemOpen
  \bibfield  {author} {\bibinfo {author} {\bibfnamefont {K.}~\bibnamefont
  {Shiraishi}}, \bibinfo {author} {\bibfnamefont {M.}~\bibnamefont {Nakagawa}},
  \bibinfo {author} {\bibfnamefont {T.}~\bibnamefont {Mori}},\ and\ \bibinfo
  {author} {\bibfnamefont {M.}~\bibnamefont {Ueda}},\ }\bibfield  {title}
  {\bibinfo {title} {Quantum master equation for many-body systems based on the
  {Lieb}-{Robinson} bound},\ }\href
  {https://doi.org/10.1103/PhysRevB.111.184311} {\bibfield  {journal} {\bibinfo
   {journal} {Physical Review B}\ }\textbf {\bibinfo {volume} {111}},\ \bibinfo
  {pages} {184311} (\bibinfo {year} {2025})}\BibitemShut {NoStop}%
\bibitem [{\citenamefont {Alicki}\ and\ \citenamefont
  {Lendi}(2007)}]{alickiLendi2011book}%
  \BibitemOpen
  \bibfield  {author} {\bibinfo {author} {\bibfnamefont {R.}~\bibnamefont
  {Alicki}}\ and\ \bibinfo {author} {\bibfnamefont {K.}~\bibnamefont {Lendi}},\
  }\href {https://doi.org/https://doi.org/10.1007/3-540-70861-8} {\emph
  {\bibinfo {title} {Quantum dynamical semigroups and applications}}},\ Vol.\
  \bibinfo {volume} {717}\ (\bibinfo  {publisher} {Springer},\ \bibinfo {year}
  {2007})\BibitemShut {NoStop}%
\bibitem [{\citenamefont {Nielsen}\ and\ \citenamefont
  {Chuang}(2010)}]{nielsenChuang}%
  \BibitemOpen
  \bibfield  {author} {\bibinfo {author} {\bibfnamefont {M.~A.}\ \bibnamefont
  {Nielsen}}\ and\ \bibinfo {author} {\bibfnamefont {I.~L.}\ \bibnamefont
  {Chuang}},\ }\href@noop {} {\emph {\bibinfo {title} {Quantum computation and
  quantum information}}}\ (\bibinfo  {publisher} {Cambridge university press},\
  \bibinfo {year} {2010})\BibitemShut {NoStop}%
\bibitem [{\citenamefont {Ashida}\ \emph {et~al.}(2020)\citenamefont {Ashida},
  \citenamefont {Gong},\ and\ \citenamefont {Ueda}}]{ueda2020review}%
  \BibitemOpen
  \bibfield  {author} {\bibinfo {author} {\bibfnamefont {Y.}~\bibnamefont
  {Ashida}}, \bibinfo {author} {\bibfnamefont {Z.}~\bibnamefont {Gong}},\ and\
  \bibinfo {author} {\bibfnamefont {M.}~\bibnamefont {Ueda}},\ }\bibfield
  {title} {\bibinfo {title} {Non-hermitian physics},\ }\href
  {https://doi.org/10.1080/00018732.2021.1876991} {\bibfield  {journal}
  {\bibinfo  {journal} {Advances in Physics}\ }\textbf {\bibinfo {volume}
  {69}},\ \bibinfo {pages} {249} (\bibinfo {year} {2020})}\BibitemShut
  {NoStop}%
\bibitem [{\citenamefont {Oganesyan}\ and\ \citenamefont
  {Huse}(2007)}]{oganesyan2007levels}%
  \BibitemOpen
  \bibfield  {author} {\bibinfo {author} {\bibfnamefont {V.}~\bibnamefont
  {Oganesyan}}\ and\ \bibinfo {author} {\bibfnamefont {D.~A.}\ \bibnamefont
  {Huse}},\ }\bibfield  {title} {\bibinfo {title} {Localization of interacting
  fermions at high temperature},\ }\href
  {https://doi.org/10.1103/PhysRevB.75.155111} {\bibfield  {journal} {\bibinfo
  {journal} {Phys. Rev. B}\ }\textbf {\bibinfo {volume} {75}},\ \bibinfo
  {pages} {155111} (\bibinfo {year} {2007})}\BibitemShut {NoStop}%
\bibitem [{\citenamefont {Atas}\ \emph {et~al.}(2013)\citenamefont {Atas},
  \citenamefont {Bogomolny}, \citenamefont {Giraud},\ and\ \citenamefont
  {Roux}}]{atas2013levels}%
  \BibitemOpen
  \bibfield  {author} {\bibinfo {author} {\bibfnamefont {Y.~Y.}\ \bibnamefont
  {Atas}}, \bibinfo {author} {\bibfnamefont {E.}~\bibnamefont {Bogomolny}},
  \bibinfo {author} {\bibfnamefont {O.}~\bibnamefont {Giraud}},\ and\ \bibinfo
  {author} {\bibfnamefont {G.}~\bibnamefont {Roux}},\ }\bibfield  {title}
  {\bibinfo {title} {Distribution of the ratio of consecutive level spacings in
  random matrix ensembles},\ }\href
  {https://doi.org/10.1103/PhysRevLett.110.084101} {\bibfield  {journal}
  {\bibinfo  {journal} {Phys. Rev. Lett.}\ }\textbf {\bibinfo {volume} {110}},\
  \bibinfo {pages} {084101} (\bibinfo {year} {2013})}\BibitemShut {NoStop}%
\bibitem [{\citenamefont {Prasad}\ \emph {et~al.}(2022)\citenamefont {Prasad},
  \citenamefont {Yadalam}, \citenamefont {Aron},\ and\ \citenamefont
  {Kulkarni}}]{prasad2022dissipative}%
  \BibitemOpen
  \bibfield  {author} {\bibinfo {author} {\bibfnamefont {M.}~\bibnamefont
  {Prasad}}, \bibinfo {author} {\bibfnamefont {H.~K.}\ \bibnamefont {Yadalam}},
  \bibinfo {author} {\bibfnamefont {C.}~\bibnamefont {Aron}},\ and\ \bibinfo
  {author} {\bibfnamefont {M.}~\bibnamefont {Kulkarni}},\ }\bibfield  {title}
  {\bibinfo {title} {Dissipative quantum dynamics, phase transitions, and
  non-hermitian random matrices},\ }\href
  {https://doi.org/10.1103/PhysRevA.105.L050201} {\bibfield  {journal}
  {\bibinfo  {journal} {Phys. Rev. A}\ }\textbf {\bibinfo {volume} {105}},\
  \bibinfo {pages} {L050201} (\bibinfo {year} {2022})}\BibitemShut {NoStop}%
\bibitem [{\citenamefont {Denisov}\ \emph {et~al.}(2019)\citenamefont
  {Denisov}, \citenamefont {Laptyeva}, \citenamefont {Tarnowski}, \citenamefont
  {Chru\ifmmode \acute{s}\else \'{s}\fi{}ci\ifmmode~\acute{n}\else
  \'{n}\fi{}ski},\ and\ \citenamefont {\ifmmode~\dot{Z}\else
  \.{Z}\fi{}yczkowski}}]{denisov2019universal}%
  \BibitemOpen
  \bibfield  {author} {\bibinfo {author} {\bibfnamefont {S.}~\bibnamefont
  {Denisov}}, \bibinfo {author} {\bibfnamefont {T.}~\bibnamefont {Laptyeva}},
  \bibinfo {author} {\bibfnamefont {W.}~\bibnamefont {Tarnowski}}, \bibinfo
  {author} {\bibfnamefont {D.}~\bibnamefont {Chru\ifmmode \acute{s}\else
  \'{s}\fi{}ci\ifmmode~\acute{n}\else \'{n}\fi{}ski}},\ and\ \bibinfo {author}
  {\bibfnamefont {K.}~\bibnamefont {\ifmmode~\dot{Z}\else
  \.{Z}\fi{}yczkowski}},\ }\bibfield  {title} {\bibinfo {title} {Universal
  spectra of random lindblad operators},\ }\href
  {https://doi.org/10.1103/PhysRevLett.123.140403} {\bibfield  {journal}
  {\bibinfo  {journal} {Phys. Rev. Lett.}\ }\textbf {\bibinfo {volume} {123}},\
  \bibinfo {pages} {140403} (\bibinfo {year} {2019})}\BibitemShut {NoStop}%
\bibitem [{\citenamefont {S\'a}\ \emph {et~al.}(2022)\citenamefont {S\'a},
  \citenamefont {Ribeiro},\ and\ \citenamefont {Prosen}}]{sa2022lindbladian}%
  \BibitemOpen
  \bibfield  {author} {\bibinfo {author} {\bibfnamefont {L.}~\bibnamefont
  {S\'a}}, \bibinfo {author} {\bibfnamefont {P.}~\bibnamefont {Ribeiro}},\ and\
  \bibinfo {author} {\bibfnamefont {T.~c.~v.}\ \bibnamefont {Prosen}},\
  }\bibfield  {title} {\bibinfo {title} {Lindbladian dissipation of
  strongly-correlated quantum matter},\ }\href
  {https://doi.org/10.1103/PhysRevResearch.4.L022068} {\bibfield  {journal}
  {\bibinfo  {journal} {Phys. Rev. Res.}\ }\textbf {\bibinfo {volume} {4}},\
  \bibinfo {pages} {L022068} (\bibinfo {year} {2022})}\BibitemShut {NoStop}%
\bibitem [{\citenamefont {Tarnowski}\ \emph {et~al.}(2021)\citenamefont
  {Tarnowski}, \citenamefont {Yusipov}, \citenamefont {Laptyeva}, \citenamefont
  {Denisov}, \citenamefont {Chru\ifmmode \acute{s}\else
  \'{s}\fi{}ci\ifmmode~\acute{n}\else \'{n}\fi{}ski},\ and\ \citenamefont
  {\ifmmode~\dot{Z}\else \.{Z}\fi{}yczkowski}}]{tarnowski2021random}%
  \BibitemOpen
  \bibfield  {author} {\bibinfo {author} {\bibfnamefont {W.}~\bibnamefont
  {Tarnowski}}, \bibinfo {author} {\bibfnamefont {I.}~\bibnamefont {Yusipov}},
  \bibinfo {author} {\bibfnamefont {T.}~\bibnamefont {Laptyeva}}, \bibinfo
  {author} {\bibfnamefont {S.}~\bibnamefont {Denisov}}, \bibinfo {author}
  {\bibfnamefont {D.}~\bibnamefont {Chru\ifmmode \acute{s}\else
  \'{s}\fi{}ci\ifmmode~\acute{n}\else \'{n}\fi{}ski}},\ and\ \bibinfo {author}
  {\bibfnamefont {K.}~\bibnamefont {\ifmmode~\dot{Z}\else
  \.{Z}\fi{}yczkowski}},\ }\bibfield  {title} {\bibinfo {title} {Random
  generators of markovian evolution: A quantum-classical transition by
  superdecoherence},\ }\href {https://doi.org/10.1103/PhysRevE.104.034118}
  {\bibfield  {journal} {\bibinfo  {journal} {Phys. Rev. E}\ }\textbf {\bibinfo
  {volume} {104}},\ \bibinfo {pages} {034118} (\bibinfo {year}
  {2021})}\BibitemShut {NoStop}%
\bibitem [{\citenamefont {Buča}\ and\ \citenamefont
  {Prosen}(2012)}]{buca2012note}%
  \BibitemOpen
  \bibfield  {author} {\bibinfo {author} {\bibfnamefont {B.}~\bibnamefont
  {Buča}}\ and\ \bibinfo {author} {\bibfnamefont {T.}~\bibnamefont {Prosen}},\
  }\bibfield  {title} {\bibinfo {title} {A note on symmetry reductions of the
  lindblad equation: transport in constrained open spin chains},\ }\href
  {https://doi.org/10.1088/1367-2630/14/7/073007} {\bibfield  {journal}
  {\bibinfo  {journal} {New Journal of Physics}\ }\textbf {\bibinfo {volume}
  {14}},\ \bibinfo {pages} {073007} (\bibinfo {year} {2012})}\BibitemShut
  {NoStop}%
\bibitem [{\citenamefont {Albert}\ and\ \citenamefont
  {Jiang}(2014)}]{albert2014symmetry}%
  \BibitemOpen
  \bibfield  {author} {\bibinfo {author} {\bibfnamefont {V.~V.}\ \bibnamefont
  {Albert}}\ and\ \bibinfo {author} {\bibfnamefont {L.}~\bibnamefont {Jiang}},\
  }\bibfield  {title} {\bibinfo {title} {Symmetries and conserved quantities in
  lindblad master equations},\ }\href
  {https://doi.org/10.1103/PhysRevA.89.022118} {\bibfield  {journal} {\bibinfo
  {journal} {Phys. Rev. A}\ }\textbf {\bibinfo {volume} {89}},\ \bibinfo
  {pages} {022118} (\bibinfo {year} {2014})}\BibitemShut {NoStop}%
\bibitem [{\citenamefont {Ba\~nuls}\ \emph {et~al.}(2011)\citenamefont
  {Ba\~nuls}, \citenamefont {Cirac},\ and\ \citenamefont
  {Hastings}}]{banuls2011isingmodel}%
  \BibitemOpen
  \bibfield  {author} {\bibinfo {author} {\bibfnamefont {M.~C.}\ \bibnamefont
  {Ba\~nuls}}, \bibinfo {author} {\bibfnamefont {J.~I.}\ \bibnamefont
  {Cirac}},\ and\ \bibinfo {author} {\bibfnamefont {M.~B.}\ \bibnamefont
  {Hastings}},\ }\bibfield  {title} {\bibinfo {title} {Strong and weak
  thermalization of infinite nonintegrable quantum systems},\ }\href
  {https://doi.org/10.1103/PhysRevLett.106.050405} {\bibfield  {journal}
  {\bibinfo  {journal} {Phys. Rev. Lett.}\ }\textbf {\bibinfo {volume} {106}},\
  \bibinfo {pages} {050405} (\bibinfo {year} {2011})}\BibitemShut {NoStop}%
\bibitem [{\citenamefont {Kim}\ \emph {et~al.}(2014)\citenamefont {Kim},
  \citenamefont {Ikeda},\ and\ \citenamefont {Huse}}]{kim2014testing}%
  \BibitemOpen
  \bibfield  {author} {\bibinfo {author} {\bibfnamefont {H.}~\bibnamefont
  {Kim}}, \bibinfo {author} {\bibfnamefont {T.~N.}\ \bibnamefont {Ikeda}},\
  and\ \bibinfo {author} {\bibfnamefont {D.~A.}\ \bibnamefont {Huse}},\
  }\bibfield  {title} {\bibinfo {title} {Testing whether all eigenstates obey
  the eigenstate thermalization hypothesis},\ }\href
  {https://doi.org/10.1103/PhysRevE.90.052105} {\bibfield  {journal} {\bibinfo
  {journal} {Phys. Rev. E}\ }\textbf {\bibinfo {volume} {90}},\ \bibinfo
  {pages} {052105} (\bibinfo {year} {2014})}\BibitemShut {NoStop}%
\bibitem [{\citenamefont {Chiba}(2024)}]{chiba2024isingProof}%
  \BibitemOpen
  \bibfield  {author} {\bibinfo {author} {\bibfnamefont {Y.}~\bibnamefont
  {Chiba}},\ }\bibfield  {title} {\bibinfo {title} {Proof of absence of local
  conserved quantities in the mixed-field ising chain},\ }\href
  {https://doi.org/10.1103/PhysRevB.109.035123} {\bibfield  {journal} {\bibinfo
   {journal} {Phys. Rev. B}\ }\textbf {\bibinfo {volume} {109}},\ \bibinfo
  {pages} {035123} (\bibinfo {year} {2024})}\BibitemShut {NoStop}%
\bibitem [{\citenamefont {Li}\ \emph {et~al.}(2024)\citenamefont {Li},
  \citenamefont {Yan}, \citenamefont {Prosen},\ and\ \citenamefont
  {Chan}}]{liProsenChan2024MultipleModelDsff}%
  \BibitemOpen
  \bibfield  {author} {\bibinfo {author} {\bibfnamefont {J.}~\bibnamefont
  {Li}}, \bibinfo {author} {\bibfnamefont {S.}~\bibnamefont {Yan}}, \bibinfo
  {author} {\bibfnamefont {T.}~\bibnamefont {Prosen}},\ and\ \bibinfo {author}
  {\bibfnamefont {A.}~\bibnamefont {Chan}},\ }\href
  {https://doi.org/10.48550/arXiv.2405.01641} {\bibinfo {title} {Spectral form
  factor in chaotic, localized, and integrable open quantum many-body systems}}
  (\bibinfo {year} {2024}),\ \Eprint {https://arxiv.org/abs/2405.01641}
  {arXiv:2405.01641 [cond-mat]} \BibitemShut {NoStop}%
\bibitem [{\citenamefont {Torres-Herrera}\ \emph {et~al.}(2016)\citenamefont
  {Torres-Herrera}, \citenamefont {Karp}, \citenamefont {T{\'a}vora},\ and\
  \citenamefont {Santos}}]{torres2016realistic}%
  \BibitemOpen
  \bibfield  {author} {\bibinfo {author} {\bibfnamefont {E.~J.}\ \bibnamefont
  {Torres-Herrera}}, \bibinfo {author} {\bibfnamefont {J.}~\bibnamefont
  {Karp}}, \bibinfo {author} {\bibfnamefont {M.}~\bibnamefont {T{\'a}vora}},\
  and\ \bibinfo {author} {\bibfnamefont {L.~F.}\ \bibnamefont {Santos}},\
  }\bibfield  {title} {\bibinfo {title} {Realistic many-body quantum systems
  vs. full random matrices: Static and dynamical properties},\ }\href
  {https://api.semanticscholar.org/CorpusID:11129849} {\bibfield  {journal}
  {\bibinfo  {journal} {Entropy}\ }\textbf {\bibinfo {volume} {18}},\ \bibinfo
  {pages} {359} (\bibinfo {year} {2016})}\BibitemShut {NoStop}%
\bibitem [{\citenamefont {Edelman}\ \emph {et~al.}(1994)\citenamefont
  {Edelman}, \citenamefont {Kostlan},\ and\ \citenamefont
  {Shub}}]{edelman1994many}%
  \BibitemOpen
  \bibfield  {author} {\bibinfo {author} {\bibfnamefont {A.}~\bibnamefont
  {Edelman}}, \bibinfo {author} {\bibfnamefont {E.}~\bibnamefont {Kostlan}},\
  and\ \bibinfo {author} {\bibfnamefont {M.}~\bibnamefont {Shub}},\ }\bibfield
  {title} {\bibinfo {title} {How many eigenvalues of a random matrix are
  real?},\ }\href@noop {} {\bibfield  {journal} {\bibinfo  {journal} {Journal
  of the American Mathematical Society}\ }\textbf {\bibinfo {volume} {7}},\
  \bibinfo {pages} {247} (\bibinfo {year} {1994})}\BibitemShut {NoStop}%
\bibitem [{\citenamefont {Can}\ \emph {et~al.}(2019)\citenamefont {Can},
  \citenamefont {Oganesyan}, \citenamefont {Orgad},\ and\ \citenamefont
  {Gopalakrishnan}}]{can2019midgap}%
  \BibitemOpen
  \bibfield  {author} {\bibinfo {author} {\bibfnamefont {T.}~\bibnamefont
  {Can}}, \bibinfo {author} {\bibfnamefont {V.}~\bibnamefont {Oganesyan}},
  \bibinfo {author} {\bibfnamefont {D.}~\bibnamefont {Orgad}},\ and\ \bibinfo
  {author} {\bibfnamefont {S.}~\bibnamefont {Gopalakrishnan}},\ }\bibfield
  {title} {\bibinfo {title} {Spectral gaps and midgap states in random quantum
  master equations},\ }\href {https://doi.org/10.1103/PhysRevLett.123.234103}
  {\bibfield  {journal} {\bibinfo  {journal} {Phys. Rev. Lett.}\ }\textbf
  {\bibinfo {volume} {123}},\ \bibinfo {pages} {234103} (\bibinfo {year}
  {2019})}\BibitemShut {NoStop}%
\bibitem [{\citenamefont {Can}(2019)}]{can2019random}%
  \BibitemOpen
  \bibfield  {author} {\bibinfo {author} {\bibfnamefont {T.}~\bibnamefont
  {Can}},\ }\bibfield  {title} {\bibinfo {title} {Random lindblad dynamics},\
  }\href {https://doi.org/10.1088/1751-8121/ab4d26} {\bibfield  {journal}
  {\bibinfo  {journal} {Journal of Physics A: Mathematical and Theoretical}\
  }\textbf {\bibinfo {volume} {52}},\ \bibinfo {pages} {485302} (\bibinfo
  {year} {2019})}\BibitemShut {NoStop}%
\bibitem [{\citenamefont {Chalker}\ and\ \citenamefont
  {Mehlig}(1998)}]{chalkerMehlig1998prlEigenvectors}%
  \BibitemOpen
  \bibfield  {author} {\bibinfo {author} {\bibfnamefont {J.~T.}\ \bibnamefont
  {Chalker}}\ and\ \bibinfo {author} {\bibfnamefont {B.}~\bibnamefont
  {Mehlig}},\ }\bibfield  {title} {\bibinfo {title} {Eigenvector statistics in
  non-hermitian random matrix ensembles},\ }\href
  {https://doi.org/10.1103/PhysRevLett.81.3367} {\bibfield  {journal} {\bibinfo
   {journal} {Phys. Rev. Lett.}\ }\textbf {\bibinfo {volume} {81}},\ \bibinfo
  {pages} {3367} (\bibinfo {year} {1998})}\BibitemShut {NoStop}%
\bibitem [{\citenamefont {Fyodorov}(2018)}]{fyodorov2018statistics}%
  \BibitemOpen
  \bibfield  {author} {\bibinfo {author} {\bibfnamefont {Y.~V.}\ \bibnamefont
  {Fyodorov}},\ }\bibfield  {title} {\bibinfo {title} {On statistics of
  bi-orthogonal eigenvectors in real and complex ginibre ensembles: combining
  partial schur decomposition with supersymmetry},\ }\href
  {https://doi.org/https://doi.org/10.1007/s00220-018-3163-3} {\bibfield
  {journal} {\bibinfo  {journal} {Communications in Mathematical Physics}\
  }\textbf {\bibinfo {volume} {363}},\ \bibinfo {pages} {579} (\bibinfo {year}
  {2018})}\BibitemShut {NoStop}%
\bibitem [{\citenamefont {Bourgade}\ and\ \citenamefont
  {Dubach}(2020)}]{bourgade2020distribution}%
  \BibitemOpen
  \bibfield  {author} {\bibinfo {author} {\bibfnamefont {P.}~\bibnamefont
  {Bourgade}}\ and\ \bibinfo {author} {\bibfnamefont {G.}~\bibnamefont
  {Dubach}},\ }\bibfield  {title} {\bibinfo {title} {The distribution of
  overlaps between eigenvectors of ginibre matrices},\ }\href
  {https://doi.org/https://doi.org/10.1007/s00440-019-00953-x} {\bibfield
  {journal} {\bibinfo  {journal} {Probability Theory and Related Fields}\
  }\textbf {\bibinfo {volume} {177}},\ \bibinfo {pages} {397} (\bibinfo {year}
  {2020})}\BibitemShut {NoStop}%
\bibitem [{\citenamefont {Hartmann}\ \emph {et~al.}(2024)\citenamefont
  {Hartmann}, \citenamefont {Li},\ and\ \citenamefont {Luitz}}]{hartmann2024}%
  \BibitemOpen
  \bibfield  {author} {\bibinfo {author} {\bibfnamefont {N.~D.}\ \bibnamefont
  {Hartmann}}, \bibinfo {author} {\bibfnamefont {J.~L.}\ \bibnamefont {Li}},\
  and\ \bibinfo {author} {\bibfnamefont {D.~J.}\ \bibnamefont {Luitz}},\
  }\bibfield  {title} {\bibinfo {title} {Fate of dissipative hierarchy of
  timescales in the presence of unitary dynamics},\ }\href
  {https://doi.org/10.1103/PhysRevB.109.054203} {\bibfield  {journal} {\bibinfo
   {journal} {Phys. Rev. B}\ }\textbf {\bibinfo {volume} {109}},\ \bibinfo
  {pages} {054203} (\bibinfo {year} {2024})}\BibitemShut {NoStop}%
\bibitem [{\citenamefont {Qi}\ and\ \citenamefont
  {Streicher}(2019)}]{qi2019quantum}%
  \BibitemOpen
  \bibfield  {author} {\bibinfo {author} {\bibfnamefont {X.-L.}\ \bibnamefont
  {Qi}}\ and\ \bibinfo {author} {\bibfnamefont {A.}~\bibnamefont {Streicher}},\
  }\bibfield  {title} {\bibinfo {title} {Quantum epidemiology: operator growth,
  thermal effects, and syk},\ }\href {https://doi.org/10.1007/JHEP08(2019)012}
  {\bibfield  {journal} {\bibinfo  {journal} {Journal of High Energy Physics}\
  }\textbf {\bibinfo {volume} {2019}} (\bibinfo {year} {2019})}\BibitemShut
  {NoStop}%
\bibitem [{\citenamefont {Shirai}\ and\ \citenamefont
  {Mori}(2024)}]{shirai2024accelerated}%
  \BibitemOpen
  \bibfield  {author} {\bibinfo {author} {\bibfnamefont {T.}~\bibnamefont
  {Shirai}}\ and\ \bibinfo {author} {\bibfnamefont {T.}~\bibnamefont {Mori}},\
  }\bibfield  {title} {\bibinfo {title} {Accelerated decay due to operator
  spreading in bulk-dissipated quantum systems},\ }\href
  {https://doi.org/10.1103/PhysRevLett.133.040201} {\bibfield  {journal}
  {\bibinfo  {journal} {Phys. Rev. Lett.}\ }\textbf {\bibinfo {volume} {133}},\
  \bibinfo {pages} {040201} (\bibinfo {year} {2024})}\BibitemShut {NoStop}%
\bibitem [{\citenamefont {Abanin}\ \emph {et~al.}(2015)\citenamefont {Abanin},
  \citenamefont {De~Roeck},\ and\ \citenamefont
  {Huveneers}}]{abanin2015floquetPrethermal}%
  \BibitemOpen
  \bibfield  {author} {\bibinfo {author} {\bibfnamefont {D.~A.}\ \bibnamefont
  {Abanin}}, \bibinfo {author} {\bibfnamefont {W.}~\bibnamefont {De~Roeck}},\
  and\ \bibinfo {author} {\bibfnamefont {F.}~\bibnamefont {Huveneers}},\
  }\bibfield  {title} {\bibinfo {title} {Exponentially {Slow} {Heating} in
  {Periodically} {Driven} {Many}-{Body} {Systems}},\ }\bibfield  {journal}
  {\bibinfo  {journal} {Physical Review Letters}\ }\textbf {\bibinfo {volume}
  {115}},\ \href {https://doi.org/10.1103/PhysRevLett.115.256803}
  {10.1103/PhysRevLett.115.256803} (\bibinfo {year} {2015}),\ \bibinfo {note}
  {arXiv: 1507.01474}\BibitemShut {NoStop}%
\bibitem [{\citenamefont {Mori}\ \emph {et~al.}(2016)\citenamefont {Mori},
  \citenamefont {Kuwahara},\ and\ \citenamefont
  {Saito}}]{mori2016floquetBound}%
  \BibitemOpen
  \bibfield  {author} {\bibinfo {author} {\bibfnamefont {T.}~\bibnamefont
  {Mori}}, \bibinfo {author} {\bibfnamefont {T.}~\bibnamefont {Kuwahara}},\
  and\ \bibinfo {author} {\bibfnamefont {K.}~\bibnamefont {Saito}},\ }\bibfield
   {title} {\bibinfo {title} {Rigorous {Bound} on {Energy} {Absorption} and
  {Generic} {Relaxation} in {Periodically} {Driven} {Quantum} {Systems}},\
  }\bibfield  {journal} {\bibinfo  {journal} {Physical Review Letters}\
  }\textbf {\bibinfo {volume} {116}},\ \href
  {https://doi.org/10.1103/PhysRevLett.116.120401}
  {10.1103/PhysRevLett.116.120401} (\bibinfo {year} {2016}),\ \bibinfo {note}
  {arXiv: 1509.03968}\BibitemShut {NoStop}%
\bibitem [{\citenamefont {Santos}\ and\ \citenamefont
  {Rigol}(2010)}]{santos2010onset}%
  \BibitemOpen
  \bibfield  {author} {\bibinfo {author} {\bibfnamefont {L.~F.}\ \bibnamefont
  {Santos}}\ and\ \bibinfo {author} {\bibfnamefont {M.}~\bibnamefont {Rigol}},\
  }\bibfield  {title} {\bibinfo {title} {Onset of quantum chaos in
  one-dimensional bosonic and fermionic systems and its relation to
  thermalization},\ }\href {https://doi.org/10.1103/PhysRevE.81.036206}
  {\bibfield  {journal} {\bibinfo  {journal} {Phys. Rev. E}\ }\textbf {\bibinfo
  {volume} {81}},\ \bibinfo {pages} {036206} (\bibinfo {year}
  {2010})}\BibitemShut {NoStop}%
\bibitem [{\citenamefont {Rigol}\ and\ \citenamefont
  {Santos}(2010)}]{rigol2010gapped}%
  \BibitemOpen
  \bibfield  {author} {\bibinfo {author} {\bibfnamefont {M.}~\bibnamefont
  {Rigol}}\ and\ \bibinfo {author} {\bibfnamefont {L.~F.}\ \bibnamefont
  {Santos}},\ }\bibfield  {title} {\bibinfo {title} {Quantum chaos and
  thermalization in gapped systems},\ }\href
  {https://doi.org/10.1103/PhysRevA.82.011604} {\bibfield  {journal} {\bibinfo
  {journal} {Phys. Rev. A}\ }\textbf {\bibinfo {volume} {82}},\ \bibinfo
  {pages} {011604} (\bibinfo {year} {2010})}\BibitemShut {NoStop}%
\bibitem [{\citenamefont {Gill}\ and\ \citenamefont
  {Sarkar}(2025)}]{gill2025speedlimitsscramblingkrylov}%
  \BibitemOpen
  \bibfield  {author} {\bibinfo {author} {\bibfnamefont {A.}~\bibnamefont
  {Gill}}\ and\ \bibinfo {author} {\bibfnamefont {T.}~\bibnamefont {Sarkar}},\
  }\bibfield  {title} {\bibinfo {title} {Speed limits and scrambling in krylov
  space},\ }\href {https://doi.org/10.1103/PhysRevB.111.184307} {\bibfield
  {journal} {\bibinfo  {journal} {Phys. Rev. B}\ }\textbf {\bibinfo {volume}
  {111}},\ \bibinfo {pages} {184307} (\bibinfo {year} {2025})}\BibitemShut
  {NoStop}%
\bibitem [{\citenamefont {Yang}\ and\ \citenamefont
  {Abanin}(2025)}]{abanin2025integralsmotionslowmodes}%
  \BibitemOpen
  \bibfield  {author} {\bibinfo {author} {\bibfnamefont {T.-H.}\ \bibnamefont
  {Yang}}\ and\ \bibinfo {author} {\bibfnamefont {D.~A.}\ \bibnamefont
  {Abanin}},\ }\bibfield  {title} {\bibinfo {title} {Integrals of motion as
  slow modes in dissipative many-body operator dynamics},\ }\href
  {https://doi.org/10.1103/d9b4-wlv6} {\bibfield  {journal} {\bibinfo
  {journal} {Phys. Rev. B}\ }\textbf {\bibinfo {volume} {112}},\ \bibinfo
  {pages} {094304} (\bibinfo {year} {2025})}\BibitemShut {NoStop}%
\bibitem [{\citenamefont {Cipolloni}\ and\ \citenamefont
  {Kudler-Flam}(2024)}]{cipolloni2024nonHermitianViolatesETH}%
  \BibitemOpen
  \bibfield  {author} {\bibinfo {author} {\bibfnamefont {G.}~\bibnamefont
  {Cipolloni}}\ and\ \bibinfo {author} {\bibfnamefont {J.}~\bibnamefont
  {Kudler-Flam}},\ }\bibfield  {title} {\bibinfo {title} {Non-hermitian
  hamiltonians violate the eigenstate thermalization hypothesis},\ }\href
  {https://doi.org/10.1103/PhysRevB.109.L020201} {\bibfield  {journal}
  {\bibinfo  {journal} {Phys. Rev. B}\ }\textbf {\bibinfo {volume} {109}},\
  \bibinfo {pages} {L020201} (\bibinfo {year} {2024})}\BibitemShut {NoStop}%
\bibitem [{\citenamefont {Singha~Roy}\ \emph {et~al.}(2025)\citenamefont
  {Singha~Roy}, \citenamefont {Bandyopadhyay}, \citenamefont {Costa~de
  Almeida},\ and\ \citenamefont {Hauke}}]{roy2025nonHermitianETH}%
  \BibitemOpen
  \bibfield  {author} {\bibinfo {author} {\bibfnamefont {S.}~\bibnamefont
  {Singha~Roy}}, \bibinfo {author} {\bibfnamefont {S.}~\bibnamefont
  {Bandyopadhyay}}, \bibinfo {author} {\bibfnamefont {R.}~\bibnamefont
  {Costa~de Almeida}},\ and\ \bibinfo {author} {\bibfnamefont {P.}~\bibnamefont
  {Hauke}},\ }\bibfield  {title} {\bibinfo {title} {Unveiling eigenstate
  thermalization for non-hermitian systems},\ }\href
  {https://doi.org/10.1103/PhysRevLett.134.180405} {\bibfield  {journal}
  {\bibinfo  {journal} {Phys. Rev. Lett.}\ }\textbf {\bibinfo {volume} {134}},\
  \bibinfo {pages} {180405} (\bibinfo {year} {2025})}\BibitemShut {NoStop}%
\bibitem [{\citenamefont {Lidar}\ \emph {et~al.}(1998)\citenamefont {Lidar},
  \citenamefont {Chuang},\ and\ \citenamefont {Whaley}}]{lidar1998dfs}%
  \BibitemOpen
  \bibfield  {author} {\bibinfo {author} {\bibfnamefont {D.~A.}\ \bibnamefont
  {Lidar}}, \bibinfo {author} {\bibfnamefont {I.~L.}\ \bibnamefont {Chuang}},\
  and\ \bibinfo {author} {\bibfnamefont {K.~B.}\ \bibnamefont {Whaley}},\
  }\bibfield  {title} {\bibinfo {title} {Decoherence-free subspaces for quantum
  computation},\ }\href {https://doi.org/10.1103/PhysRevLett.81.2594}
  {\bibfield  {journal} {\bibinfo  {journal} {Phys. Rev. Lett.}\ }\textbf
  {\bibinfo {volume} {81}},\ \bibinfo {pages} {2594} (\bibinfo {year}
  {1998})}\BibitemShut {NoStop}%
\bibitem [{\citenamefont {Islam}\ \emph {et~al.}(2015)\citenamefont {Islam},
  \citenamefont {Ma}, \citenamefont {Preiss}, \citenamefont {Eric~Tai},
  \citenamefont {Lukin}, \citenamefont {Rispoli},\ and\ \citenamefont
  {Greiner}}]{islam2015purity}%
  \BibitemOpen
  \bibfield  {author} {\bibinfo {author} {\bibfnamefont {R.}~\bibnamefont
  {Islam}}, \bibinfo {author} {\bibfnamefont {R.}~\bibnamefont {Ma}}, \bibinfo
  {author} {\bibfnamefont {P.~M.}\ \bibnamefont {Preiss}}, \bibinfo {author}
  {\bibfnamefont {M.}~\bibnamefont {Eric~Tai}}, \bibinfo {author}
  {\bibfnamefont {A.}~\bibnamefont {Lukin}}, \bibinfo {author} {\bibfnamefont
  {M.}~\bibnamefont {Rispoli}},\ and\ \bibinfo {author} {\bibfnamefont
  {M.}~\bibnamefont {Greiner}},\ }\bibfield  {title} {\bibinfo {title}
  {Measuring entanglement entropy in a quantum many-body system},\ }\href
  {https://doi.org/https://doi.org/10.1038/nature15750} {\bibfield  {journal}
  {\bibinfo  {journal} {Nature}\ }\textbf {\bibinfo {volume} {528}},\ \bibinfo
  {pages} {77} (\bibinfo {year} {2015})}\BibitemShut {NoStop}%
\bibitem [{\citenamefont {Kaufman}\ \emph {et~al.}(2016)\citenamefont
  {Kaufman}, \citenamefont {Tai}, \citenamefont {Lukin}, \citenamefont
  {Rispoli}, \citenamefont {Schittko}, \citenamefont {Preiss},\ and\
  \citenamefont {Greiner}}]{kaufman2016thermalizationExperiment}%
  \BibitemOpen
  \bibfield  {author} {\bibinfo {author} {\bibfnamefont {A.~M.}\ \bibnamefont
  {Kaufman}}, \bibinfo {author} {\bibfnamefont {M.~E.}\ \bibnamefont {Tai}},
  \bibinfo {author} {\bibfnamefont {A.}~\bibnamefont {Lukin}}, \bibinfo
  {author} {\bibfnamefont {M.}~\bibnamefont {Rispoli}}, \bibinfo {author}
  {\bibfnamefont {R.}~\bibnamefont {Schittko}}, \bibinfo {author}
  {\bibfnamefont {P.~M.}\ \bibnamefont {Preiss}},\ and\ \bibinfo {author}
  {\bibfnamefont {M.}~\bibnamefont {Greiner}},\ }\bibfield  {title} {\bibinfo
  {title} {Quantum thermalization through entanglement in an isolated many-body
  system},\ }\href {https://doi.org/10.1126/science.aaf6725} {\bibfield
  {journal} {\bibinfo  {journal} {Science}\ }\textbf {\bibinfo {volume}
  {353}},\ \bibinfo {pages} {794} (\bibinfo {year} {2016})}\BibitemShut
  {NoStop}%
\bibitem [{\citenamefont {Zhou}\ \emph {et~al.}(2021)\citenamefont {Zhou},
  \citenamefont {Mao},\ and\ \citenamefont {Zhai}}]{zhou2021renyi}%
  \BibitemOpen
  \bibfield  {author} {\bibinfo {author} {\bibfnamefont {Y.~N.}\ \bibnamefont
  {Zhou}}, \bibinfo {author} {\bibfnamefont {L.}~\bibnamefont {Mao}},\ and\
  \bibinfo {author} {\bibfnamefont {H.}~\bibnamefont {Zhai}},\ }\bibfield
  {title} {\bibinfo {title} {Rényi entropy dynamics and {Lindblad} spectrum
  for open quantum systems},\ }\bibfield  {journal} {\bibinfo  {journal}
  {Physical Review Research}\ }\textbf {\bibinfo {volume} {3}},\ \href
  {https://doi.org/10.1103/PhysRevResearch.3.043060}
  {10.1103/PhysRevResearch.3.043060} (\bibinfo {year} {2021})\BibitemShut
  {NoStop}%
\bibitem [{\citenamefont {Wang}\ \emph {et~al.}(2024)\citenamefont {Wang},
  \citenamefont {Liu}, \citenamefont {Zhang},\ and\ \citenamefont
  {Garc\'{\i}a-Garc\'{\i}a}}]{hantengWang2024sykLindbladEntanglement}%
  \BibitemOpen
  \bibfield  {author} {\bibinfo {author} {\bibfnamefont {H.}~\bibnamefont
  {Wang}}, \bibinfo {author} {\bibfnamefont {C.}~\bibnamefont {Liu}}, \bibinfo
  {author} {\bibfnamefont {P.}~\bibnamefont {Zhang}},\ and\ \bibinfo {author}
  {\bibfnamefont {A.~M.}\ \bibnamefont {Garc\'{\i}a-Garc\'{\i}a}},\ }\bibfield
  {title} {\bibinfo {title} {Entanglement transition and replica wormholes in
  the dissipative sachdev-ye-kitaev model},\ }\href
  {https://doi.org/10.1103/PhysRevD.109.046005} {\bibfield  {journal} {\bibinfo
   {journal} {Phys. Rev. D}\ }\textbf {\bibinfo {volume} {109}},\ \bibinfo
  {pages} {046005} (\bibinfo {year} {2024})}\BibitemShut {NoStop}%
\bibitem [{\citenamefont {Zyczkowski}\ and\ \citenamefont
  {Sommers}(2001)}]{zyczkowski2001induced}%
  \BibitemOpen
  \bibfield  {author} {\bibinfo {author} {\bibfnamefont {K.}~\bibnamefont
  {Zyczkowski}}\ and\ \bibinfo {author} {\bibfnamefont {H.-J.}\ \bibnamefont
  {Sommers}},\ }\bibfield  {title} {\bibinfo {title} {Induced measures in the
  space of mixed quantum states},\ }\href
  {https://doi.org/10.1088/0305-4470/34/35/335} {\bibfield  {journal} {\bibinfo
   {journal} {Journal of Physics A: Mathematical and General}\ }\textbf
  {\bibinfo {volume} {34}},\ \bibinfo {pages} {7111} (\bibinfo {year}
  {2001})}\BibitemShut {NoStop}%
\bibitem [{\citenamefont {{\.Z}yczkowski}\ \emph {et~al.}(2011)\citenamefont
  {{\.Z}yczkowski}, \citenamefont {Penson}, \citenamefont {Nechita},\ and\
  \citenamefont {Collins}}]{zyczkowski2011generating}%
  \BibitemOpen
  \bibfield  {author} {\bibinfo {author} {\bibfnamefont {K.}~\bibnamefont
  {{\.Z}yczkowski}}, \bibinfo {author} {\bibfnamefont {K.~A.}\ \bibnamefont
  {Penson}}, \bibinfo {author} {\bibfnamefont {I.}~\bibnamefont {Nechita}},\
  and\ \bibinfo {author} {\bibfnamefont {B.}~\bibnamefont {Collins}},\
  }\bibfield  {title} {\bibinfo {title} {Generating random density matrices},\
  }\href {https://doi.org/10.1063/1.3595693} {\bibfield  {journal} {\bibinfo
  {journal} {Journal of Mathematical Physics}\ }\textbf {\bibinfo {volume}
  {52}} (\bibinfo {year} {2011})}\BibitemShut {NoStop}%
\bibitem [{\citenamefont {Lidar}\ \emph {et~al.}(2006)\citenamefont {Lidar},
  \citenamefont {Shabani},\ and\ \citenamefont
  {Alicki}}]{lidar2006purityMonotonicProof}%
  \BibitemOpen
  \bibfield  {author} {\bibinfo {author} {\bibfnamefont {D.}~\bibnamefont
  {Lidar}}, \bibinfo {author} {\bibfnamefont {A.}~\bibnamefont {Shabani}},\
  and\ \bibinfo {author} {\bibfnamefont {R.}~\bibnamefont {Alicki}},\
  }\bibfield  {title} {\bibinfo {title} {Conditions for strictly
  purity-decreasing quantum markovian dynamics},\ }\href
  {https://doi.org/https://doi.org/10.1016/j.chemphys.2005.06.038} {\bibfield
  {journal} {\bibinfo  {journal} {Chemical physics}\ }\textbf {\bibinfo
  {volume} {322}},\ \bibinfo {pages} {82} (\bibinfo {year} {2006})}\BibitemShut
  {NoStop}%
\bibitem [{\citenamefont {Bartsch}\ and\ \citenamefont
  {Gemmer}(2009)}]{bartsch2009typicality}%
  \BibitemOpen
  \bibfield  {author} {\bibinfo {author} {\bibfnamefont {C.}~\bibnamefont
  {Bartsch}}\ and\ \bibinfo {author} {\bibfnamefont {J.}~\bibnamefont
  {Gemmer}},\ }\bibfield  {title} {\bibinfo {title} {Dynamical typicality of
  quantum expectation values},\ }\href
  {https://doi.org/10.1103/PhysRevLett.102.110403} {\bibfield  {journal}
  {\bibinfo  {journal} {Phys. Rev. Lett.}\ }\textbf {\bibinfo {volume} {102}},\
  \bibinfo {pages} {110403} (\bibinfo {year} {2009})}\BibitemShut {NoStop}%
\bibitem [{\citenamefont {Bao}(2026)}]{bao2026lindbladTypicality}%
  \BibitemOpen
  \bibfield  {author} {\bibinfo {author} {\bibfnamefont {R.}~\bibnamefont
  {Bao}},\ }\bibfield  {title} {\bibinfo {title} {Initial-state typicality in
  quantum relaxation},\ }\href {https://doi.org/10.1103/wgr5-lb6b} {\bibfield
  {journal} {\bibinfo  {journal} {Phys. Rev. Lett.}\ }\textbf {\bibinfo
  {volume} {136}},\ \bibinfo {pages} {070402} (\bibinfo {year}
  {2026})}\BibitemShut {NoStop}%
\bibitem [{\citenamefont {Lessa}\ \emph {et~al.}(2025)\citenamefont {Lessa},
  \citenamefont {Ma}, \citenamefont {Zhang}, \citenamefont {Bi}, \citenamefont
  {Cheng},\ and\ \citenamefont {Wang}}]{lessa2025strongtoweak}%
  \BibitemOpen
  \bibfield  {author} {\bibinfo {author} {\bibfnamefont {L.~A.}\ \bibnamefont
  {Lessa}}, \bibinfo {author} {\bibfnamefont {R.}~\bibnamefont {Ma}}, \bibinfo
  {author} {\bibfnamefont {J.-H.}\ \bibnamefont {Zhang}}, \bibinfo {author}
  {\bibfnamefont {Z.}~\bibnamefont {Bi}}, \bibinfo {author} {\bibfnamefont
  {M.}~\bibnamefont {Cheng}},\ and\ \bibinfo {author} {\bibfnamefont
  {C.}~\bibnamefont {Wang}},\ }\bibfield  {title} {\bibinfo {title}
  {Strong-to-{Weak} {Spontaneous} {Symmetry} {Breaking} in {Mixed} {Quantum}
  {States}},\ }\href {https://doi.org/10.1103/PRXQuantum.6.010344} {\bibfield
  {journal} {\bibinfo  {journal} {PRX Quantum}\ }\textbf {\bibinfo {volume}
  {6}},\ \bibinfo {pages} {010344} (\bibinfo {year} {2025})}\BibitemShut
  {NoStop}%
\bibitem [{\citenamefont {Sala}\ \emph {et~al.}(2024)\citenamefont {Sala},
  \citenamefont {Gopalakrishnan}, \citenamefont {Oshikawa},\ and\ \citenamefont
  {You}}]{sala2024spontaneous}%
  \BibitemOpen
  \bibfield  {author} {\bibinfo {author} {\bibfnamefont {P.}~\bibnamefont
  {Sala}}, \bibinfo {author} {\bibfnamefont {S.}~\bibnamefont
  {Gopalakrishnan}}, \bibinfo {author} {\bibfnamefont {M.}~\bibnamefont
  {Oshikawa}},\ and\ \bibinfo {author} {\bibfnamefont {Y.}~\bibnamefont
  {You}},\ }\bibfield  {title} {\bibinfo {title} {Spontaneous strong symmetry
  breaking in open systems: {Purification} perspective},\ }\href
  {https://doi.org/10.1103/PhysRevB.110.155150} {\bibfield  {journal} {\bibinfo
   {journal} {Physical Review B}\ }\textbf {\bibinfo {volume} {110}},\ \bibinfo
  {pages} {155150} (\bibinfo {year} {2024})}\BibitemShut {NoStop}%
\bibitem [{\citenamefont {Lee}\ \emph {et~al.}(2023)\citenamefont {Lee},
  \citenamefont {Jian},\ and\ \citenamefont
  {Xu}}]{leeCenke2023decoherenceCriticality}%
  \BibitemOpen
  \bibfield  {author} {\bibinfo {author} {\bibfnamefont {J.~Y.}\ \bibnamefont
  {Lee}}, \bibinfo {author} {\bibfnamefont {C.-M.}\ \bibnamefont {Jian}},\ and\
  \bibinfo {author} {\bibfnamefont {C.}~\bibnamefont {Xu}},\ }\bibfield
  {title} {\bibinfo {title} {Quantum {Criticality} {Under} {Decoherence} or
  {Weak} {Measurement}},\ }\href {https://doi.org/10.1103/PRXQuantum.4.030317}
  {\bibfield  {journal} {\bibinfo  {journal} {PRX Quantum}\ }\textbf {\bibinfo
  {volume} {4}},\ \bibinfo {pages} {030317} (\bibinfo {year}
  {2023})}\BibitemShut {NoStop}%
\bibitem [{\citenamefont {Bao}\ \emph {et~al.}(2021)\citenamefont {Bao},
  \citenamefont {Choi},\ and\ \citenamefont {Altman}}]{bao2021symmetry}%
  \BibitemOpen
  \bibfield  {author} {\bibinfo {author} {\bibfnamefont {Y.}~\bibnamefont
  {Bao}}, \bibinfo {author} {\bibfnamefont {S.}~\bibnamefont {Choi}},\ and\
  \bibinfo {author} {\bibfnamefont {E.}~\bibnamefont {Altman}},\ }\bibfield
  {title} {\bibinfo {title} {Symmetry enriched phases of quantum circuits},\
  }\href {https://doi.org/https://doi.org/10.1016/j.aop.2021.168618} {\bibfield
   {journal} {\bibinfo  {journal} {Annals of Physics}\ }\textbf {\bibinfo
  {volume} {435}},\ \bibinfo {pages} {168618} (\bibinfo {year}
  {2021})}\BibitemShut {NoStop}%
\bibitem [{\citenamefont {Zhang}\ \emph {et~al.}(2025)\citenamefont {Zhang},
  \citenamefont {Hsieh}, \citenamefont {Kim},\ and\ \citenamefont
  {Zou}}]{hsieh2025probingRenyi}%
  \BibitemOpen
  \bibfield  {author} {\bibinfo {author} {\bibfnamefont {Y.}~\bibnamefont
  {Zhang}}, \bibinfo {author} {\bibfnamefont {T.~H.}\ \bibnamefont {Hsieh}},
  \bibinfo {author} {\bibfnamefont {Y.~B.}\ \bibnamefont {Kim}},\ and\ \bibinfo
  {author} {\bibfnamefont {Y.}~\bibnamefont {Zou}},\ }\bibfield  {title}
  {\bibinfo {title} {Probing mixed-state phases on a quantum computer via renyi
  correlators and variational decoding},\ }\href
  {https://arxiv.org/abs/2505.02900v1} {\bibfield  {journal} {\bibinfo
  {journal} {arXiv.org}\ } (\bibinfo {year} {2025})}\BibitemShut {NoStop}%
\bibitem [{\citenamefont {Roberts}\ and\ \citenamefont
  {Stanford}(2015)}]{roberts2015otocCFT}%
  \BibitemOpen
  \bibfield  {author} {\bibinfo {author} {\bibfnamefont {D.~A.}\ \bibnamefont
  {Roberts}}\ and\ \bibinfo {author} {\bibfnamefont {D.}~\bibnamefont
  {Stanford}},\ }\bibfield  {title} {\bibinfo {title} {Diagnosing chaos using
  four-point functions in two-dimensional conformal field theory},\ }\href
  {https://doi.org/10.1103/PhysRevLett.115.131603} {\bibfield  {journal}
  {\bibinfo  {journal} {Phys. Rev. Lett.}\ }\textbf {\bibinfo {volume} {115}},\
  \bibinfo {pages} {131603} (\bibinfo {year} {2015})}\BibitemShut {NoStop}%
\bibitem [{\citenamefont {Maldacena}\ \emph {et~al.}(2016)\citenamefont
  {Maldacena}, \citenamefont {Shenker},\ and\ \citenamefont
  {Stanford}}]{maldacena2016bound}%
  \BibitemOpen
  \bibfield  {author} {\bibinfo {author} {\bibfnamefont {J.}~\bibnamefont
  {Maldacena}}, \bibinfo {author} {\bibfnamefont {S.~H.}\ \bibnamefont
  {Shenker}},\ and\ \bibinfo {author} {\bibfnamefont {D.}~\bibnamefont
  {Stanford}},\ }\bibfield  {title} {\bibinfo {title} {A bound on chaos},\
  }\href {https://doi.org/http://dx.doi.org/10.1007/JHEP08(2016)106} {\bibfield
   {journal} {\bibinfo  {journal} {Journal of High Energy Physics}\ }\textbf
  {\bibinfo {volume} {2016}},\ \bibinfo {pages} {1} (\bibinfo {year}
  {2016})}\BibitemShut {NoStop}%
\bibitem [{\citenamefont {Bohrdt}\ \emph {et~al.}(2017)\citenamefont {Bohrdt},
  \citenamefont {Mendl}, \citenamefont {Endres},\ and\ \citenamefont
  {Knap}}]{bohrdt2017scrambling}%
  \BibitemOpen
  \bibfield  {author} {\bibinfo {author} {\bibfnamefont {A.}~\bibnamefont
  {Bohrdt}}, \bibinfo {author} {\bibfnamefont {C.~B.}\ \bibnamefont {Mendl}},
  \bibinfo {author} {\bibfnamefont {M.}~\bibnamefont {Endres}},\ and\ \bibinfo
  {author} {\bibfnamefont {M.}~\bibnamefont {Knap}},\ }\bibfield  {title}
  {\bibinfo {title} {Scrambling and thermalization in a diffusive quantum
  many-body system},\ }\bibfield  {journal} {\bibinfo  {journal} {New Journal
  of Physics}\ }\textbf {\bibinfo {volume} {19}},\ \href
  {https://doi.org/10.1088/1367-2630/aa719b} {10.1088/1367-2630/aa719b}
  (\bibinfo {year} {2017}),\ \Eprint {https://arxiv.org/abs/1612.02434}
  {1612.02434} \BibitemShut {NoStop}%
\bibitem [{\citenamefont {von Keyserlingk}\ \emph {et~al.}(2018)\citenamefont
  {von Keyserlingk}, \citenamefont {Rakovszky}, \citenamefont {Pollmann},\ and\
  \citenamefont {Sondhi}}]{keyserlingk2018otoc}%
  \BibitemOpen
  \bibfield  {author} {\bibinfo {author} {\bibfnamefont {C.~W.}\ \bibnamefont
  {von Keyserlingk}}, \bibinfo {author} {\bibfnamefont {T.}~\bibnamefont
  {Rakovszky}}, \bibinfo {author} {\bibfnamefont {F.}~\bibnamefont
  {Pollmann}},\ and\ \bibinfo {author} {\bibfnamefont {S.~L.}\ \bibnamefont
  {Sondhi}},\ }\bibfield  {title} {\bibinfo {title} {Operator hydrodynamics,
  otocs, and entanglement growth in systems without conservation laws},\ }\href
  {https://doi.org/10.1103/PhysRevX.8.021013} {\bibfield  {journal} {\bibinfo
  {journal} {Phys. Rev. X}\ }\textbf {\bibinfo {volume} {8}},\ \bibinfo {pages}
  {021013} (\bibinfo {year} {2018})}\BibitemShut {NoStop}%
\bibitem [{\citenamefont {Nahum}\ \emph {et~al.}(2018)\citenamefont {Nahum},
  \citenamefont {Vijay},\ and\ \citenamefont {Haah}}]{nahum2018randomunitary}%
  \BibitemOpen
  \bibfield  {author} {\bibinfo {author} {\bibfnamefont {A.}~\bibnamefont
  {Nahum}}, \bibinfo {author} {\bibfnamefont {S.}~\bibnamefont {Vijay}},\ and\
  \bibinfo {author} {\bibfnamefont {J.}~\bibnamefont {Haah}},\ }\bibfield
  {title} {\bibinfo {title} {Operator spreading in random unitary circuits},\
  }\href {https://doi.org/10.1103/PhysRevX.8.021014} {\bibfield  {journal}
  {\bibinfo  {journal} {Phys. Rev. X}\ }\textbf {\bibinfo {volume} {8}},\
  \bibinfo {pages} {021014} (\bibinfo {year} {2018})}\BibitemShut {NoStop}%
\bibitem [{\citenamefont {Khemani}\ \emph {et~al.}(2018)\citenamefont
  {Khemani}, \citenamefont {Vishwanath},\ and\ \citenamefont
  {Huse}}]{khemani2018conserved}%
  \BibitemOpen
  \bibfield  {author} {\bibinfo {author} {\bibfnamefont {V.}~\bibnamefont
  {Khemani}}, \bibinfo {author} {\bibfnamefont {A.}~\bibnamefont
  {Vishwanath}},\ and\ \bibinfo {author} {\bibfnamefont {D.~A.}\ \bibnamefont
  {Huse}},\ }\bibfield  {title} {\bibinfo {title} {Operator spreading and the
  emergence of dissipative hydrodynamics under unitary evolution with
  conservation laws},\ }\href {https://doi.org/10.1103/PhysRevX.8.031057}
  {\bibfield  {journal} {\bibinfo  {journal} {Phys. Rev. X}\ }\textbf {\bibinfo
  {volume} {8}},\ \bibinfo {pages} {031057} (\bibinfo {year}
  {2018})}\BibitemShut {NoStop}%
\bibitem [{\citenamefont {Parker}\ \emph {et~al.}(2019)\citenamefont {Parker},
  \citenamefont {Cao}, \citenamefont {Avdoshkin}, \citenamefont {Scaffidi},\
  and\ \citenamefont {Altman}}]{parker2019universalOGH}%
  \BibitemOpen
  \bibfield  {author} {\bibinfo {author} {\bibfnamefont {D.~E.}\ \bibnamefont
  {Parker}}, \bibinfo {author} {\bibfnamefont {X.}~\bibnamefont {Cao}},
  \bibinfo {author} {\bibfnamefont {A.}~\bibnamefont {Avdoshkin}}, \bibinfo
  {author} {\bibfnamefont {T.}~\bibnamefont {Scaffidi}},\ and\ \bibinfo
  {author} {\bibfnamefont {E.}~\bibnamefont {Altman}},\ }\bibfield  {title}
  {\bibinfo {title} {A {{Universal Operator Growth Hypothesis}}},\ }\href
  {https://doi.org/10.1103/PhysRevX.9.041017} {\bibfield  {journal} {\bibinfo
  {journal} {Physical Review X}\ }\textbf {\bibinfo {volume} {9}},\ \bibinfo
  {pages} {041017} (\bibinfo {year} {2019})}\BibitemShut {NoStop}%
\bibitem [{\citenamefont {Fisher}\ \emph {et~al.}(2023)\citenamefont {Fisher},
  \citenamefont {Khemani}, \citenamefont {Nahum},\ and\ \citenamefont
  {Vijay}}]{fisher2023circuitreview}%
  \BibitemOpen
  \bibfield  {author} {\bibinfo {author} {\bibfnamefont {M.~P.}\ \bibnamefont
  {Fisher}}, \bibinfo {author} {\bibfnamefont {V.}~\bibnamefont {Khemani}},
  \bibinfo {author} {\bibfnamefont {A.}~\bibnamefont {Nahum}},\ and\ \bibinfo
  {author} {\bibfnamefont {S.}~\bibnamefont {Vijay}},\ }\bibfield  {title}
  {\bibinfo {title} {Random quantum circuits},\ }\href
  {https://doi.org/https://doi.org/10.1146/annurev-conmatphys-031720-030658}
  {\bibfield  {journal} {\bibinfo  {journal} {Annual Review of Condensed Matter
  Physics}\ }\textbf {\bibinfo {volume} {14}},\ \bibinfo {pages} {335}
  (\bibinfo {year} {2023})}\BibitemShut {NoStop}%
\bibitem [{\citenamefont {Zanardi}\ and\ \citenamefont
  {Anand}(2021)}]{zanardi2021}%
  \BibitemOpen
  \bibfield  {author} {\bibinfo {author} {\bibfnamefont {P.}~\bibnamefont
  {Zanardi}}\ and\ \bibinfo {author} {\bibfnamefont {N.}~\bibnamefont
  {Anand}},\ }\bibfield  {title} {\bibinfo {title} {Information scrambling and
  chaos in open quantum systems},\ }\href
  {https://doi.org/10.1103/PhysRevA.103.062214} {\bibfield  {journal} {\bibinfo
   {journal} {Phys. Rev. A}\ }\textbf {\bibinfo {volume} {103}},\ \bibinfo
  {pages} {062214} (\bibinfo {year} {2021})}\BibitemShut {NoStop}%
\bibitem [{\citenamefont {Yoshimura}\ and\ \citenamefont
  {S\'a}(2026)}]{yoshimuraSa2025dynamics}%
  \BibitemOpen
  \bibfield  {author} {\bibinfo {author} {\bibfnamefont {T.}~\bibnamefont
  {Yoshimura}}\ and\ \bibinfo {author} {\bibfnamefont {L.}~\bibnamefont
  {S\'a}},\ }\bibfield  {title} {\bibinfo {title} {Dynamics of loschmidt echoes
  from operator growth in noisy quantum many-body systems},\ }\href
  {https://doi.org/10.1103/fmr4-14vd} {\bibfield  {journal} {\bibinfo
  {journal} {Phys. Rev. B}\ }\textbf {\bibinfo {volume} {113}},\ \bibinfo
  {pages} {144306} (\bibinfo {year} {2026})}\BibitemShut {NoStop}%
\bibitem [{\citenamefont {Liu}\ \emph {et~al.}(2024)\citenamefont {Liu},
  \citenamefont {Meyer},\ and\ \citenamefont
  {Xian}}]{liu2024lindbladSykOperatorGrowth}%
  \BibitemOpen
  \bibfield  {author} {\bibinfo {author} {\bibfnamefont {J.}~\bibnamefont
  {Liu}}, \bibinfo {author} {\bibfnamefont {R.}~\bibnamefont {Meyer}},\ and\
  \bibinfo {author} {\bibfnamefont {Z.-Y.}\ \bibnamefont {Xian}},\ }\bibfield
  {title} {\bibinfo {title} {Operator size growth in {{Lindbladian SYK}}},\
  }\href {https://doi.org/10.1007/JHEP08(2024)092} {\bibfield  {journal}
  {\bibinfo  {journal} {Journal of High Energy Physics}\ }\textbf {\bibinfo
  {volume} {2024}},\ \bibinfo {pages} {92} (\bibinfo {year}
  {2024})}\BibitemShut {NoStop}%
\bibitem [{\citenamefont {Bhattacharya}\ \emph {et~al.}(2022)\citenamefont
  {Bhattacharya}, \citenamefont {Nandy}, \citenamefont {Nath},\ and\
  \citenamefont {Sahu}}]{bhattacharya2022operator}%
  \BibitemOpen
  \bibfield  {author} {\bibinfo {author} {\bibfnamefont {A.}~\bibnamefont
  {Bhattacharya}}, \bibinfo {author} {\bibfnamefont {P.}~\bibnamefont {Nandy}},
  \bibinfo {author} {\bibfnamefont {P.~P.}\ \bibnamefont {Nath}},\ and\
  \bibinfo {author} {\bibfnamefont {H.}~\bibnamefont {Sahu}},\ }\bibfield
  {title} {\bibinfo {title} {Operator growth and {Krylov} construction in
  dissipative open quantum systems},\ }\bibfield  {journal} {\bibinfo
  {journal} {Journal of High Energy Physics}\ }\textbf {\bibinfo {volume}
  {2022}},\ \href {https://doi.org/10.1007/JHEP12(2022)081}
  {10.1007/JHEP12(2022)081} (\bibinfo {year} {2022})\BibitemShut {NoStop}%
\bibitem [{\citenamefont {Mori}(2024)}]{mori2024liouvilliangap}%
  \BibitemOpen
  \bibfield  {author} {\bibinfo {author} {\bibfnamefont {T.}~\bibnamefont
  {Mori}},\ }\bibfield  {title} {\bibinfo {title} {Liouvillian-gap analysis of
  open quantum many-body systems in the weak dissipation limit},\ }\href
  {https://doi.org/10.1103/PhysRevB.109.064311} {\bibfield  {journal} {\bibinfo
   {journal} {Physical Review B}\ }\textbf {\bibinfo {volume} {109}},\ \bibinfo
  {pages} {064311} (\bibinfo {year} {2024})}\BibitemShut {NoStop}%
\bibitem [{\citenamefont {Jacoby}\ \emph {et~al.}(2025)\citenamefont {Jacoby},
  \citenamefont {Huse},\ and\ \citenamefont
  {Gopalakrishnan}}]{jacoby2025spectralGaps}%
  \BibitemOpen
  \bibfield  {author} {\bibinfo {author} {\bibfnamefont {J.~A.}\ \bibnamefont
  {Jacoby}}, \bibinfo {author} {\bibfnamefont {D.~A.}\ \bibnamefont {Huse}},\
  and\ \bibinfo {author} {\bibfnamefont {S.}~\bibnamefont {Gopalakrishnan}},\
  }\bibfield  {title} {\bibinfo {title} {Spectral gaps of local quantum
  channels in the weak-dissipation limit},\ }\href
  {https://doi.org/10.1103/PhysRevB.111.104303} {\bibfield  {journal} {\bibinfo
   {journal} {Physical Review B}\ }\textbf {\bibinfo {volume} {111}},\ \bibinfo
  {pages} {104303} (\bibinfo {year} {2025})}\BibitemShut {NoStop}%
\bibitem [{\citenamefont {Zhang}\ \emph {et~al.}(2026)\citenamefont {Zhang},
  \citenamefont {Nie},\ and\ \citenamefont
  {Keyserlingk}}]{zhang2025ruellePollicot}%
  \BibitemOpen
  \bibfield  {author} {\bibinfo {author} {\bibfnamefont {C.}~\bibnamefont
  {Zhang}}, \bibinfo {author} {\bibfnamefont {L.}~\bibnamefont {Nie}},\ and\
  \bibinfo {author} {\bibfnamefont {C.~v.}\ \bibnamefont {Keyserlingk}},\
  }\bibfield  {title} {\bibinfo {title} {Thermalization rates and quantum
  ruelle-pollicott resonances: Insights from operator hydrodynamics},\ }\href
  {https://doi.org/10.1103/r8qy-dcgq} {\bibfield  {journal} {\bibinfo
  {journal} {Phys. Rev. Res.}\ } (\bibinfo {year} {2026})}\BibitemShut
  {NoStop}%
\bibitem [{\citenamefont {Lieb}\ and\ \citenamefont
  {Robinson}(1972)}]{lieb1972finite}%
  \BibitemOpen
  \bibfield  {author} {\bibinfo {author} {\bibfnamefont {E.~H.}\ \bibnamefont
  {Lieb}}\ and\ \bibinfo {author} {\bibfnamefont {D.~W.}\ \bibnamefont
  {Robinson}},\ }\bibfield  {title} {\bibinfo {title} {The finite group
  velocity of quantum spin systems},\ }\href@noop {} {\bibfield  {journal}
  {\bibinfo  {journal} {Communications in mathematical physics}\ }\textbf
  {\bibinfo {volume} {28}},\ \bibinfo {pages} {251} (\bibinfo {year}
  {1972})}\BibitemShut {NoStop}%
\bibitem [{\citenamefont {Jonay}\ \emph {et~al.}(2018)\citenamefont {Jonay},
  \citenamefont {Huse},\ and\ \citenamefont
  {Nahum}}]{jonay2018operatorentangle}%
  \BibitemOpen
  \bibfield  {author} {\bibinfo {author} {\bibfnamefont {C.}~\bibnamefont
  {Jonay}}, \bibinfo {author} {\bibfnamefont {D.~A.}\ \bibnamefont {Huse}},\
  and\ \bibinfo {author} {\bibfnamefont {A.}~\bibnamefont {Nahum}},\ }\href
  {https://arxiv.org/abs/1803.00089} {\bibinfo {title} {Coarse-grained dynamics
  of operator and state entanglement}} (\bibinfo {year} {2018}),\ \Eprint
  {https://arxiv.org/abs/1803.00089} {arXiv:1803.00089 [cond-mat.stat-mech]}
  \BibitemShut {NoStop}%
\bibitem [{\citenamefont {Rakovszky}\ \emph {et~al.}(2018)\citenamefont
  {Rakovszky}, \citenamefont {Pollmann},\ and\ \citenamefont {von
  Keyserlingk}}]{rakovszky2018conserved}%
  \BibitemOpen
  \bibfield  {author} {\bibinfo {author} {\bibfnamefont {T.}~\bibnamefont
  {Rakovszky}}, \bibinfo {author} {\bibfnamefont {F.}~\bibnamefont
  {Pollmann}},\ and\ \bibinfo {author} {\bibfnamefont {C.~W.}\ \bibnamefont
  {von Keyserlingk}},\ }\bibfield  {title} {\bibinfo {title} {Diffusive
  hydrodynamics of out-of-time-ordered correlators with charge conservation},\
  }\href {https://doi.org/10.1103/PhysRevX.8.031058} {\bibfield  {journal}
  {\bibinfo  {journal} {Phys. Rev. X}\ }\textbf {\bibinfo {volume} {8}},\
  \bibinfo {pages} {031058} (\bibinfo {year} {2018})}\BibitemShut {NoStop}%
\bibitem [{\citenamefont {Swingle}\ and\ \citenamefont
  {Yunger~Halpern}(2018)}]{swingle2018scramblingmeasurements}%
  \BibitemOpen
  \bibfield  {author} {\bibinfo {author} {\bibfnamefont {B.}~\bibnamefont
  {Swingle}}\ and\ \bibinfo {author} {\bibfnamefont {N.}~\bibnamefont
  {Yunger~Halpern}},\ }\bibfield  {title} {\bibinfo {title} {Resilience of
  scrambling measurements},\ }\href
  {https://doi.org/10.1103/PhysRevA.97.062113} {\bibfield  {journal} {\bibinfo
  {journal} {Phys. Rev. A}\ }\textbf {\bibinfo {volume} {97}},\ \bibinfo
  {pages} {062113} (\bibinfo {year} {2018})}\BibitemShut {NoStop}%
\bibitem [{\citenamefont {Martinez-Azcona}\ \emph {et~al.}(2023)\citenamefont
  {Martinez-Azcona}, \citenamefont {Kundu}, \citenamefont {del Campo},\ and\
  \citenamefont {Chenu}}]{martinez2023opertorvariance}%
  \BibitemOpen
  \bibfield  {author} {\bibinfo {author} {\bibfnamefont {P.}~\bibnamefont
  {Martinez-Azcona}}, \bibinfo {author} {\bibfnamefont {A.}~\bibnamefont
  {Kundu}}, \bibinfo {author} {\bibfnamefont {A.}~\bibnamefont {del Campo}},\
  and\ \bibinfo {author} {\bibfnamefont {A.}~\bibnamefont {Chenu}},\ }\bibfield
   {title} {\bibinfo {title} {Stochastic operator variance: An observable to
  diagnose noise and scrambling},\ }\href
  {https://doi.org/10.1103/PhysRevLett.131.160202} {\bibfield  {journal}
  {\bibinfo  {journal} {Phys. Rev. Lett.}\ }\textbf {\bibinfo {volume} {131}},\
  \bibinfo {pages} {160202} (\bibinfo {year} {2023})}\BibitemShut {NoStop}%
\bibitem [{\citenamefont {Liu}\ \emph {et~al.}(2023)\citenamefont {Liu},
  \citenamefont {Tang},\ and\ \citenamefont {Zhai}}]{liu2023krylovopen}%
  \BibitemOpen
  \bibfield  {author} {\bibinfo {author} {\bibfnamefont {C.}~\bibnamefont
  {Liu}}, \bibinfo {author} {\bibfnamefont {H.}~\bibnamefont {Tang}},\ and\
  \bibinfo {author} {\bibfnamefont {H.}~\bibnamefont {Zhai}},\ }\bibfield
  {title} {\bibinfo {title} {Krylov complexity in open quantum systems},\
  }\href {https://doi.org/10.1103/PhysRevResearch.5.033085} {\bibfield
  {journal} {\bibinfo  {journal} {Phys. Rev. Res.}\ }\textbf {\bibinfo {volume}
  {5}},\ \bibinfo {pages} {033085} (\bibinfo {year} {2023})}\BibitemShut
  {NoStop}%
\bibitem [{\citenamefont {Srivatsa}\ and\ \citenamefont {von
  Keyserlingk}(2024)}]{srivatsa2024growthhypothesis}%
  \BibitemOpen
  \bibfield  {author} {\bibinfo {author} {\bibfnamefont {N.~S.}\ \bibnamefont
  {Srivatsa}}\ and\ \bibinfo {author} {\bibfnamefont {C.}~\bibnamefont {von
  Keyserlingk}},\ }\bibfield  {title} {\bibinfo {title} {Operator growth
  hypothesis in open quantum systems},\ }\href
  {https://doi.org/10.1103/PhysRevB.109.125149} {\bibfield  {journal} {\bibinfo
   {journal} {Phys. Rev. B}\ }\textbf {\bibinfo {volume} {109}},\ \bibinfo
  {pages} {125149} (\bibinfo {year} {2024})}\BibitemShut {NoStop}%
\bibitem [{\citenamefont {Bhattacharya}\ \emph {et~al.}(2023)\citenamefont
  {Bhattacharya}, \citenamefont {Nandy}, \citenamefont {Nath},\ and\
  \citenamefont {Sahu}}]{bhattacharya2023biLanczos}%
  \BibitemOpen
  \bibfield  {author} {\bibinfo {author} {\bibfnamefont {A.}~\bibnamefont
  {Bhattacharya}}, \bibinfo {author} {\bibfnamefont {P.}~\bibnamefont {Nandy}},
  \bibinfo {author} {\bibfnamefont {P.~P.}\ \bibnamefont {Nath}},\ and\
  \bibinfo {author} {\bibfnamefont {H.}~\bibnamefont {Sahu}},\ }\bibfield
  {title} {\bibinfo {title} {On {Krylov} complexity in open systems: an
  approach via bi-{Lanczos} algorithm},\ }\bibfield  {journal} {\bibinfo
  {journal} {Journal of High Energy Physics}\ }\textbf {\bibinfo {volume}
  {2023}},\ \href {https://doi.org/10.1007/JHEP12(2023)066}
  {10.1007/JHEP12(2023)066} (\bibinfo {year} {2023})\BibitemShut {NoStop}%
\bibitem [{\citenamefont {Bhattacharjee}\ \emph {et~al.}(2024)\citenamefont
  {Bhattacharjee}, \citenamefont {Nandy},\ and\ \citenamefont
  {Pathak}}]{bhattacharjee2024perspective}%
  \BibitemOpen
  \bibfield  {author} {\bibinfo {author} {\bibfnamefont {B.}~\bibnamefont
  {Bhattacharjee}}, \bibinfo {author} {\bibfnamefont {P.}~\bibnamefont
  {Nandy}},\ and\ \bibinfo {author} {\bibfnamefont {T.}~\bibnamefont
  {Pathak}},\ }\bibfield  {title} {\bibinfo {title} {Operator dynamics in
  {Lindbladian} {SYK}: a {Krylov} complexity perspective},\ }\bibfield
  {journal} {\bibinfo  {journal} {Journal of High Energy Physics}\ }\textbf
  {\bibinfo {volume} {2024}},\ \href {https://doi.org/10.1007/JHEP01(2024)094}
  {10.1007/JHEP01(2024)094} (\bibinfo {year} {2024})\BibitemShut {NoStop}%
\bibitem [{\citenamefont {Carolan}\ \emph {et~al.}(2024)\citenamefont
  {Carolan}, \citenamefont {Kiely}, \citenamefont {Campbell},\ and\
  \citenamefont {Deffner}}]{carolan2024operator}%
  \BibitemOpen
  \bibfield  {author} {\bibinfo {author} {\bibfnamefont {E.}~\bibnamefont
  {Carolan}}, \bibinfo {author} {\bibfnamefont {A.}~\bibnamefont {Kiely}},
  \bibinfo {author} {\bibfnamefont {S.}~\bibnamefont {Campbell}},\ and\
  \bibinfo {author} {\bibfnamefont {S.}~\bibnamefont {Deffner}},\ }\bibfield
  {title} {\bibinfo {title} {Operator growth and spread complexity in open
  quantum systems},\ }\href {https://doi.org/10.1209/0295-5075/ad5b17}
  {\bibfield  {journal} {\bibinfo  {journal} {Europhysics Letters}\ }\textbf
  {\bibinfo {volume} {147}},\ \bibinfo {pages} {38002} (\bibinfo {year}
  {2024})}\BibitemShut {NoStop}%
\bibitem [{\citenamefont {Lee}\ \emph {et~al.}(2024)\citenamefont {Lee},
  \citenamefont {Oh}, \citenamefont {Wong}, \citenamefont {Chen},\ and\
  \citenamefont {Jiang}}]{leeJiang2024cmi}%
  \BibitemOpen
  \bibfield  {author} {\bibinfo {author} {\bibfnamefont {S.-u.}\ \bibnamefont
  {Lee}}, \bibinfo {author} {\bibfnamefont {C.}~\bibnamefont {Oh}}, \bibinfo
  {author} {\bibfnamefont {Y.}~\bibnamefont {Wong}}, \bibinfo {author}
  {\bibfnamefont {S.}~\bibnamefont {Chen}},\ and\ \bibinfo {author}
  {\bibfnamefont {L.}~\bibnamefont {Jiang}},\ }\bibfield  {title} {\bibinfo
  {title} {Universal spreading of conditional mutual information in noisy
  random circuits},\ }\href {https://doi.org/10.1103/PhysRevLett.133.200402}
  {\bibfield  {journal} {\bibinfo  {journal} {Phys. Rev. Lett.}\ }\textbf
  {\bibinfo {volume} {133}},\ \bibinfo {pages} {200402} (\bibinfo {year}
  {2024})}\BibitemShut {NoStop}%
\bibitem [{\citenamefont {Zhang}\ and\ \citenamefont
  {Gopalakrishnan}(2024)}]{gopalakrishnan2024cmi}%
  \BibitemOpen
  \bibfield  {author} {\bibinfo {author} {\bibfnamefont {Y.}~\bibnamefont
  {Zhang}}\ and\ \bibinfo {author} {\bibfnamefont {S.}~\bibnamefont
  {Gopalakrishnan}},\ }\bibfield  {title} {\bibinfo {title} {Nonlocal growth of
  quantum conditional mutual information under decoherence},\ }\href
  {https://doi.org/10.1103/PhysRevA.110.032426} {\bibfield  {journal} {\bibinfo
   {journal} {Phys. Rev. A}\ }\textbf {\bibinfo {volume} {110}},\ \bibinfo
  {pages} {032426} (\bibinfo {year} {2024})}\BibitemShut {NoStop}%
\bibitem [{\citenamefont {Touil}\ and\ \citenamefont
  {Deffner}(2021)}]{touil2021entropysinks}%
  \BibitemOpen
  \bibfield  {author} {\bibinfo {author} {\bibfnamefont {A.}~\bibnamefont
  {Touil}}\ and\ \bibinfo {author} {\bibfnamefont {S.}~\bibnamefont
  {Deffner}},\ }\bibfield  {title} {\bibinfo {title} {Information scrambling
  versus decoherence---two competing sinks for entropy},\ }\href
  {https://doi.org/10.1103/PRXQuantum.2.010306} {\bibfield  {journal} {\bibinfo
   {journal} {PRX Quantum}\ }\textbf {\bibinfo {volume} {2}},\ \bibinfo {pages}
  {010306} (\bibinfo {year} {2021})}\BibitemShut {NoStop}%
\bibitem [{\citenamefont {Cipolloni}\ and\ \citenamefont
  {{Kudler-Flam}}(2023)}]{cipolloni2023ginibreEvecEntanglement}%
  \BibitemOpen
  \bibfield  {author} {\bibinfo {author} {\bibfnamefont {G.}~\bibnamefont
  {Cipolloni}}\ and\ \bibinfo {author} {\bibfnamefont {J.}~\bibnamefont
  {{Kudler-Flam}}},\ }\bibfield  {title} {\bibinfo {title} {Entanglement
  {{Entropy}} of {{Non-Hermitian Eigenstates}} and the {{Ginibre Ensemble}}},\
  }\href {https://doi.org/10.1103/PhysRevLett.130.010401} {\bibfield  {journal}
  {\bibinfo  {journal} {Physical Review Letters}\ }\textbf {\bibinfo {volume}
  {130}},\ \bibinfo {pages} {010401} (\bibinfo {year} {2023})}\BibitemShut
  {NoStop}%
\bibitem [{\citenamefont {Shivam}\ \emph {et~al.}(2023)\citenamefont {Shivam},
  \citenamefont {De~Luca}, \citenamefont {Huse},\ and\ \citenamefont
  {Chan}}]{Shivam2024}%
  \BibitemOpen
  \bibfield  {author} {\bibinfo {author} {\bibfnamefont {S.}~\bibnamefont
  {Shivam}}, \bibinfo {author} {\bibfnamefont {A.}~\bibnamefont {De~Luca}},
  \bibinfo {author} {\bibfnamefont {D.~A.}\ \bibnamefont {Huse}},\ and\
  \bibinfo {author} {\bibfnamefont {A.}~\bibnamefont {Chan}},\ }\bibfield
  {title} {\bibinfo {title} {Many-body quantum chaos and emergence of ginibre
  ensemble},\ }\href {https://doi.org/10.1103/PhysRevLett.130.140403}
  {\bibfield  {journal} {\bibinfo  {journal} {Phys. Rev. Lett.}\ }\textbf
  {\bibinfo {volume} {130}},\ \bibinfo {pages} {140403} (\bibinfo {year}
  {2023})}\BibitemShut {NoStop}%
\bibitem [{\citenamefont {Chirame}\ and\ \citenamefont
  {Burnell}(2026)}]{zenodoRMT}%
  \BibitemOpen
  \bibfield  {author} {\bibinfo {author} {\bibfnamefont {S.}~\bibnamefont
  {Chirame}}\ and\ \bibinfo {author} {\bibfnamefont {F.~J.}\ \bibnamefont
  {Burnell}},\ }\href {https://doi.org/10.5281/zenodo.20486041} {\bibinfo
  {title} {Simulation data for "open-system dynamics in local lindbladians with
  chaotic spectra"}} (\bibinfo {year} {2026})\BibitemShut {NoStop}%
\end{thebibliography}%

\end{document}